\documentclass[11pt,a4paper]{article}

\usepackage{ascmac}
\usepackage{amsmath}
\usepackage{amssymb}
\usepackage{bm}
\usepackage[footnotesize,bf]{caption}
\usepackage{fullpage}
\usepackage{color}
\usepackage{float}
\usepackage{graphicx}
\usepackage[hidelinks]{hyperref} 
\usepackage[]{natbib}
\usepackage{subfigure}
\usepackage{txfonts}
\usepackage{threeparttable}
\usepackage{wrapfig}
\usepackage[]{multicol}
\usepackage{wrapfig}
\usepackage{authblk}

\usepackage{floatflt}

\title{\large Complete list of the ASTRO-H Science Working Group}
\date{\vspace{-0.5cm}}
\newcommand{\MakeWhitePaperTitle}{
	\begin{center}
		\begin{figure}
			\vspace{1cm}
			\begin{center}
				\includegraphics[width=0.2\hsize]{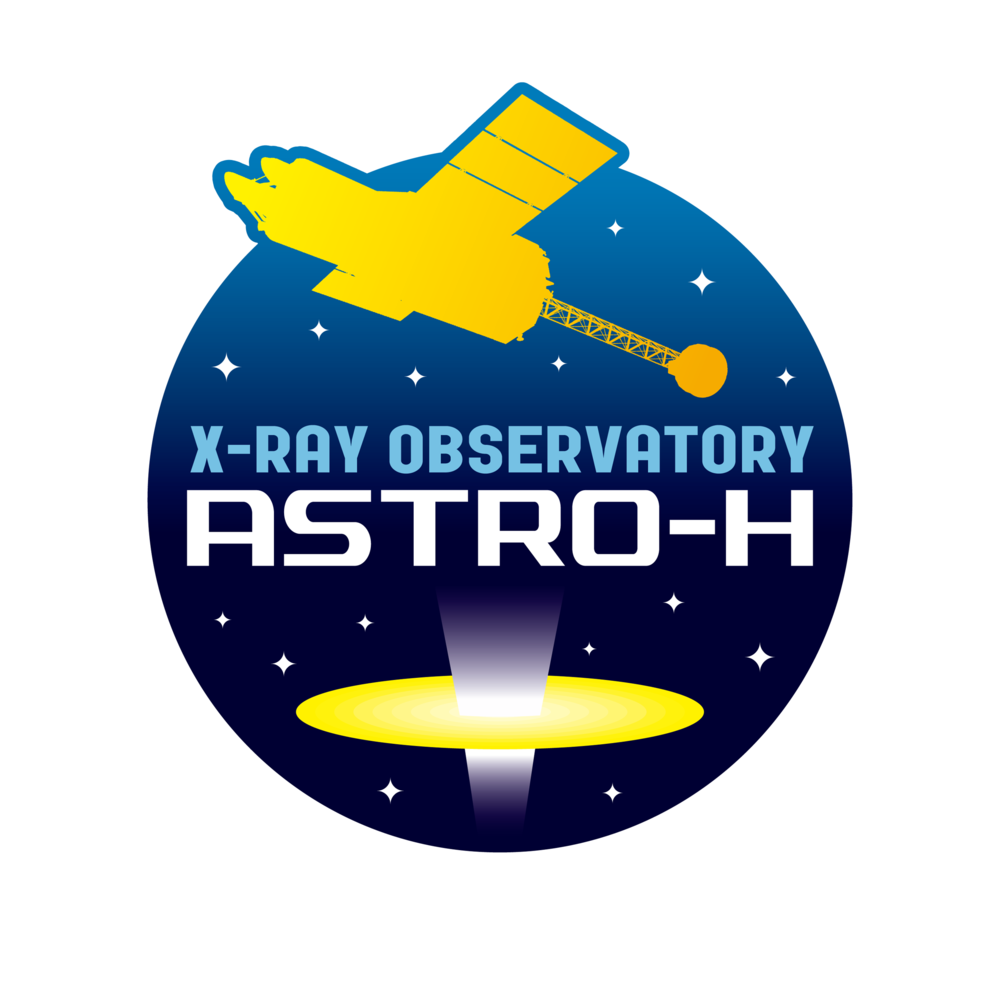}
			\end{center}
		\end{figure}
		\vspace{1cm}
		{\LARGE
		ASTRO-H Space X-ray Observatory\\
		White Paper\\
		}
		\vspace{5mm}
		{\large
		\WhitePaperTitle\\
		}
		\vspace{1cm}
		{
		\WhitePaperAuthors\\
		on behalf of the ASTRO-H Science Working Group
		}
	\end{center}
}

\usepackage{authblk}
\author[a]{Tadayuki~Takahashi}
\author[a]{Kazuhisa~Mitsuda}
\author[b]{Richard~Kelley}
\author[c]{Felix~Aharonian}
\author[d]{Hiroki~Akamatsu}
\author[e]{Fumie~Akimoto}
\author[f]{Steve~Allen}
\author[g]{Naohisa~Anabuki}
\author[b]{Lorella~Angelini}
\author[h]{Keith~Arnaud}
% ...
\affil[a]{Institute of Space and Astronautical Science (ISAS), Japan Aerospace Exploration Agency (JAXA), Kanagawa 252-5210, Japan}
\affil[b]{NASA/Goddard Space Flight Center, MD 20771, USA}
\affil[c]{Astronomy and Astrophysics Section, Dublin Institute for Advanced Studies, Dublin 2, Ireland}
\affil[d]{SRON Netherlands Institute for Space Research, Utrecht, The Netherlands}
\affil[e]{Department of Physics, Nagoya University, Aichi 338-8570, Japan}
\affil[f]{Kavli Institute for Particle Astrophysics and Cosmology, Stanford University, CA 94305, USA}
\affil[g]{Department of Earth and Space Science, Osaka University, Osaka 560-0043, Japan}
\affil[h]{Department of Astronomy, University of Maryland, MD 20742, USA}
% ...

\begin{document}

\newcommand{\WhitePaperTitle}{Stellar-Mass Black Holes}
\newcommand{\WhitePaperAuthors}{
	J.~M.~Miller~(University~of~Michigan), S.~Mineshige~(Kyoto~University),
	A.~Kubota~(Shibaura~Inst.~of~Tech.), S.~Yamada~(Tokyo~Metropolitan~University),
	F.~Aharonian~(MPI, Heidelberg), C.~Done~(University~of~Durham), 
	N.~Kawai~(Tokyo~Institute~of~Technology), 
	K.~Hayashida~(Osaka~University), R.~Reis~(University~of~Michigan), 
	T.~Mizuno~(Hiroshima~University), H.~Noda~(RIKEN),
	Y.~Ueda~(Kyoto~University), and M.~Shidatsu~(Kyoto~University)
}
\MakeWhitePaperTitle

\def\physrep{Phys. Rep. }
\def\apj{ApJ}
\def\mnras{MNRAS}
\def\nat{Nat}
\def\physrevB{Phys. Rev. B}
\def\araa{ARA\&A}                
\def\aap{A\&A}                   
\def\aapr{A\&AR}                   
\def\aaps{A\&AS}                 
\def\aj{AJ}                      
\def\apjs{ApJS}                  
\def\pasp{PASP}                  
\def\apjl{ApJ}                   
\def\pasj{PASJ}
\def\araa{ARA\&A}                
\def\gca{Geochimica et Cosmochimica Acta}
\def\ssr{Space Science Reviews}
\def\minus{-}

\begin{abstract}
Thanks to extensive observations with X-ray missions and facilities
working in other wavelengths, as well as rapidly--advancing numerical
simulations of accretion flows, our knowledge of
astrophysical black holes has been remarkably enriched.  Rapid
progress has opened new areas of enquiry, including measurements of
black hole spin, the properties and driving mechanisms of jets and
disk winds, the impact of feedback into local environments, the origin
of periodic and aperiodic X-ray variations, and the nature of
super-Eddington accretion flows, among others.  The goal of this White
Paper is to illustrate how {\it ASTRO-H} can make dramatic progress in the
study of astrophysical black holes, particularly the study of black
hole X-ray binaries.
\end{abstract}

\maketitle
\clearpage

\tableofcontents
\clearpage

\clearpage

\section{Introduction}

Black hole X-ray binaries are laboratories for the study of
fundamental physics, including strong gravitation, the nature of gas
accretion and ejection over many orders of magnitude in accretion rate,
and feedback between black holes and their host environments.  Though
the known population of these sources in the Milky Way and Magellanic
Clouds is much smaller than the population of massive black holes that
power active galactic nuclei (AGN), the proximity of Galactic X-ray
binaries enables very sensitive observations.  Moreover, the short
time scales intrinsic to X-ray binaries make it possible to study the
evolution of accretion flows into distinct phases or ``states'' that
may connect with and explain different AGN classes.

Studies of black hole X-ray binaries with current and recent missions,
including, e.g., {\it RXTE}, {\it Chandra}, {\it XMM-Newton}, {\it Swift}, and {\it Suzaku}, have
made enormous progress.  Observations have begun to constrain the
angular momenta or ``spin'' of some black holes
\citep[e.g.][]{miller02j1650, miller09spin, hiemstra1652,
  reisspin, reis1752, reismaxi, shafee06, mcclintock06, Steiner2011},
and to explore connections between accretion disks and jets
\citep[e.g.][]{fenderetal04}. Some spectra have hinted at the very
nature of disk accretion itself \citep[e.g.][]{Miller06Natur,
  Miller2008j1655wind}, while others point to complexity or
structure in hard X-ray coronae
\citep[e.g.][]{makishimacygx108}. Ionized X-ray disk winds have been
discovered; these winds may bear close analogy to the X-ray warm
absorbers observed in Seyfert AGN \citep{KingMiller2013}.  At the same
time, robust detections of quasi-periodic oscillations have finally
been obtained from massive black holes \citep{rej1034qpo,
  Reis2012qpo}.  ``Ultra-luminous" X-ray sources have been studied in
some detail; some modeling suggests that ultra-luminous X-ray sources
(ULXs) are examples super-Eddington accretion
\citep[e.g.][]{Gladstone09ULS, 2006PASJ...58..915V}, while others treatments suggest that a
subset of the most extreme ULXs may harbor intermediate-mass black
holes \citep[e.g.][]{2001ApJ...547L..25M,ulxqpo22003,
  Farrell09Natur, StrohmayerMushotzky2009, Sutton2012ulx}.

It is particularly noteworthy that many of these advances have bridged
the mass scale, revealing a host of phenomena to actually be present
in both stellar-mass black holes and AGN.  It is also fortuitous that
much of this observational progress has happened concurrently with
rapid advances in numerical simulations of accretion flows
 \citep[e.g.][]{OhsugaMineshige11, ONeillreynolds2011, Schnittman2012states, McKinneyqpo2012}.  In
this sense, the last several years have been something of a golden
moment in the study of black hole X-ray binaries.  However, progress
on both observational and theoretical fronts has opened new areas of
enquiry, and highlighted the need to make further progress in some
emerging and established areas.  A necessarily incomplete list of
developing questions might include:

\begin{itemize}
\parskip=-0.2cm
\item What is the distribution of stellar-mass black hole spin parameters?
\item Can systematic errors associated with spin constraints be reduced?
\item What are the processes by which disk winds and jets are powered?
\item How much mass and energy do winds and jets carry away to which direction?
\item What is the geometry of the hard X-ray corona, and what emission
  mechanisms power hard X-ray production?
\item Which emission mechanisms dominate at low Eddington fraction?
\item Is gas bound to black holes at very low Eddington fraction?
\item How can we understand the origin of the complex X-ray variability?
\item What does super-Eddington accretion look like?
\end{itemize}
\parskip=0.0cm
In each of these cases, {\it ASTRO-H}  \citep{astroh2012} will be able
to make progress or make revolutionary observations.  Its bandpass,
sensitivity, and instrument complement are well suited to addressing
each of these problems.  The impact of these advances will be
registered in many distinct areas of astrophysics.  The goal of this
White Paper is to explain how the instrumentation aboard {\it ASTRO-H} -
coupled with carefully-planned observations - can make dramatic
progress in the study of black hole X-ray binaries in particular,
and black hole accretion generally.

\section{Black hole spin}
\subsection{Current Black Hole Spin Measurements}

X-ray measurements of black hole spin amount to explorations of some
of the most fundamental predictions of General Relativity.  Indeed,
X-ray measurements and sub-mm VLBI imaging \citep{Doeleman2008Natur,Doeleman2012Sci} 
are currently the only two means of probing gravitation in the strong
field limit, and the latter will only work in two cases (Sgr. A* and
M87).  Even binary radio pulsars are typically separated by large
multiples of $GM/c^{2}$; in contrast, the accretion disk around a
maximally-spinning "Kerr" black hole can extend to just $\sim
GM/c^{2}$.  Apart from revealing a fundamental physical theory, black
hole spin measurements likely hold the keys to understanding the birth
and evolution of black holes \citep{millersn11}, and
the processes by which relativistic jets are launched (\citealt{NarayanMcClintock2012, Steiner2013spinjet}; however see \citealt{fender2010jets,RussellGallofender2013jets,KingMiller2013}).

Affecting a large change to the spin of a black hole requires a doubling
of its mass.  The spin of a black hole can therefore reveal
the nature of its formation and evolution.  In the case of massive
black holes in galaxy centers, the spin of the central black hole is
determined by accretion and by mergers.  In general, these processes
likely work in opposite directions, with accretion increasing the spin
of the hole, and black hole mergers reducing the spin as spin vectors
are not typically aligned \citep[see, e.g.,][]{BertiVolonteri2008}.  In
contrast, a stellar-mass black hole with a low-mass companion star
cannot double its mass, and a high-mass companion does not live long
enough for mass doubling (nor is it clear that the accretion is not
chaotic).  This means that the spin of stellar-mass black holes is a
rare glimpse into the inner workings of supernovae and/or gamma-ray
bursts \citep[e.g.][]{gammie04}.

Studies of massive black holes - particularly those at the center of
galaxy clusters - have shown that relativistic jets are able to
strongly influence large scale structure \citep[e.g.][]{FabianSanders06cluster}.
The cavities blown by jets can be used as coarse bolometers to trace
the power of the jet \citep[e.g.][]{Allen06}, which is otherwise
difficult since radio luminosity is not a precise
trace of jet power \citep[e.g.][]{2007MNRAS.381..589M}.  The power requirements implied by most extreme
cavities require tapping the spin energy of a near-maximal Kerr black
hole with a very high mass \citep{McNamara2009}.  Of course, this
is consistent with the predictions of \citet{BlandfordZnajek1977}, who
showed that black hole spin energy could be tapped to power jets
through magnetic connections to the ergosphere.  Note, however, that recent theoretical work points to the importance of magnetic flux, not just spin \citep{2013ApJ...764L..24S}; moreover, a broad range of spins may be difficult to reproduce in current cosmological simulations \citep{2013ApJ...775...94V}.
Again, though, the
timescales natural to massive black holes can complicate efforts to
connect jet power and spin, whereas stellar-mass black holes may
enable an answer to the problem of jet production.

All current means of measuring the spin of a black hole actually
measure the inner radius of the accretion disk.  The measurements rely
on the assumption that the disk is truncated at the innermost stable
circular orbit, or ISCO, which is set by General Relativity and
sensitive to the spin parameter of the black hole \citep[see e.g.][]{Bardeenetal1972}. 
 In practice, this amounts to an assumption
that there is a marked contrast in the emissivity of the plunging gas
within the ISCO, relative to gas on stable orbits in the disk outside
of the ISCO.  There is no observational means of testing this
assumption, but the latest numerical simulations specifically aimed at
examining this assumption (those that analyze a large $\phi$ angle and
numerous orbital timescales) suggest that the assumption is robust as
long as radiative cooling is efficient \citep{Shafee2008, NobleKrolik09, reynoldsfabian08}.  
The assumption of a disk that remains
at the ISCO must break down at a low fraction of the Eddington
luminosity ($L_{\rm Edd}$).  However, the exact Eddington fraction at
which this occurs is not yet clear.  Some results suggest that the
disk remains close to the ISCO for $L \geq 0.001~L_{\rm Edd}$ and recedes
at lower luminosity \citep[e.g.][]{tomsick09gx, reislhs, reis1752}, while other results suggest that the
disk may recede at a higher Eddington fraction \citep[e.g.][]{makishimacygx108, takahashi165508, Shidatsu2011lhs}.

There are currently three viable means of constraining or measuring
the spin parameter of a black hole: quasi-periodic oscillations
(QPOs), thermal continuum emission from the disk, and atomic emission
and absorption features due to disk reflection.  The QPOs that are
most likely to reveal spin are the so-called high-frequency QPOs
($few\times 100$~Hz, and sometimes seen in 2:3 frequency ratios,
e.g. Strohmayer 2000).  The detection of such QPOs may be within the
capability of the {\it ASTRO-H} HXI instrument, but their detection has
relied upon high-cadence monitoring of an outburst and careful data
screening.  Such studies are really the domain of proposed future
missions such as Athena+ and WF-MAXI.  Another problem with the QPOs is that there
are many different models that point to different radii and spin, and
there is no consensus on which model is correct.  For these reasons,
QPOs lie beyond the scope of this White Paper.  It may also be
possible to constrain or measure black hole spin using X-ray
polarization \citep[e.g.][]{Dovciak08polorization,
  SchnittmanKrolik09polarization}, but this technique has not yet been
attempted, and it is also beyond the scope of this White Paper.

Measurements of spin using thermal continuum emission from the
accretion disk have the advantage of utilizing a major component of
the flux seen in stellar-mass black holes.  Said differently, the
potential signal is ample.  When the distance, mass, and inner disk
inclination to a source are known, and when applied in phases where
the disk emission is strongly dominant over non-thermal emission, new
models enable the measurement of the black hole spin parameter \citep[see,][]{shafee06, mcclintock06}.  In the past,
measurements of distance have typically carried large fractional
errors, but parallax techniques - made more powerful with sensitivity
upgrades to radio telescopes such as MERLIN and the VLA - hold much
promise \citep[e.g][]{MillerJonesv404cyg2009}.  A remaining difficulty is
that the inclination of the inner disk need not be the same as that of
the outer disk, and the alignment timescale may be quite long
\citep[e.g.][]{Maccarone2002misalignment}.  However, this too may be remedied with radio
observations: in cases where jet emission may be resolved, the axis of
the jet may be taken as normal to the plane of the inner disk
(e.g. \citealt{HjellmingRupen1995j1655}).  Last, continuum-derived spins rely on
correctly encapsulating the effects of scattering in the disk
atmosphere; currently, this is done using an overall multiplicative
constant that corrects the disk flux and temperature \citep{ShimuraTakahara1995,merlonifabianross00}.

The basic theory of X-ray disk ``reflection" was rapidly developed
after the detection of a broad Fe K line in Cygnus X-1 that could
potentially have originated in the inner disk (e.g. \citealt{Barrwhitepage85, George91, pexrav}).  A
broad, sometimes skewed Fe line is merely the most distinctive part of
a disk reflection spectrum that is now understood to likely include a
forest of low-energy atomic emission lines, the Fe line, and the
Compton back-scattering hump peaking in the 20-30~keV range (for
reviews, see \citealt{reynoldsnowak03, miller07review}).  The {\it ASCA} mission
flew CCDs capable of handling modest fluxes, leading to the detection
of skewed disk lines in many Seyfert AGN \citep{tanaka1995}.  It
was not until CCDs (plus or minus gratings) capable of handling high
count rates aboard {\it Chandra}, {\it XMM-Newton}, and {\it Suzaku}
were flown that the prevalence of relativistic disk lines in
stellar-mass black holes became clear.  Arguably, the first strong
constraint or measurement of black hole spin using disk reflection was
obtained in 2002 with {\it XMM-Newton} \citep{miller02j1650}.  Such
lines are now observed in neutron star X-ray binaries as well
(e.g. \citealt{bhattacharyya07, cackett08, cackett10, Papitto09}).

Disk reflection occurs in the atmosphere of an accretion disk.  Models
typically assume either a constant gas density, or an atmosphere in
vertical hydrostatic equilibrium (e.g. \citealt{NayakshinKallman2001}).
Most models only include external irradiation, but at least one new
model includes X-ray emission from the midplane (e.g. \citealt{refbhb}).
  Observations do not appear to be sensitive to such variations
\citep{reisgx}; this may be consistent with a scenario in which
magnetic pressure sets a constant gas density \citep{magneticpressureblaes2006} and
renders midplane emission relatively unimportant.  The observed
reflection spectrum depends strongly on the ionization of the disk,
enabling robust measurements of this parameter (this is a more careful
treatment of the hardening factor that is also important when
measuring spins via the disk continuum).  Reflection models are
calculated in the fluid frame and must be convolved with a
relativistic smearing function to match observations.  Smearing models
are based on ray-tracing, and are fairly robust \citep{BeckwithDone2004}.  
Indeed, these models are even able to measure the inclination
of the inner disk owing to the strong influence of viewing angle on
line profiles.  The emissivity of the accretion disk, $J(r) \propto
r^{-q}$, is important to measuring spin accurately, but this too can
be measured using smearing models.  New theoretical efforts appear to
be making firm predictions concerning emissivity profiles \citep{wilkins2012}, 
and both quasar microlensing observations \citep[e.g.][]{sizepg1115, size1104, ChartasKochanek2012quasar, size2237} and X-ray reverberation studies in Seyferts \citep[e.g.][]{FabZog09, Fabian2012iras13224, Zoghbi2012lag, Cackett2012ESOlag}  imply {\it very} compact black hole hard X-ray coronae through measurements of very short lag time scales.

{\it ASTRO-H} is arguably best-suited to improving the measurement of spin
via disk reflection spectra.  The largest advantage to this technique
is that the measurement is a relative one: the {\it degree} to which
Special Relativity (broadening lines and enhancing blue wings through
beaming) and General Relativity (creating extended red wings as
photons lose energy escaping from a very deep potential) shape the
spectrum forms the basis of the measurement.  Whereas the absolute
flux of a spectral feature can be difficult to measure (since
calibration enters), the width of an atomic line is arguably easier to
measure.  And whereas an absolute measurement requires the mass and
distance of the black hole to be known in advance, the mass and
distance to the black hole are not required for reflection-derived
spins.  A secondary advantage of disk reflection is that it is
readily observed in both massive black holes and stellar-mass black
holes, enabling comparisons across the mass scale.

Employing modest quality metrics, 14 stellar-mass black hole spins
have been measured using relativistic disk reflection techniques, and
7 have been measured using the disk continuum \citep[see][]{miller09spin, millersn11, McClintock2011review}.  
Although the number measured using each method is
small, the peaks of the spin distributions from the independent
techniques agree and point to high spin.  The fact that independent
methods with partially independent systematics are arriving at a
commensurate peak spin value may point to a situation where
astrophysics dominates over systematic errors.  Taken literally, and
when compared to the implied natal spins of neutron stars, these
results imply that stellar-mass black holes could be born in gamma-ray
bursts rather than in typical supernovae \citep{millersn11}. 
 Of course, more spin measurements are urgently required in
order to increase the statistics in each distribution, and to better
sample the wings of each distribution.

Currently, there is no compelling evidence in favor of black hole spin
as the source of jet power, though the utility of stellar-mass black
holes for such purposes is increasingly clear.  Claims of a
correlation between spin and jet power have been made \cite{NarayanMcClintock2012},
 but these appear to be statistically insignificant,
and/or at least partially driven by data selection \citep{RussellGallofender2013jets}. 
 Weak correlations, with slightly better statistical
significance, may be found when comparing spin and jet power across
the black hole mass scale (\citealt{KingMiller2013}).  Here especially,
additional spin measurements will drive future progress.

{\it ASTRO-H} will be able to advance studies of black hole spin in the
following ways:
\begin{itemize}
\parskip=-0.2cm
\item The mission should be able to obtain excellent
  broad-band spectra of 1-2 transients per year.  Particularly if
  observations are made both in the "high/soft" state (in order to
  utilize the disk continuum) and bright phases of harder states (in
  order to utilize disk reflection), it will be possible to measure
  1-2 spins per year.  In five years, then, {\it ASTRO-H} should gather 5-10
  spin values, or 10-20 spins over 10 years of mission operation.
  {\it Thus, it is likely that {\it ASTRO-H} can double the current number
  of spin measurements.}
\item For both the disk continuum and disk reflection, intervening
  disk winds can pose serious complications.  A strong wind can alter
  the implied mass accretion rate, which is important to continuum
  measurements.  It can also distort the shape of the continuum, which
  is important to both continuum and reflection techniques.  The
  sensitivity of the SXS will permit the detection and
  characterization of even very weak absorption lines from a disk
  wind, and enable continuum and reflection components to be measured
  accurately.
\item The extraordinary sensitivity and bandpass of the HXI will
  permit unprecedented studies of the hard X-ray continuum.  This owes
  partly to the fact that the HXI sits behind a focusing hard X-ray
  telescope, and partly to innovative background rejection techniques.
  The HXI will facilitate the detection of even weak hard X-ray
  components, enabling better separation of the disk flux for
  continuum-based spin measurements.  The improvements for
  reflection-based spins are likely to be at least as marked: the
  Compton back-scattering hump will be revealed extremely well using
  the HXI, enabling improved broad-band modeling.
\end{itemize}
 
\subsection{Disk reflection spin measurements with {\it ASTRO-H}}

A typical 100~ks observation of a black hole binary similar to MAXI
J1836$-$194 in the ``intermediate'' state (Reis et al.\ 2012) will
reveal a relativistic line and Compton hump in a manner that has yet
to be seen with current observatories (see Figure 1). The
unprecedented sensitivity of the SXS in energies up to and beyond the
iron line region (4--8~keV), together with the overlap with the HXI
will provide strong constraints on the inclination and emissivity of
the accretion disk as well and on the black hole spin, as these
parameters are strongly dependent on the overall shape of the iron
line and absorption depth complex and on the strength of the Compton
hump at $\sim30$~keV. Simulations show that {\it ASTRO-H} will determine the
inclination and emissivity profile of systems like MAXI J1836$-$194,
with 90\% errors less than 1\% and 3\% respectively.  These strong
constraints on the inclination and emissivity directly lead to precise
spin measurements, with {\it statistical} errors of approximately 1\%
for spin $\gtrsim0.9$ for a source like MAXI J1836$-$194, and approximately 5\%
for spins $\lesssim0.5$ for sources such as XTE~J1752--223. For comparison, the
current 90\% statistical errors on the spin of MAXI J1836$-$194
($a\sim0.9$) and XTE~J1752$-$223 ($a\sim0.5$), both obtained with 
{\it Suzaku}, are 3.5\% and 35\% respectively.

\begin{figure}
\begin{center}
{\hspace*{-0.2cm}
 \rotatebox{0}{
{\includegraphics[width=8.5cm]{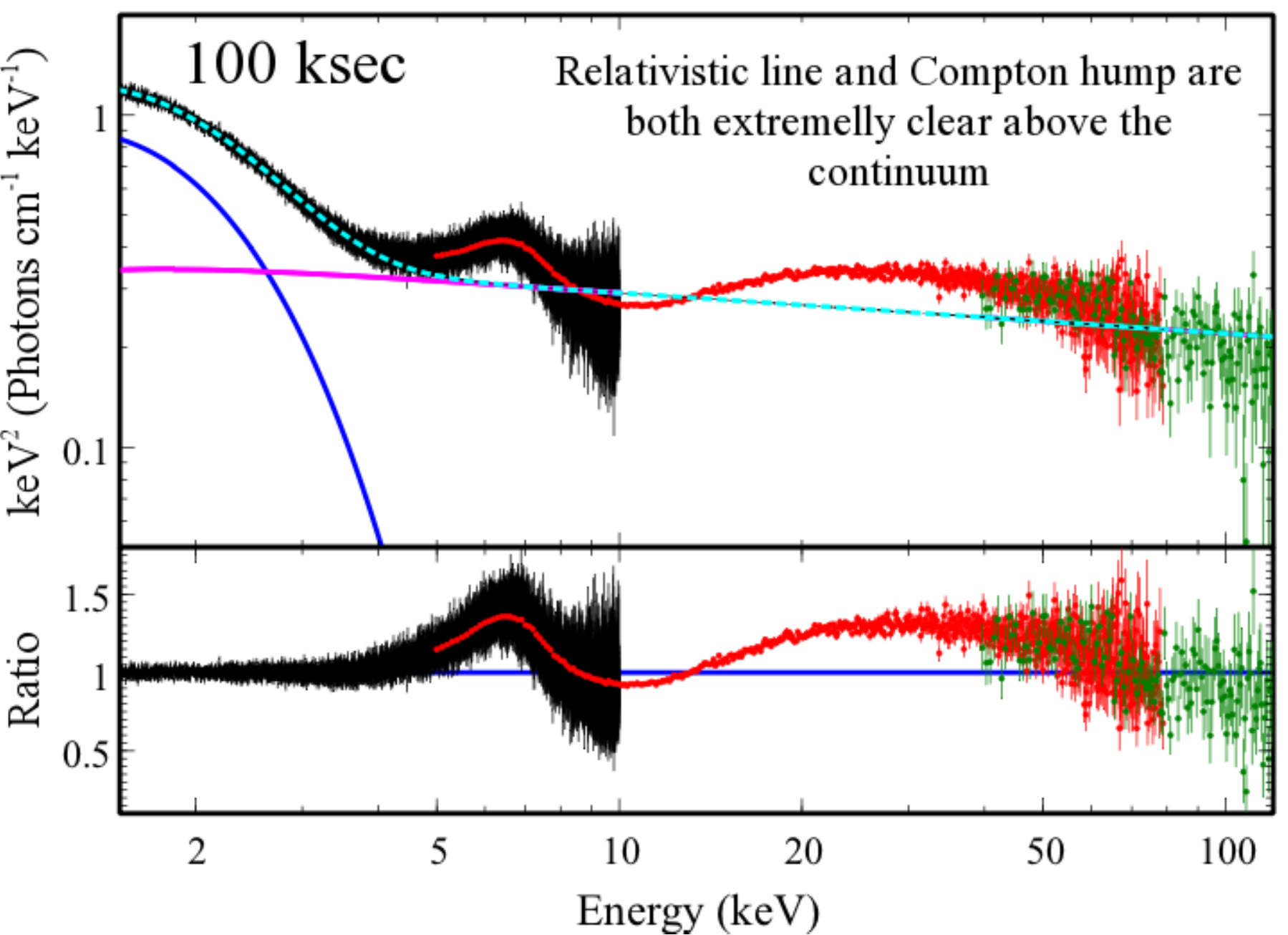}  
}}}\vspace*{-0.15cm}
\end{center}
\caption{The presence of relativistic features including
  the broad Fe line and Compton hump will be revealed in detail with
  {\it ASTRO-H}.  (Top) Simulated 100~ks spectrum of MAXI J1836$-$194
  assuming a full reflection model and subsequently fit with a simple
  continuum consisting of an absorbed power law (magenta) and a disk
  blackbody (blue).  (Bottom) Simulated data-to-model ratio
  emphasising the excellent sensitivity and energy coverage provided
  by {\it ASTRO-H}.}
\end{figure}

\subsection{Connections to jet production}
X-ray binaries are often treated as {\em thermal} sources,
effectively transforming the gravitational energy of the compact
object (a neutron star or a black hole) into thermal X-ray emission
radiated away by the hot accretion plasma.  However, since the
discovery of compact Galactic sources with relativistic jets (dubbed
as microquasars) the general view on the role of nonthermal processes
in X-ray binaries has significantly changed.  It is now recognized
that non-thermal processes do play a non-negligible role in these
accretion-driven objects.  Approximately 20 per cent of the $\sim 250$
known X-ray binaries show synchrotron radio emission, and observations
in recent years have revealed the presence of radio jets in several
classes of X-ray binary sources (e.g. \citealt{Fender2001jets}).  The high
brightness temperature and the polarization of the radio emission from
X-ray binaries are indicators of the synchrotron origin of radiation.
The non-thermal power of synchrotron jets (in the form of accelerated
electrons and kinetic energy of the relativistic outflow) during
strong radio flares could be comparable with, or even exceed, the
thermal X-ray luminosity of the central compact object.

If the acceleration of electrons proceeds at a very high rate, the
spectrum of synchrotron radiation of the jet can extend to the hard
X-ray and soft $\gamma$-ray domain \citep{AtoyanAharonian99,Markoff2001J1118}. 
 In addition, the high density photon fields supplied by
the accretion disk and by the companion star, as well as produced by
the jet itself, create favorable conditions for effective production
of X- and $\gamma$-rays of inverse Compton origin inside the jet
\citep{LevinsonBlandford96,AtoyanAharonian99, Georganopoulos02}.  Generally, this radiation is expected
to have an episodic character associated with strong radio flares in
objects like GRS~1915+105.

A large fraction of microquasars are associated with Galactic black
hole X-ray binaries.  The previous observations (OSSE and COMPTEL) show
that the spectra of these highly variable objects, in particular
GRS~1915+105 and Cyg X-1, extend to the domain of very hard X-rays and
soft gamma-rays.  For any reasonable temperature of the accretion
plasma, models of thermal Comptonization cannot explain the MeV
radiation, even when one invokes the so-called bulk-motion
Comptonization.  To explain this excess, the so-called ``hybrid
thermal/non-thermal Comptonization'' model has been proposed; it
assumes that the radiation consists of two components -- (i) the
thermal Comptonization component with a conventional temperature of
the accretion plasma $k T_{\rm e} \sim 20-30 \rm \ keV$ and (ii) a
nonthermal high energy component produced during the development of a
linear pair cascade initiated by relativistic particles in the
accretion plasma surrounding the black hole (for a review see \citealt{eqpair}).  This model requires existence of a relativistic electron
population in the accretion plasma, resulting from either direct
electron acceleration or pion-production processes in the
two-temperature accretion disk with $T_{\rm i} \sim 10^{12} \ K$
(\citealt{Mahadevan97}).

An alternative site for production of hard X-rays and low energy
gamma-rays could be the synchrotron jets.  In particular, it has been
proposed that the synchrotron emission of microquasars might extend to
X-ray energies, either in the extended jet structure \citep{AtoyanAharonian99} or close to the base of the jet \citep{Markoff2001J1118}. Recently, a significant contribution of the nonthermal
X-ray emission to the total X-ray luminosity of Cyg X-3 has been
suggested by \citet{Zdziarski2011}, based on detection of gamma-rays
by {\it Fermi} LAT and {\it AGILE}.  Confirming the existence of a synchrotron
X-ray component from the jets in microquasars will not only help to
understand the acceleration mechanisms in these objects, but also add
to an emerging picture of disk--jet coupling.

The most promising energy band for the extraction of the synchrotron
component is the hard X-ray to the soft gamma-ray band, where the
radiation from the accretion plasma is suppressed.  The performance of
{\it ASTRO-H} the HXI and SGD are suited to spectroscopic and temporal
studies of the most prominent representatives of microquasars like GRS
1915$+$105, Cygnus X-1, and Cygnus X-3.  A detection of polarization
by the soft gamma-ray detectors would provide crucial test of the
synchrotron origin of radiation. In this regard, one should mention
the claim of detection of polarization of hard X-ray emission above
400 keV by {\it INTEGRAL} which can be explained only by synchrotron
emission \citep{Laurent2011Sci}.

\section{Disk winds}
\subsection{Current picture based on {\it Chandra}, {\it XMM-Newton}, {\it Suzaku}}

During the last decade, a growing number of X-ray binaries have been
found to exhibit absorption lines from highly ionized elements
\citep[e.g.,]{church05, boirin04}.  These
systems range from microquasars such as GRO~J1655$-$40
\citep{Ueda1998j1655,yamaoka01,Miller06Natur}, GRS~1915+105
\citep{kotani00,lee02,ueda09}, H~$1743-322$\citep{miller06b} and
4U~$1630-47$ \citep{kubota07} to low-mass X-ray binaries such as
GX~13+1 \citep{ueda01,sidoli02}, X~1658$-$298 \citep{sidoli01} and
X~1254$-$690 \citep{boirin03}.  Blue-shifts indicative of winds are
especially prominent in the black hole systems.  In all cases, the
sources are viewed at high inclination angles, and the absorption
structure is visible throughout the orbital period
\citep[e.g.,][]{yamaoka01,sidoli01,sidoli02}.  The absorption features
are therefore thought to originate in material that is associated with
and extends above the accretion disk. This is illuminated by the
X-rays produced from the innermost regions of the accretion flow (both
the disk and hard X-ray coronal emission). The reprocessed emission
and scattered flux from the extended wind can be seen directly in the
``accretion disk corona'' (or, ADC) sources, wherein the intrinsic X-rays are obscured
\citep[e.g.][]{kallman03}, but for the majority of highly inclined
sources, the wind material is seen in absorption against the much
brighter X-ray central engine.  Multiple absorption lines give an excellent probe of the physical conditions in the wind
\citep[e.g.,][]{ueda04,Miller06Natur}, where the spectra indicate the
presence of significant columns of highly-ionized outflowing material.

Several scenarios have been proposed to explain how winds are driven.
One of the candidates is the radiation pressure on electrons.  This
can be made much more efficient if the cross section for interaction
between the matter and radiation is enhanced by line opacity.  There
are multiple line transitions in the UV region of the spectrum, so UV
emitting disks can drive a powerful wind at luminosities below
Eddington.  Such line--driven disk winds are seen in cataclysmic
variables (CVs; \citep{Pereyra00}) and are probably also responsible
for the broad absorption line (BAL) outflows seen in AGN
\citep{proga00,2013PASJ...65...40N}.  However, the disk temperature
for black hole binaries is in the X-ray region, so line driving is
probably unimportant \citep{proga02}.  The momentum absorbed in these
observed transitions is very small, and insufficient to drive a wind.

Another type of outflow from a disk is a thermally--driven wind
\citep{begelman83}.  Here again, the central illumination is
important, but the process is less direct.  The illumination heats the
upper layers of the disk to a temperature of order the Compton
temperature, $ T_{\rm C}$. The atmosphere will expand due to the
pressure gradient; at sufficiently large radii, the thermal energy
driving the expansion is larger than the binding energy, leading to a
wind being driven from the outer disk.  Simple estimates of the
launching radius of this wind give $R=10^{12} \cdot (M/10M_\odot)\cdot
( T_{\rm C}/10^7~{\rm K})^{-1}~{\rm cm}$ \citep{begelman83}.  A
separate analysis found that thermal winds can potentially be launched
at a radius a factor 5--10 smaller than this
\citep{begelman83,woods96}.  {\it Chandra} grating data of
GRS~$1915+105$ and {\it Suzaku} spectra of 4U 1630$-$47 (Kubota et
al.\ 2007) may be consistent with a thermally driven wind: the data
indicate a launching radius of $R\sim 10^{11}~{\rm cm}$
\citep{ueda09}.

The last type of outflow is a magnetically--driven wind.  These are
much harder to quantitatively study as the magnetic field
configuration is not known, yet they are almost certainly present at
some level as the underlying angular momentum transport is known to be
due to magnetic fields \citep[see e.g.][]{balbus02}.  Winds (and jets)
are clearly present in magnetohydrodynamical (MHD) simulations which
include these magnetic stresses self-consistently.  These generically
show that the mass loss is stochastic, with large fluctuations both
spatially and temporally, but that the time averaged properties are
well defined, so this magnetic wind is quasi-continuous
\citep[e.g.,][]{hawley01, machida04}.  These calculations are still in
their infancy, especially for describing the properties of a
geometrically--thin disk.  An approximation to the properties of the
self-consistent magnetic wind from an accretion disk can be made by
imposing an external field geometry \citep{proga00,proga03}.  The mass
loss rates depend on this field configuration, but in general these
allow steady, powerful winds to be launched from any radius.
Similarly, imposing poloidal field lines in the disk can give rise to
magnetocentrifugal winds \citep{Blandfordpayne82}.  Such winds
transport angular momentum from the disk without the aid of internal
viscosity, and they are known to exist in FU Orionis and T Tauri
stellar systems based on evidence of rotation in absorption line
profiles \citep{Calvet1993}.

A very small launching radius was obtained from {\it Chandra} grating
spectra of GRO~J$1655-40$ ($r\sim 10^{8-10}~{\rm
  cm}$); this wind may be driven by magnetic pressure or via
magnetocentrifugal acceleration \citep{Miller06Natur}.  This small launching
radius was derived partly through the detection of density-sensitive
Fe XXII lines, allowing for an accurate determination of the radius
via $r = \sqrt{L/n\xi}$ (where $L$ is the ionizing luminosity, $n$ is the number density, and $\xi$ is the ionization parameter).  The observed absorption spectrum was later
analyzed using a number of photoionization models constructed using
Cloudy \citep{1998PASP..110..761F}  and XSTAR \citep{1982ApJS...50..263K}; these models confirm a very small lauching radius,
and a very large mass outflow rate \citep{Miller2008j1655wind,kallman09}.  Even newer, more detailed treatments of thermal wind properties show that they cannot account for the wind in GRO J1655$-$40, again signaling a magnetic component \citep{Luketic2010}.

\subsection{Wind Outflow Rates, Launching Sites, and Driving Mechanisms}
X-ray gratings spectroscopy has revolutionized the study of black hole
accretion.  In particular, the resolution afforded by the {\it
  Chandra}/HETG and the {\it XMM-Newton}/RGS has made it possible to
detect blue shifts in highly ionized X-ray absorption spectra,
signaling the presence of disk winds in stellar-mass black holes.  The
consequences of winds can be far-reaching:

\begin{itemize}
\item Winds may remove angular momentum from the accreting gas,
  potentially enabling the basic operation of the accretion disk.
  Disk winds may hold the key to understanding the physics of disk
  accretion itself.
\item Winds may remove more gas from a system than actually accretes
  onto the black hole, affecting both the growth of the black hole,
  and the evolution of the binary system.
\item It is not yet clear if a single mechanism drives winds or
  dominates wind production, if different mechanisms may work at
  different times, or if multiple mechanisms may work concurrently.
  Detecting multiple ionization zones, constraining launching radii,
  and accrurate measurements of total outflow rates can potentially
  reveal the answers.
\item Stellar-mass black hole winds may be closely related to X-ray
  warm absorbers in Seyfert AGN, but the relationship is not yet
  clear.
\item Wind and jets appear to be state--dependent; the disk may
  alternate its outflow mode.  The nature of this phenomenon -- and the
  clues it may provide to the physics of jet and wind production --
  remain to be discovered.
\end{itemize}

Spectroscopy of black hole X-ray binary winds with the {\it ASTRO-H}/SXS
calorimeter will offer some clear advantages compared to dispersive
spectrometers.  Most important, perhaps, are the significant gains in
resolution in the Fe K band.  The baseline resolution of the SXS is
just 5~eV, and this is fixed across its bandpass.  Unlike a grating
instrument, then, {\it the resolution of a calorimeter improves with
  increasing energy}.  Whereas current instruments excel at detecting
lines such as Fe XXV He-$\alpha$ and Fe XXVI Ly-$\alpha$, the SXS will
also do an excellent job of detecting their associated $\beta$ lines,
He-like and H-like Ni lines, and any lines or edges that are
significantly blue-shifted.  As a result, {\it ASTRO-H} will be sensitive to
the most powerful components of outflows, and do a superior job of
revealing the power in wind feedback from black holes into their local
environments.

In this white paper, we have conducted a number of simulations to
demonstrate the potential of {\it ASTRO-H}, and particularly the SXS, to
greatly improve our view of accretion onto black holes.  In these
efforts, we have assumed a baseline calorimeter resolution of 5~eV.  It is also
important for the reader to note that a tunable onboard X-ray source
will periodically be used to calibrate and measure the performance of
the SXS.  This means that the incredible intrinsic resolution of the
calorimeter to measure small velocity shifts will be fully realized.

The mass outflow rate in a wind is given by:
\begin{equation}
\dot{M}_{wind} = \Omega C_{v} \mu m_{p} n_{e} r^{2} v
\end{equation}
where $\Omega$ is the covering factor ($0\leq \Omega \leq 4\pi$),
$C_{v}$ is the filling factor, $\mu$ is the mean atomic weight
(usually $\mu = 1.23$), $m_{p}$ is the proton mass, $n_{e}$ is the
electron number density, $r$ is the distance from the source of ionizing
flux to the detected absorption zone, and $v$ is the observed outflow
velocity.

Stellar-mass black hole winds may represent an simpler environment
than X-ray ``warm absorber'' winds in Seyfert AGN.  For instance,
optical and infrared constraints on the binary inclination and
comparisons of emission and absorption lines strongly suggest {\it
  equatorial} winds, so $\Omega$ is fairly well constrained ($\Omega
= 0.3-0.5$ is reasonable, see Miller et al.\ 2008, Ueda et al.\ 2009).
Stellar-mass black hole disk winds are very highly ionized, in
general, and have not shown evidence of a low ionization component.
Therefore, they are not likely to be clumpy, so $C_{v} = 1$ is
reasonable.  As noted above, the resolution of the SXS will enable
extremely accurate measurements of the observed wind velocity $v$.  Of
course, the observed velocity $v$ is effectively a lower limit, since
a vertically-launched wind will only have a component of its velocity
along an equatorial line of sight, and any disk wind that retains its
local Keplerian velocity will automatically be within $\sqrt{2}$ of
its local escape velocity.

Although He-like triplets can provide useful density diagnostics,
these features can arise whenever gas is illuminated; such lines are
not necessarily tied to the disk wind.  Recently, it has been shown
that Fe XXII absorption lines at 11.92~\AA~ and 11.77~\AA~
(1.040~keV and 1.053~keV, respectively) can act as direct diagnostics
of the wind density (Miller et al.\ 2006, 2008; \citealt{KingMillerRaymond2012}; also see \citealt{MaucheRaymond2000}).  At this energy, the
resolution of the SXS is lower than that of the {\it Chandra}/HETG.
Can the SXS confidently detect and resolve such lines?

To answer this question, we have simulated an 100 ks SXS spectrum,
using the best-fit broadband XSTAR photoionization model of GRO
J1655$-$40 (Miller et al.\ 2008) as a template.  This source was
observed at a flux of 1~Crab, which will give approximately 2000 counts/s
in the SXS.  After silencing the central 5 pixels in the array, the
recorded high and medium count rate will be 60~counts/s.  Figure 2 shows
the observed {\it Chandra}/HETG spectrum of GRO J1655$-$40 and the
simulated {\it ASTRO-H}/SXS spectrum.  While the resolution of the HETG is
better at 1 keV, the Fe XXII lines are clearly detected and resolved
in the SXS spectrum.  

\begin{figure}
\begin{center} 
\includegraphics[width=0.5\hsize]{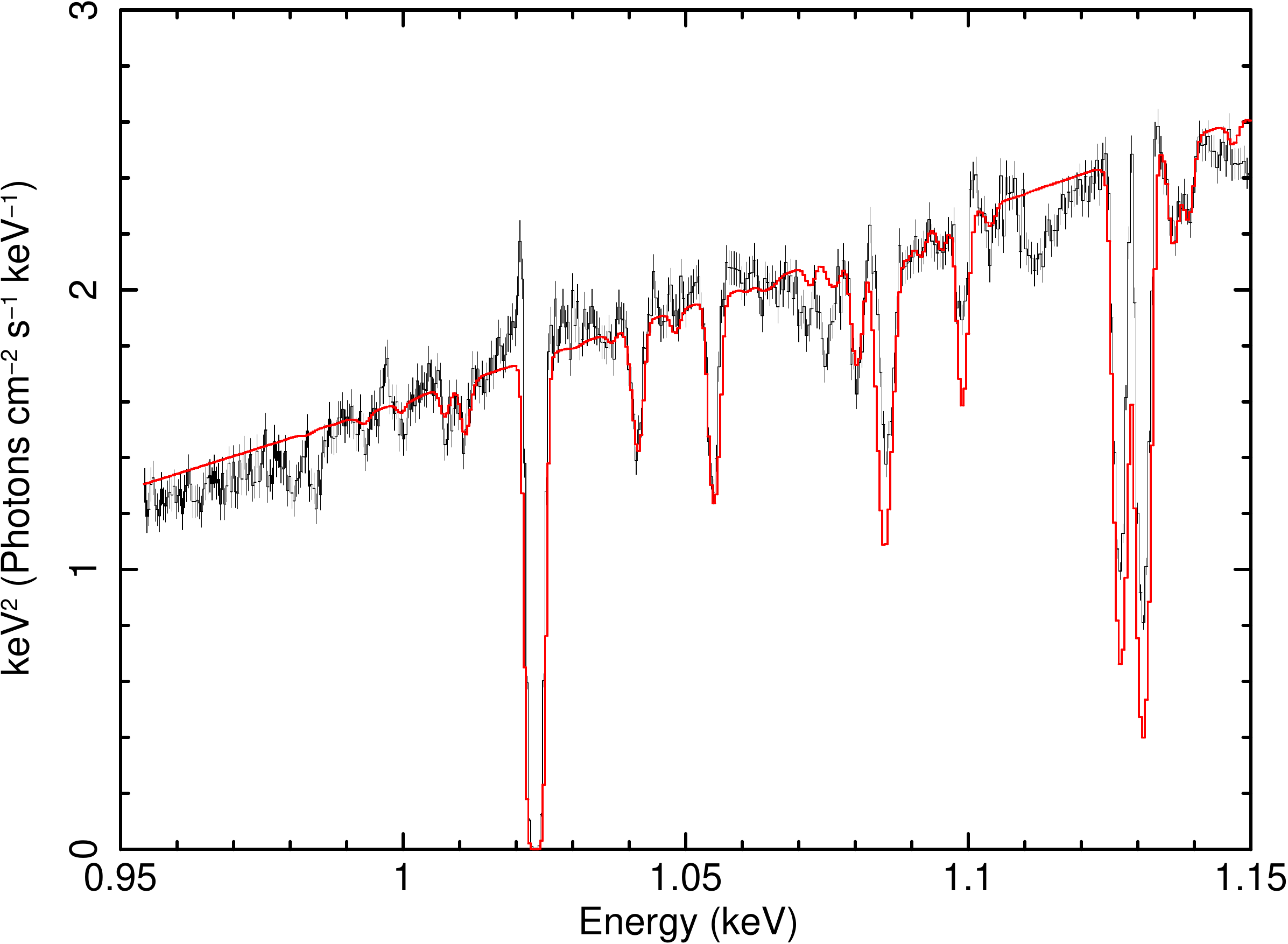}
\includegraphics[width=0.5\hsize]{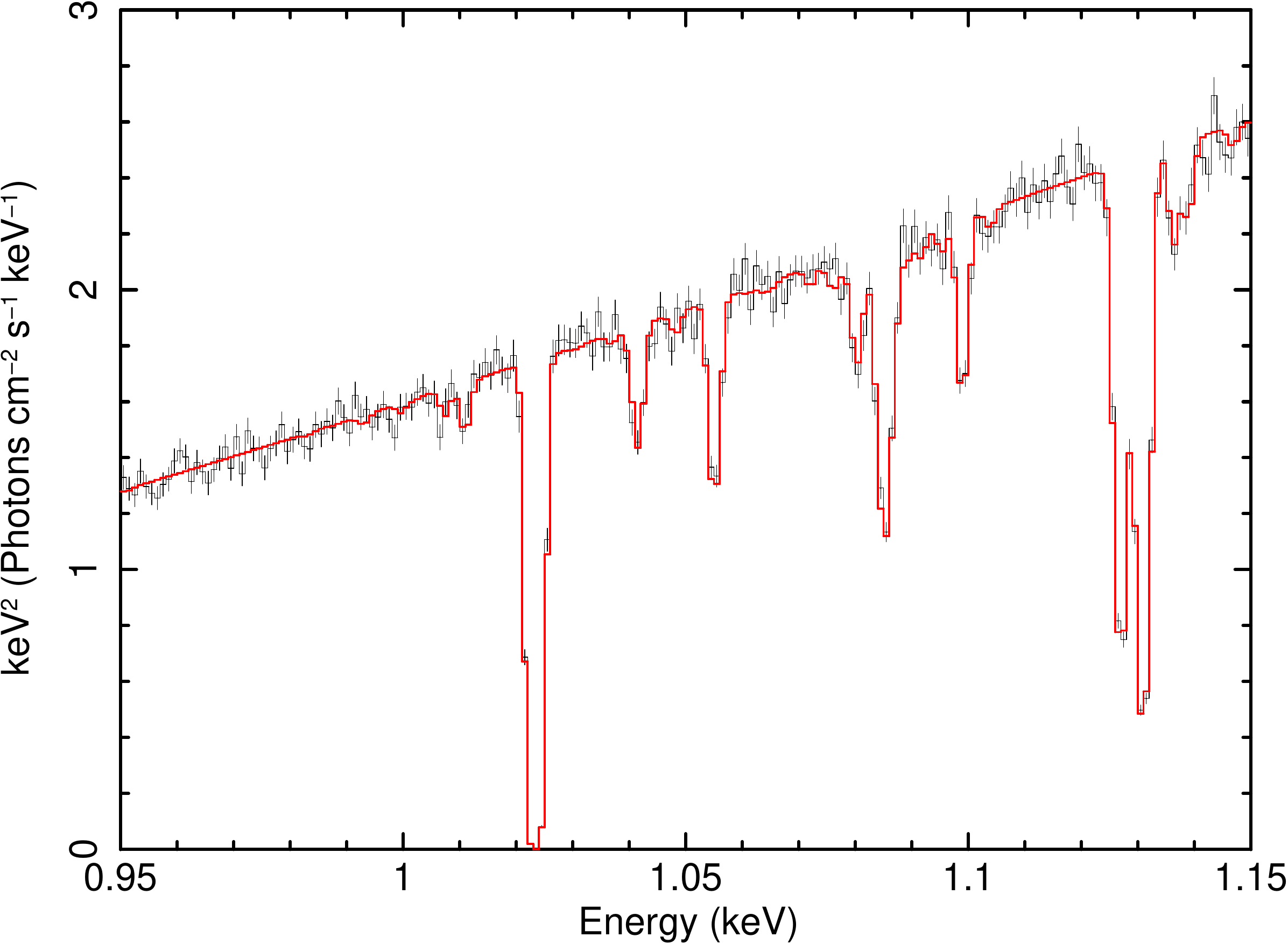}
\end{center}
\caption{Top: The {\it Chandra}/HETG spectrum of GRO J1655$-$40 is
  shown here.  The Fe XXII lines at 1.040~keV and 1.053~keV are
  clearly detected, and their flux ratio gives a density of $n =
  10^{14}~ {cm}^{-3}$.  Bottom: Although the resolution of the HETG
  is superior to the {\it ASTRO-H}/SXS at 1~keV, this simulated 100~ks SXS
  spectrum demonstrates that the density-sensitive Fe XXII pair can
  easily be detected and resolved using the SXS.  (Both spectra were
  binned for visual clarity.)}
\end{figure}


This simulation should be regarded as a proof of principle, because
the spectrum observed from GRO J1655$-$40 is atypically rich.
However, when such density diagnostics are available, one can obtain a
very strong constraint on the wind launching radius (since $r =
\sqrt{L/n\xi}$).  This is important because within the Compton radius,
highly ionized winds can only be driven magnetically (Begelman, McKee,
\& Shields 1983; also see Miller et al.\ 2008, Kallman et al.\ 2009,
Luketic et al.\ 2010)

In other cases -- particularly for highly obscured sources -- it may
not be possible to detect the Fe XXII pair near to 1 keV.  In this
case, the capabilities of the calorimeter will again help to determine
wind properties accurately.  When $n_{e}$ cannot be obtained directly,
it is typical to assume that $N = n_{e} r$ (where $N$ is the column density).  Because the SXS can
readily detect higher order lines from He-like and H-like Fe and Ni,
it can assess the extent to which such absorption is saturated, and
provide much stronger constraints on the column density of disk winds
than has previously been possible.  Even in the absence of a direct
density constraint, the total mass outflow rate may still be estimated
via:
\begin{equation}
\dot{M}_{wind} = \Omega C_{V} \mu m_{p} L_{ion} v / \xi.
\end{equation}
Here again, the ability of the SXS to measure $v$ is important, but so
too is the broad bandpass of {\it ASTRO-H}, and the sensitivity of the HXI.
Since Fe XXV and Fe XXVI are ionized by X-rays above 8.8~keV and
9.3~keV, respectively, it is important to know the continuum up to
high energy in order to accurately estimate the outflow rate in the
hottest part of the disk wind.  

\subsubsection{Detecting Wind Circulation}
The radii from which winds are launched in FU Ori and T Tauri stellar
systems are not large compared to the source of ionizing flux.  In
other words, local flux is able to imprint on local gas, and the
observed absorption profiles show the hallmarks of a rotating disk
wind (e.g Calvet, Hartmann, \& Kenyon 1993).  This strongly signals
that the disk winds in these systems are driven through a
magnetocentrifugal process, as described by Blandford \& Payne (1982).

The magnetocentrifugal wind model was conceived as an answer to the
question of how gas in a disk might lose angular momentum, and where
the angular momentum must go, in the absence of a companion star and
binary orbit that may act as a reservoir.  Thus, it may be
particularly important in stellar systems and in AGN.  But in those
cases, too, thermal disk spectra suggest that internal magnetic
viscosity is also at work in transporting angular momentum.  It is
entirely possible that the mechanisms work jointly.  Evidence of
magnetically-driven winds is emerging in black holes such as NGC 4151 \citep{Kraemer2005}, GRO J1655$-$40 (Miller et al.\ 2006, 2008),
NGC 4051 \citep{KingMiller2011}, and possibly also IGR J17091$-$3624
\citep{KingMiller2012} and H~1743$-$322 \citep{miller06b}.
Circulation may not be a signature unique to magnetocentrifugal winds,
but it is worth addressing the ability of the {\it ASTRO-H} to reveal
circulation.

Let us assume that a wind is being launched through magnetocentrifugal
acceleration, and that it is executing local Keplerian motion as it also
flows radially.  An observed line will have a net blue-shift owing to
its velocity into our line of sight, but will have small variations
about the net shift owing to orbital motion, of approximately $v
\simeq v_{Kep} r_{h}/r_{w}$ assuming a small angle approximation
(where $r_{h}$ is the radius of the central engine producing the hard
ionizing flux, and $r_{w}$ is the radius from the central engine at
which the wind is detected).  The matter is particularly simple when
scaled in gravitational radii since $v_{Kep} = c / \sqrt{w}$, where $w$
is the number of gravitational radii.  Then $v_{w}$ is also at $w$
radii, and we can scale $r_{h}$ to be at $h GM/c^{2}$ as well.
Finally, we expect symmetric wings about the line centroid of $v = \pm
c h/w^{3/2}$.

The size of the hard X-ray corona is not known well, especially in
soft states wherein disk winds are detected.  Values of
10--100~$GM/c^{2}$ may be reasonable.  Winds in sources such as GRO
J1655$-$40 and H 1743$-$322 may originate as close as $r\simeq 10^{9}$~cm;
this is of order 1000~$GM/c^{2}$.  Even for a coronal radius of just
$10~GM/c^{2}$, we expect wings offset from the line centroid by $v =
\pm 100$~km/s.  This is an order of magnitude below the measured FWHM
of the Fe XXV and Fe XXVI lines detected in stellar-mass black hole
winds.  Of course, if the corona has a typical radius of
$50~GM/c^{2}$, then we expect wings offset by $v = \pm 500$~km/s,
which is agrees better with measured line widths.

As an example, we consider the rich absorption spectrum observed from
GRS~1915$+$105 in a soft state at a flux of about 1 Crab, using the
{\it Chandra}/HETG \citep{NeilsenLee2009,ueda09}.  The Fe
XXV He-$\alpha$ line has a FWHM of 1200--1900~km/s (Neilsen \& Lee
2009).  Figure 3 shows this spectrum; it is interesting to note that
neither line profile appears to be exactly symmetric, but there is
stronger evidence of structure in the Fe XXV line.  The figure shows
the Fe XXV and Fe XXVI lines fit with two Gaussians, as a simplified
case of rotation-induced line profiles.  The composite line profile is
not clear at the resolution of the {\it Chandra}/HETG.  Using this
model, however, we simulated a 100~ks {\it ASTRO-H}/SXS spectrum.  Again, at
a flux of 1 Crab, we expect an incident count rate of 2000 counts/s, and an effective rate of 60 counts/s.  Figure 3 also shows this simulated spectrum,
and the composite nature of the line profile is extremely clear at SXS
resolution.

This is a simplified treatment of the problem of detecting orbital
motion through absorption spectra.  It shows that plausible coronal
sizes and wind launching radii may encode Keplerian motion that could
potentially be detected.  Realistic line profiles will not be as clear
as the simplified two-Gaussian models used in this simulation.  The
differences in Keplerian orbital velocities over a range in radii will
naturally give rise to a significant line width, and signatures of
rotation will be an addition to this broadening, not an alternative.
The chances of a detection are improved if $\delta r/r$ is small for a
given wind.  Moreover, it is possible that other factors may
complicate the detection of orbital motion; for instance, the
structure in the Fe XXV line in GRS 1915$+$105 may be due to the
intercombination line being detected in absorption at very high
density.  Nevertheless, it appears that rotational motion is
potentially within the grasp of {\it ASTRO-H}.

\begin{figure}
\begin{center} 
\includegraphics[width=0.5\hsize]{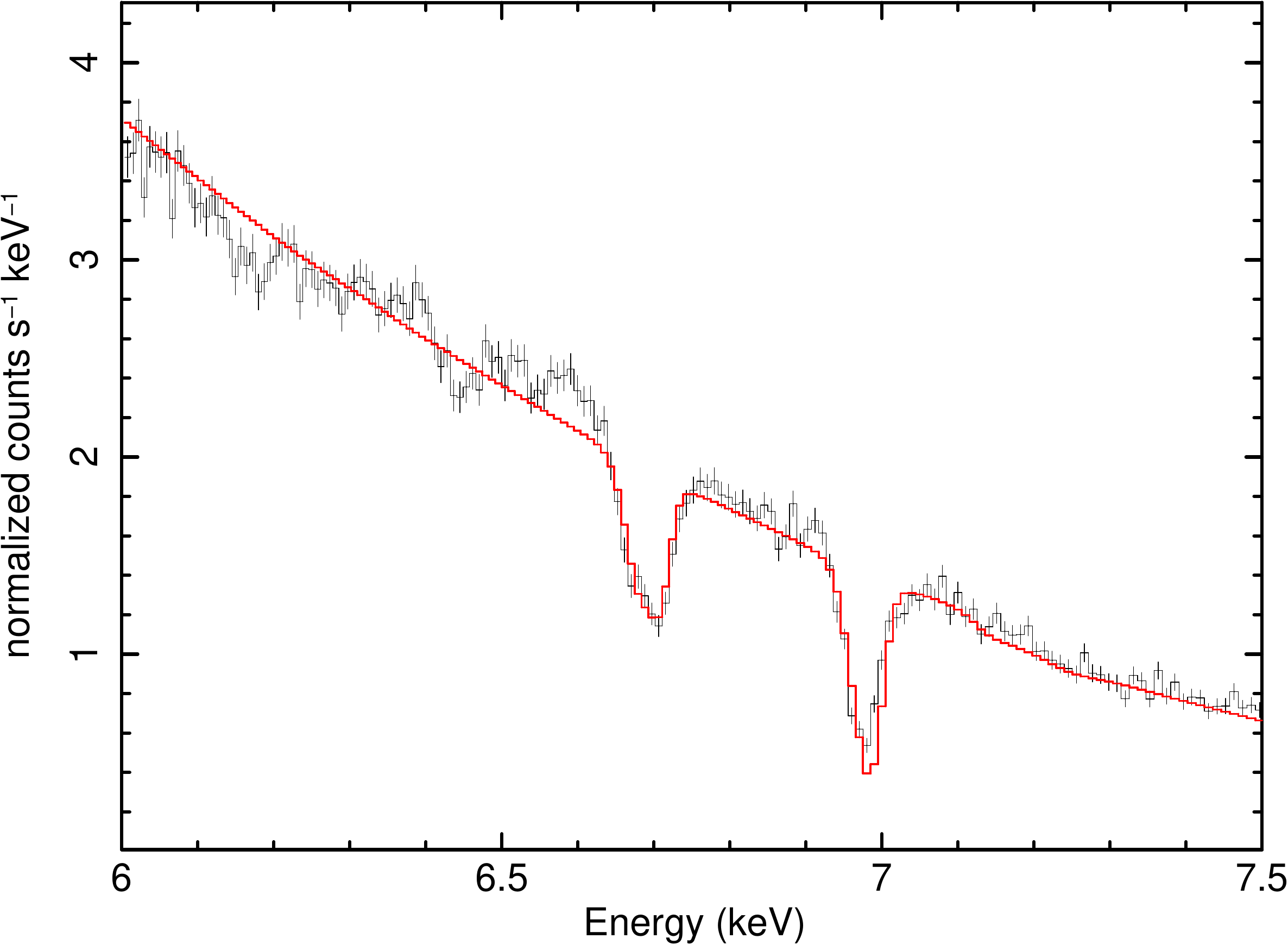}
\includegraphics[width=0.5\hsize]{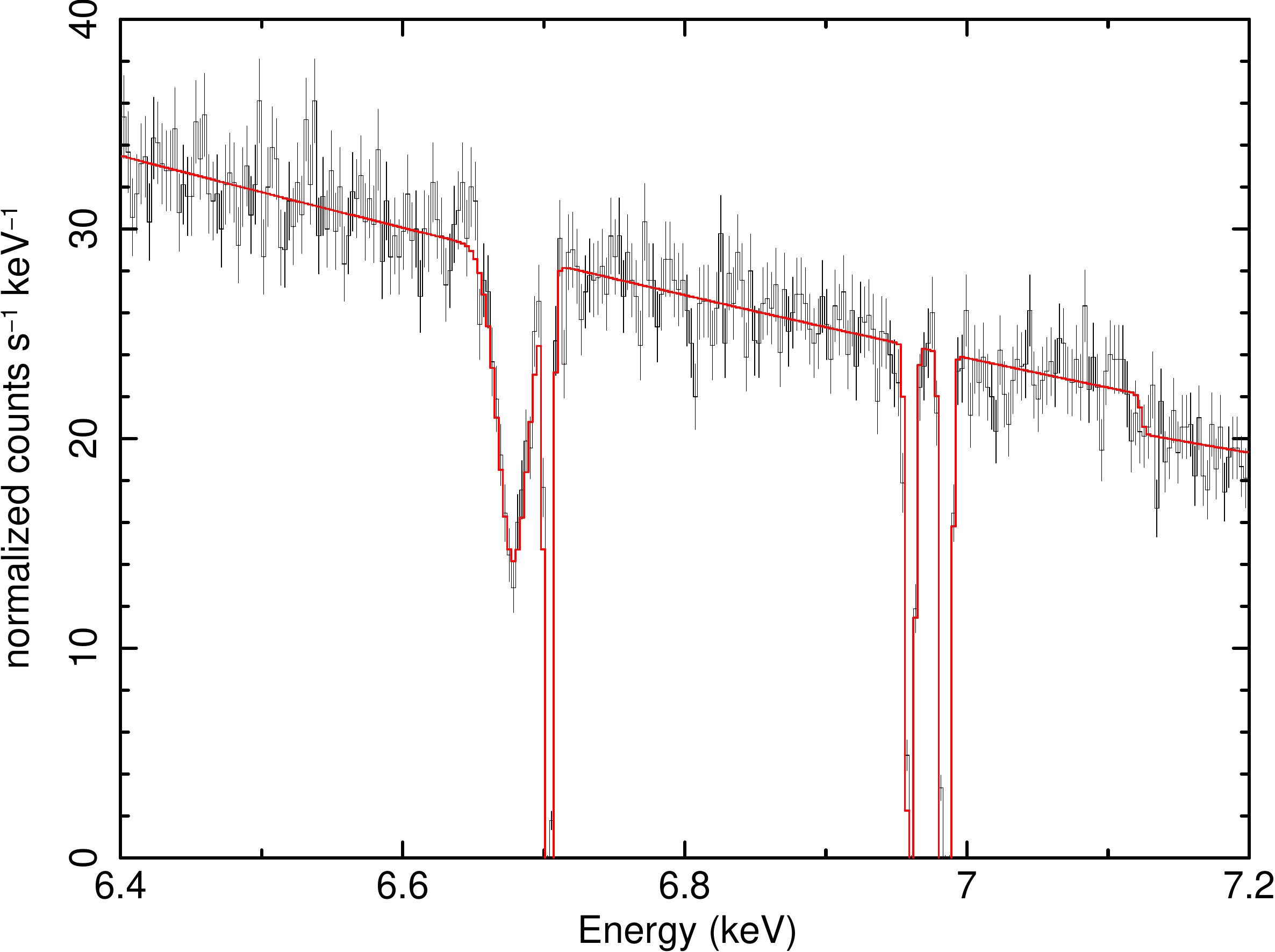}
\end{center}
\caption{Top: A {\it Chandra}/HETG spectrum of GRS 1915$+$105 is shown
  here.  The Fe XXV He-$\alpha$ and Fe XXVI Ly-$\alpha$ lines at
  6.700~keV and 6.970~keV show some evidence of structure.  The model
  shown in red fits each line with a composition of two Gaussians.
  Bottom: A simulated 100~ks {\it ASTRO-H}/SXS spectrum is shown here,
  generated using the model shown above.  The SXS is clearly able to
  separate the individual components that are blended at lower
  resolution. Plausible size scales for the central engine, and the
  smallest wind launching radii that have been inferred, could
  give rise to spectra where rotation is encoded into
  absorption.  This exercise shows that detecting a circulating,
  magnetocentrifugal wind is within the reach of the SXS.}
\end{figure}


\subsubsection{Probing Dynamical Timescales}
The sensitivity and resolution of the SXS may make it possible to
probe and utilize the dynamical timescale of winds in stellar-mass
black holes, in order to better understand their origin.  The
dynamical timescale is simply given by $t_{dyn} = r / v_{out}$.  Prior
studies of GRO J1655$-$40 find that $r$ is likely quite small, about
 $10^{9}$~cm, with a characteristic outflow velocity of
$v_{out} = 400$~km/s (Miller et al.\ 2008, Kallman et al.\ 2009),
making $t_{dyn} \simeq 25$~s.  This time scale is likely too short to
be probed with the SXS, even for sources as bright as 1 Crab.

However, other sources may offer different opportunities.  The wind in
GRS 1915$+$105 may be driven thermally from the outer disk.  \citet{NeilsenRemillardLee2011} estimate that the wind originates at 46
light-seconds ($1.4\times 10^{12}$~cm) from the black hole.  For a
characteristic outflow velocity of $v = 1000$~km/s, this gives
$t_{dyn} = 13.8$~ks, which is fairly long.

Here again,we consider the line-rich wind absorption spectrum that was
observed in GRS 1915$+$105 in a soft state at a flux of approximately
1 Crab.  The spectrum is treated as "S1" in Neilsen \& Lee (2009) and
Ueda et al.\ (2009), and it is possible to model it with a single
higly-ionized XSTAR absorption zone (e.g. King et al.\ 2013).  This
flux level will give $\sim$2000~c/s in the SXS, or approximately 60
H$+$M c/s after silencing the central five pixels in the array.

Using the best-fit XSTAR photoionization grid in King et al.\ (2013),
we simulated SXS spectra assuming the same column density measured in
the {\it Chandra}/HETG spectrum ($1.1\times 10^{23}~ {\rm cm}^{-2}$
in our fits), and half of that column density.  Figure 4 shows that a
50\% variation in the column density becomes visible in about 3~ks,
which is shorter than the nominal dynamical time scale in the wind.
This suggests that, at least in some sources, the SXS will be able to
set upper limits on wind launching radii simply by detecting
variations in e.g. column density or outflow velocity that exceed any
variations in the source luminosity.

\begin{figure}
\begin{center} 
\includegraphics[width=0.7\hsize]{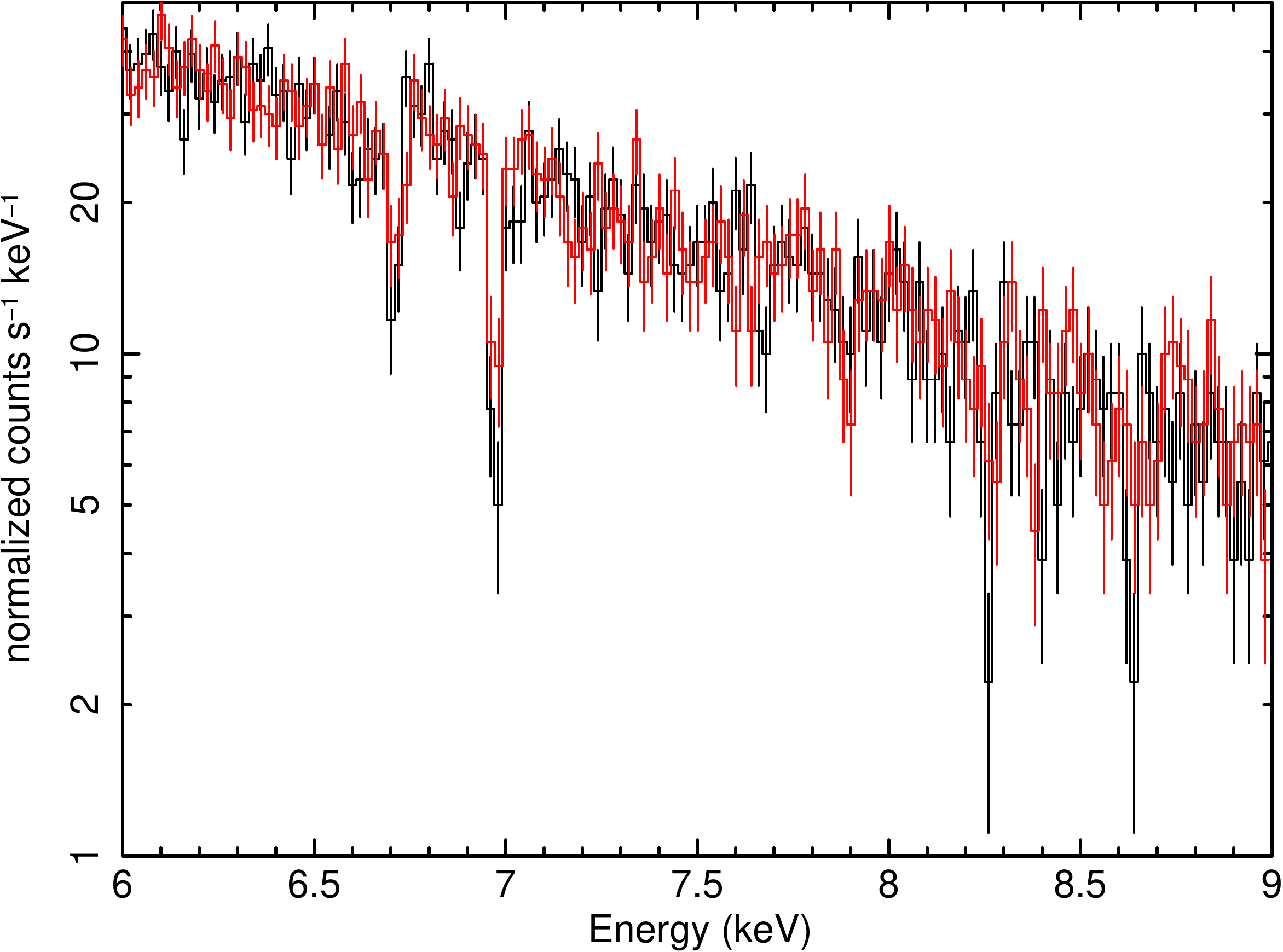}
\end{center}
\caption{The plot above shows simulated 3~ks {\it ASTRO-H}/SXS spectra of
  GRS 1915$+$105, based on the {\it Chandra}/HETG spectrum shown in
  the previous figure.  The spectrum in black has a 50\% higher column
  density than the spectrum showing in red.  Both spectra have been
  binned for visual clarity.  For published values of the outflow
  velocity and launching radius of the wind in GRS~1915$+$105 (Neilsen
  \& Lee 2009), 3~ks is well below the dynamical timescale in the
  wind ($t_{dyn} = r_{wind}/v_{wind}$).  This simulation indicates
  that at least in the case of bright stellar-mass black holes with
  strong absorption spectra, dynamical timescales in the wind can be
  probed using the {\it ASTRO-H}/SXS.}
\end{figure}


\subsection{The Disk-Wind-Jet Connection}
Compact, steady, relativistic jets are ubiquitous in the low/hard
state of stellar-mass black holes \citep{fenderetal04}.
Sensitivity in the radio band has recently increased dramatically, and
it is clear that jets are quenched in the disk--dominated high/soft
states wherein winds are detected in stellar-mass black holes \citep{Russell2011softstate}.  But is the reverse true?

It {\it appears} that disk winds are state--dependent, and quenched in
the low/hard state (\citealt{Miller06Natur, Miller2008j1655wind};  Neilsen \& Lee
2009; \citealt{Blum2010, KingMiller2012, Ponti2012diskjet}).  This
is particularly interesting because Seyfert AGN are radio-quiet and
outflowing X-ray warm absorber winds are seen in these systems;
however, warm absorbers are not observed in e.g. LLAGN, which can be
radio-loud and bear a closer resemblance to the low/hard state.
Recent work has also noted that the power in wind and jes may be
regulated in a common way (King et al.\ 2013), potentially hinting at
some common physics in their driving mechanisms.

It is possible, though, that very fast, highly ionized winds with a
low column density are hiding in the low/hard state.  Such flows might
be missed in a CCD spectrum with only moderate resolution, or even in
a gratings spectrum.  If such flows can be ruled-out, it would affirm
that winds and jets are fundamentally anti-correlated.  This would be
particularly hard to reconcile with thermal driving models, because
the outer disk should always be irradiated by the central engine, at
least in the absence of an additional obscuring geometry.

To illustrate the power of {\it ASTRO-H} to address this science, Figure 5
shows a simulated 200~ks observation of a black hole in the low/hard
state.  The continuum spectrum and flux was taken from the deepest
{\it Chandra}/HETG spectrum of H~1743$-$322 in that state, in which
90\% confidence upper limits of just 2 eV are obtained for Fe XXV and
Fe XXVI absorption lines.  The same power-law continuum and absorbed
flux (0.05 Crab, giving approximately 100 c/s and 50 H$+$M c/s in the
SXS) was assumed, as well as line equivalent widths of just 2~eV.  In
200~ks, these incredibly weak lines can be detected at the 5$\sigma$
level of confidence.  If winds persist in the low/hard state, even
with a very low column density, they will be detected with {\it ASTRO-H}.

\begin{figure}
\begin{center} 
\includegraphics[width=0.6\hsize]{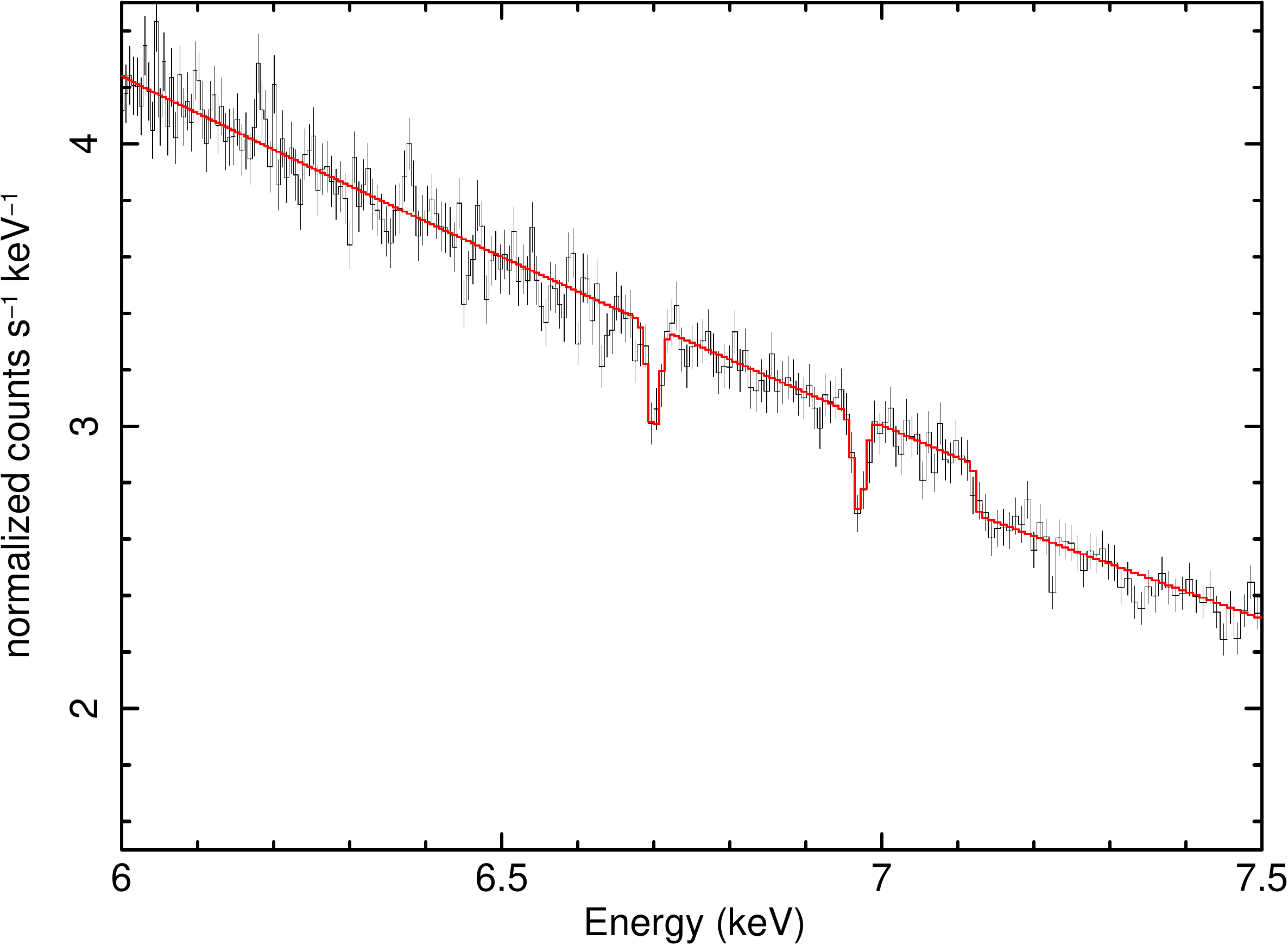}
\end{center}
\caption{X-ray disk winds and relativistic radio jets in stellar-mass
  black holes are observed to be state-dependent: the accretion flow
  appears to alternate the outflow mode.  This could point to a
  fundamental connection between the flows, and connections to the
  physical state of the inner disk.  This is not easily explained in
  terms of thermal winds, and may require that winds have a strong
  magnetic component, as jets are suspected to have.  Ruling out winds
  in the low/hard state, where jets are observed, is therefore very
  important.  The plot above shows a simulated 200~ks {\it ASTRO-H}/SXS
  spectrum of a 0.1~Crab source in the low/hard state, assuming Fe XXV
  and Fe XXVI line equivalent width of just 2~eV, which is within
  current upper limits.  Clearly, {\it ASTRO-H} can
  detect very weak disk winds that may persist in the low/hard state.}
\end{figure}


\subsection{Connections to jets and winds in AGN}

As noted previously, the advent of gratings spectroscopy made it
possible to detect blue-shifted X-ray absorption in equatorial disk
winds.  Absorption of this sort is at least superficially similar to
X-ray "warm absorbers" in Seyfert AGN.  It is possible
that there are fundamental connections between X-ray binary disk winds
and Seyfert warm absorbers, and a number of questions immediately
emerge: Are the outflows in both stellar-mass black holes and Seyfert
AGN driven by the same process?  Are they launched at the same radius
(in a $GM/c^{2}$ sense)?  An early comparison of winds across the mass
scale suggests that the answer to these questions may be ``yes'' (King
et al.\ 2013).  However, this work also points to the need for
improved observations, of exactly the sort that {\it ASTRO-H} can deliver.

\subsubsection{Detection of Multiple Ionization Zones}
X-ray ``warm absorbers'' in Seyferts consist of at least $2-3$
ionization zones, although some authors adopt a continuous ionization
gradient.  It is particularly important to detect and measure the
highest ionization components, since these appear to carry the most
mass \citep{CrenshawKraemer2012, KingMiller2013}.  If massive black
hole wind feedback into host galaxies is able to significantly affect
galactic evolution, the most highly ionized wind components -- which
carry the bulk of the mass (Crenshaw \& Kraemer 2012; King et
al.\ 2013) -- will have mattered the most.  By virtue of its
resolution and collecting area, {\it ASTRO-H} will be ideally able to detect
highly ionized absorption in the Fe K band.  Comparing these highly
ionized components to the hot, ionized outflows observed in
stellar-mass black holes will facilitate the most consistent
comparisons of winds across the mass scale.  This will greatly further
attempts to understand if aspects of winds scale with mass in the way
that jet properties appear to scale.

In comparing X-ray binaries and AGN, it becomes important to
understand whether or not stellar-mass black hole winds are really
described by a single absorption component.  Even the most complex
wind spectrum yet observed -- that seen in GRO J1655-40 (Miller et
al.\ 2006, 2008) -- can be modeled fairly well using only one
component.  Modeling by Kallman et al. \ (2009) suggests that
an improved fit is possible with an additional very highly ionized
zone that is particularly prominent in the Fe K band.  Moreover,
recent spectroscopy of IGR J17091$-$3624 has also found evidence of
two zones separated more in velocity than in ionization (King et al.\ 2012).

Figure 6 shows the {\it Chandra}/HETG data reported in Miller et al.\
(2006a, 2008) and Kallman et al.\ (2009).  The data have been fit with a
dominant absorption zone described by ${\rm log}(\xi) = 4.1$, ${\rm
N}_{\rm H} = 3.0\times 10^{23}~ {\rm cm}^{-2}$, and $v = 400$~km/s,
and by a second possible zone with ${\rm log}(\xi) = 6.0$, ${\rm
N}_{\rm H} = 2.0\times 10^{23}~ {\rm cm}^{-2}$, and $v = 1400$~km/s
(Kallman et al.\ 2009).  Scattering within the instrument makes
it difficult to determine which lines are optically thick, and the
resolution of the HETG makes it difficult to detect the putative
higher ionization, faster wind component.

Also shown in Figure 6 is a simulated SXS spectrum.  The observed
continuum flux from GRO J1655$-$40 was approximately 1 Crab, which
will give approximately 2000 c/s in the SXS.  Turning off the central
5 pixels to optimize the number of high and medium resolution events
that feed through the electronics will result in a H$+$M event rate of
60 c/s.  In a 100 ks observation, the second ionization zone is easily
detected with {\it ASTRO-H}, as can be seen from the blue wing on the Fe
XXVI line that is not fit well by the one-zone model, and by an
absorption line at 8~keV that is not addressed.  Note also that the
black lines are actually black, and incredibly well-separated.

\begin{figure}
\begin{center} 
\includegraphics[width=0.5\hsize]{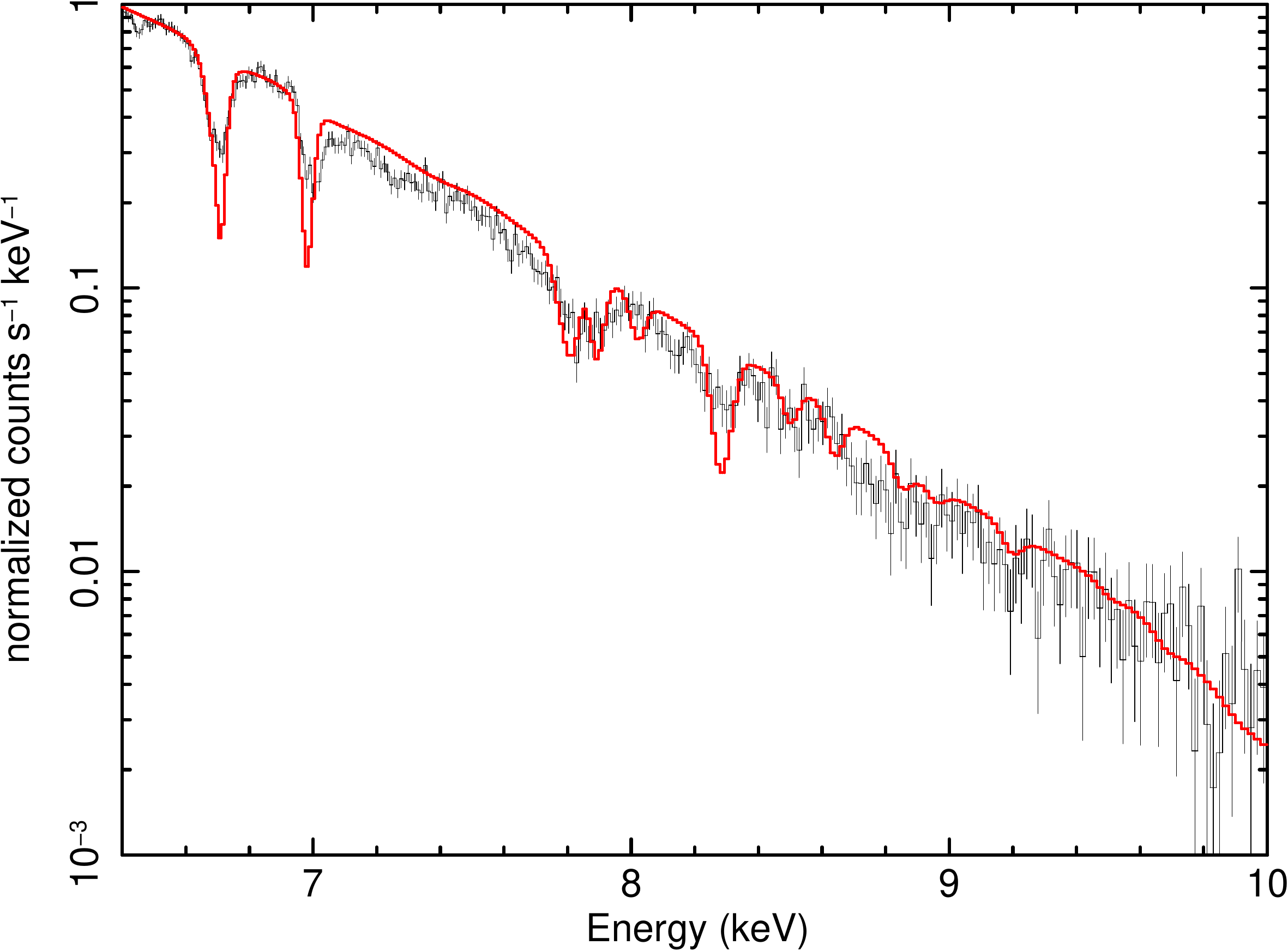}
\hspace{1.0cm}
\includegraphics[width=0.5\hsize]{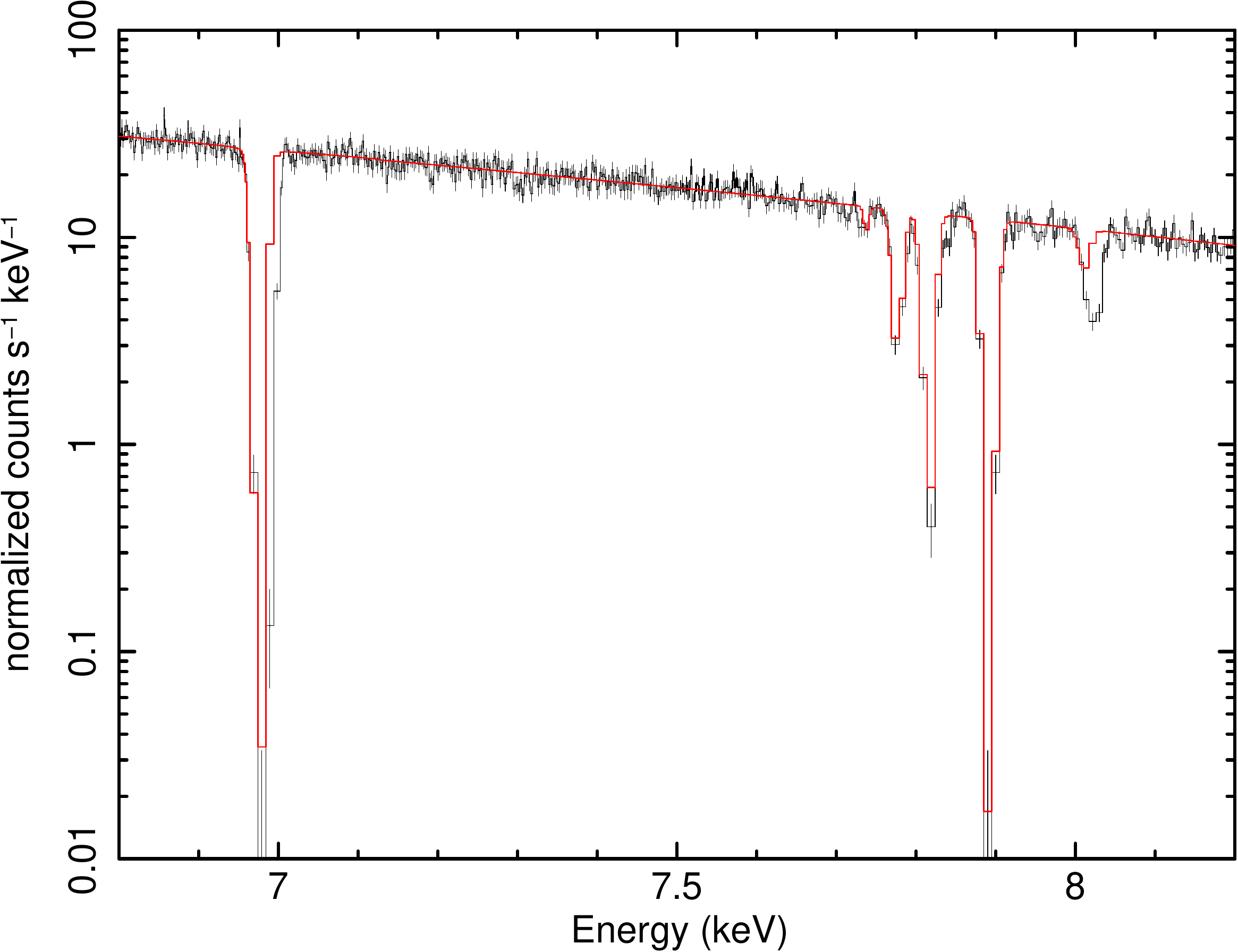}
\end{center}
\caption{Top: The {\it Chandra}/HETG spectrum of GRO J1655$-$40 is
  shown here, fit with two XSTAR photoionization models: a dominant
  highly-ionization zone with a velocity of 400~km/s, plus a tentative
  even more highly ionized zone that is outflowing at an even higher
  velocity (1400~km/s).  Bottom: A simulated 100~ks {\it ASTRO-H}/SXS
  spectrum is shown here, based on the model fit to the {\it Chandra}
  spectrum in the panel above.  The model in red shows only the
  dominant zone in the panel above.  The remaining blue flux in the Fe
  XXV He-$\alpha$ line at 6.700~keV and the strong blue flux in the
  line at 8~keV clearly show the need for the second zone.  X-ray warm
  absorbers in Seyfert AGN are generally thought to be componsed of at
  least two ionization zones; this simulation indicates that the
  {\it ASTRO-H}/SXS will be able to distinguish different zones within very
  ionized outflows, potentially building a stronger connection to
  X-ray warm absorbers in AGN.}
\end{figure}


\subsubsection{Do Ultra-Fast Outflows (UFOs) Span the Mass Scale?}
In deep spectra of Seyfert-1 AGN observed with {\it XMM-Newton} and
{\it Suzaku}, a number of absorption lines may be detected that would
signal winds with jet--like velocities (0.1--0.2$c$ in some cases \citep{Tombesi2010, Tombesi2010b}. 
 These lines are of modest statistical significance
after rigorous assessment via Monte Carlo techniques.  These putative
UFOs would surely have enormous impacts on their host environments,
and confirming such outflows is therefore of high importance.

An extremely fast outflow has also been found in IGR J17091$-$3624,
and at higher statistical significance than the AGN examples ($v = 0.03c$, King et al.\
2012).  Indeed, there is some evidence for two velocity systems, at
0.03$c$ and $0.05c$.  This source is a "twin" of the near-Eddington
microquasar GRS 1915$+$105, and if it also accretes near the Eddington
limit, it would have another factor in common with AGN in which UFOs
have been observed (the best case is likely PG 1211$+$143; \citealt{Tombesi2010}).
  It is possible that high velocity outflows have previously
eluded detection simply owing to a lack of sensitivity with increasing
energy, both in CCD and gratings spectrometers.  The energy resolution
of the {\it ASTRO-H} SXS increases with energy, and will aid the detection
of weak lines with strong velocity shifts.

Figure 7 shows a simulated 50~ksec SXS spectrum of IGR J17091$-$3624,
next to the observed {\it Chandra} spectrum (King et al.\ 2012).  The
continuum measured in the HETG spectrum and XSTAR models used to
analyze those data were used to simulate the SXS spectrum.  The
lower-velocity system is characterized by ${\rm log}(\xi) = 3.3$ and
${\rm N}_{\rm H} = 4.7\times 10^{21}~ {\rm cm}^{-2}$.  The
higher-velocity system is characterized by ${\rm log}(\xi) = 3.9$ and
${\rm N}_{\rm H} = 1.7\times 10^{22}~ {\rm cm}^{-2}$.  The flux
measured with {\it Chandra} was approximately 0.1~Crab; this flux is
likely to give about 200~c/s in the SXS.  In this case, mitigations
are not needed to cope with the count rate, and a H+M count rate of
60~c/s is anticipated.  It is clear that both velocity systems are
easily detected, and easily separated, in a modest observation.

\begin{figure}
\begin{center} 
\includegraphics[width=0.5\textwidth]{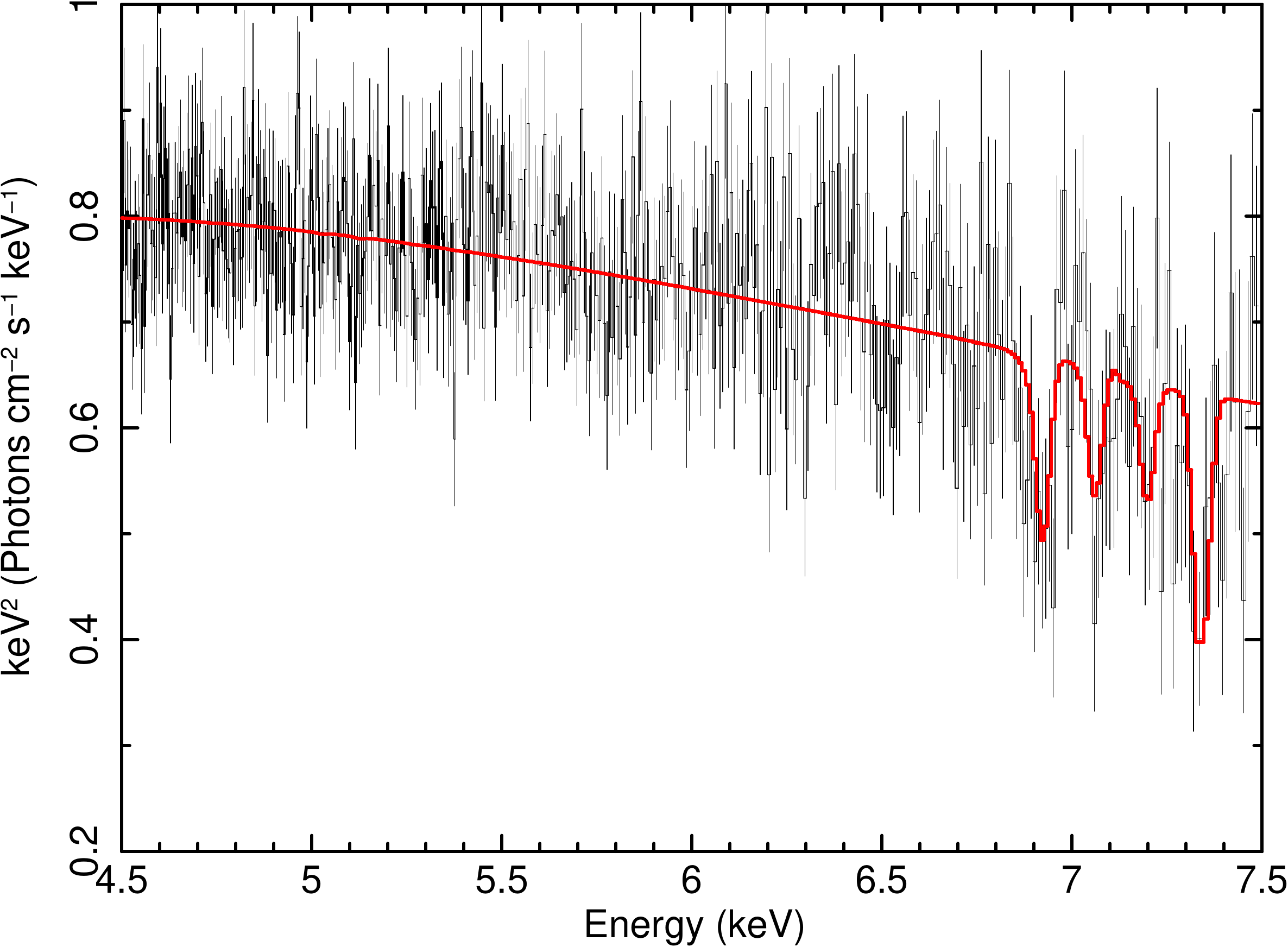}
\hspace{1.0cm}
\includegraphics[width=0.5\textwidth]{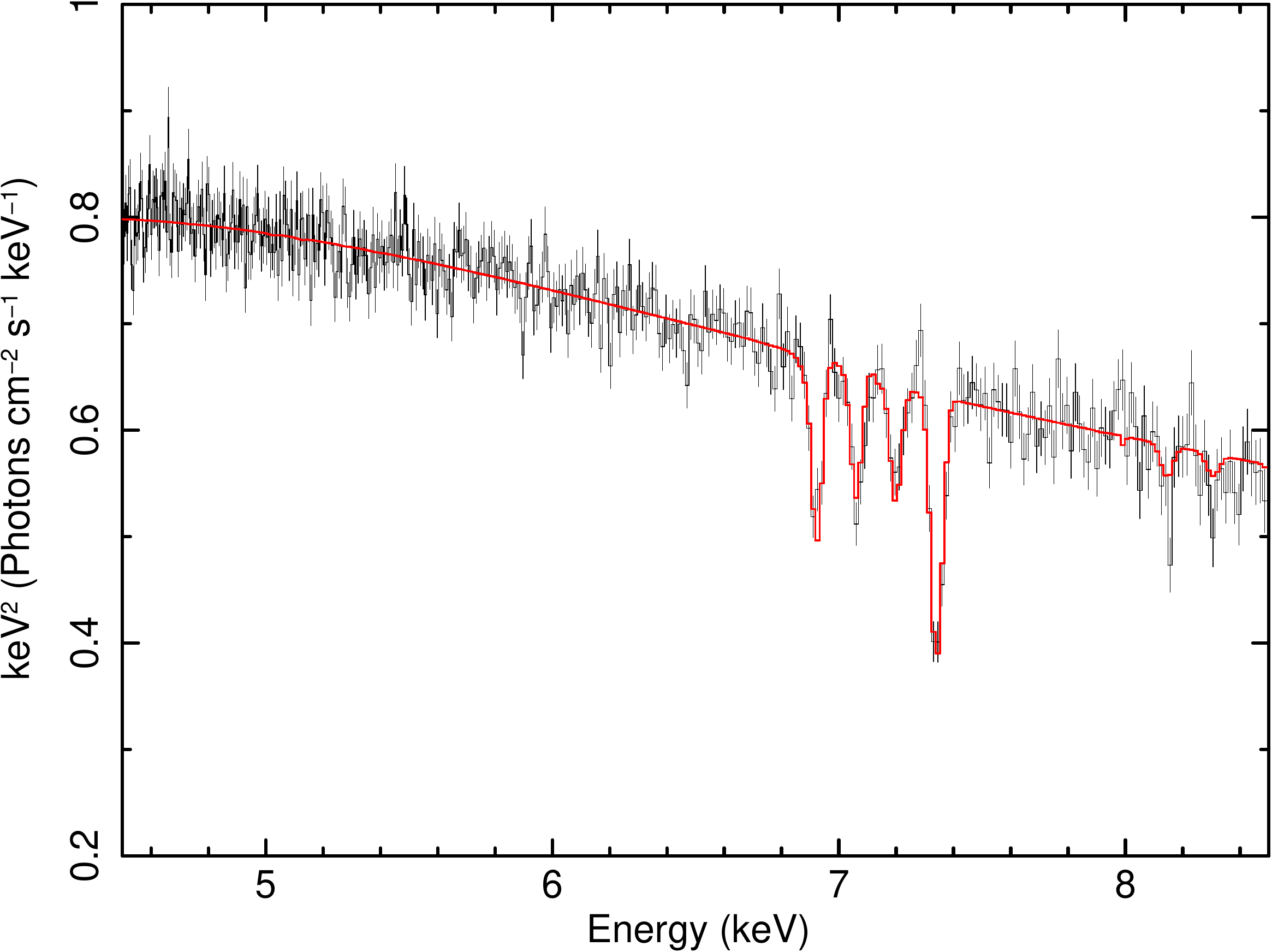}
\end{center}
\caption{Top: A {\it Chandra}/HETG spectrum of IGR J170914$-$3624 is
  shown here, fit with two XSTAR photoionization zones.  The outflow in
  this source is extreme; the components have velocities of $0.03c$
  and $0.05c$ \citep{KingMiller2012}.  Bottom: This plot shows a simulated 50~ks spectrum,
  based on the model shown in the panel above.  Whereas the components
  are difficult to detect with the {\it Chandra}/HETG, multiple fast
  outflow components can easily be detected, resolved, and separated
  at high resolution using the {\it ASTRO-H}/SXS.  }
\end{figure}


\section{Low-frequency variability}
\subsection{Links to accretion geometries} 
One of the best--known and most intriguing phenomena observed in both
black-hole X-ray binaries and AGN is short-term X-ray variability.
The fluctuations observed in many light curves are not periodic, nor
random around some mean; rather, they seem to comprise numerous peaks
(or shots) with different amplitudes and durations.  Such variability
may be a sort of $1/f$ noise, since its power spectrum exhibits a
$f^{-\beta}$ decline at high frequencies (where $f$ is the frequency
and $1 < \beta < 2$ is a constant).

At low frequencies (below a few Hz), the power spectra are roughly
flat.  For this reason, the broad band variability observed in black
holes is sometimes called low-frequency variability or low-frequency
noise.  Since similar variability is commonly observed in other
accretion systems, they are thought to be generic features of the
accretion process, but there are still no widely accepted models
despite intensive efforts having been done over many decades since its
discovery.

The normalized power spectra are quite similar among different
sources.  As the total fractional variability increases (or
decreases), the level of the flat part in the normalized power spectral densities (PSDs) also
increases (decreases), while neither that of the power-law decline
part nor its power-law slope changes significantly.  That is, the
break frequency separating the flat and power-law decline parts
decreases (increases) accordingly.  As a result, there is a linear
correlation between the fractional variability and the break frequency
\citep{BellonHasinger90i}.

It is important to note that low frequencies mean long timescales.
The timescale corresponding to a few Hz is a few tenths of a second,
much longer than the dynamical timescale,
\begin{equation}
 \tau_{\rm dyn} \equiv \sqrt{\frac{r^3}{GM}}
    \simeq 4~\left(\frac{M}{10~M_\odot}\right)
             \left(\frac{r}{10~r_{\rm S}}\right)^{3/2}~ {\rm ms}
\end{equation}
(with $M$ and $r$ being the mass of a black hole and the distance from
the black hole, respectively) or the thermal timescale ($= \tau_{\rm
  dyn}/\alpha$ with $\alpha\sim 0.1$ being the viscosity parameter) of
an accretion disk.  By contrast, the viscous (accretion) timescale is
much longer;
\begin{equation}
  \tau_{\rm vis}
       \equiv \frac{1}{\alpha} \left(\frac{r}{H} \right)^2 \sqrt{\frac{r^3}{GM}}   
       \simeq  4 \left(\frac{\alpha}{0.1}\right)^{-1}                           
                         \left(\frac{M}{10~M_\odot}\right)
                         \left(\frac{T}{10^{10}~{\rm K}}\right)^{-1}
                         \left(\frac{r}{10~r_{\rm S}}\right)^{1/2} {\rm s}    
\end{equation}
where $H$ denotes the half-thickness of the flow, and $T$ denotes its
(ion) temperature.  This means that X-ray variability is not a local
phenomenon, but somehow reflects broader trends in the accretion rate.
Indeed, the basic properties of the X-ray light curves can be
reproduced by propagation of density fluctuations (e.g.,
\citealt{Mineshige1994,Lyubarskii97}).

Further, the character of the observed X-ray variability is closely
related to the spectral state of the accretion disk.  Rapid X-ray
fluctuations are more pronounced during the hard (low) state and the
``very high'' state, although there exist small fluctuations during
the soft state as well \citep{Miyamoto1992}.  In contrast,
variability is always observed in AGN.  In both of the low/hard and
very high states, radiation mainly originates in the hot accretion
flow, rather than in the (relatively) cool, geometrically thin disk,
which dominates in the high/soft state.  Since magnetic activities are
more enhanced in the hot accretion flow, X-ray variability could be
caused by magnetic flares and the rapid release of magnetic energy,
perhaps akin to solar flares.

\subsection{Low-frequency QPOs and Lense-Thirring precession}

Low frequency QPOs are commonly observed in the X-ray flux of both
neutron star and black hole X-ray binaries. They are most clearly
observed as strong, {\it in}coherent features in the power spectral
density (PSD) which are Lorentzian in shape and so can be described by
amplitude (i.e. fractional rms variability), centroid frequency
($f_{QPO}$) and width ($\Delta f$). These properties are observed to
be correlated with the spectral properties of the source and with
other noise components (see e.g. \citealt{vanderKlisbook,Bellonibook2010}).

\smallskip
\begin{figure}
\begin{center}
  \includegraphics[width=0.3\hsize]{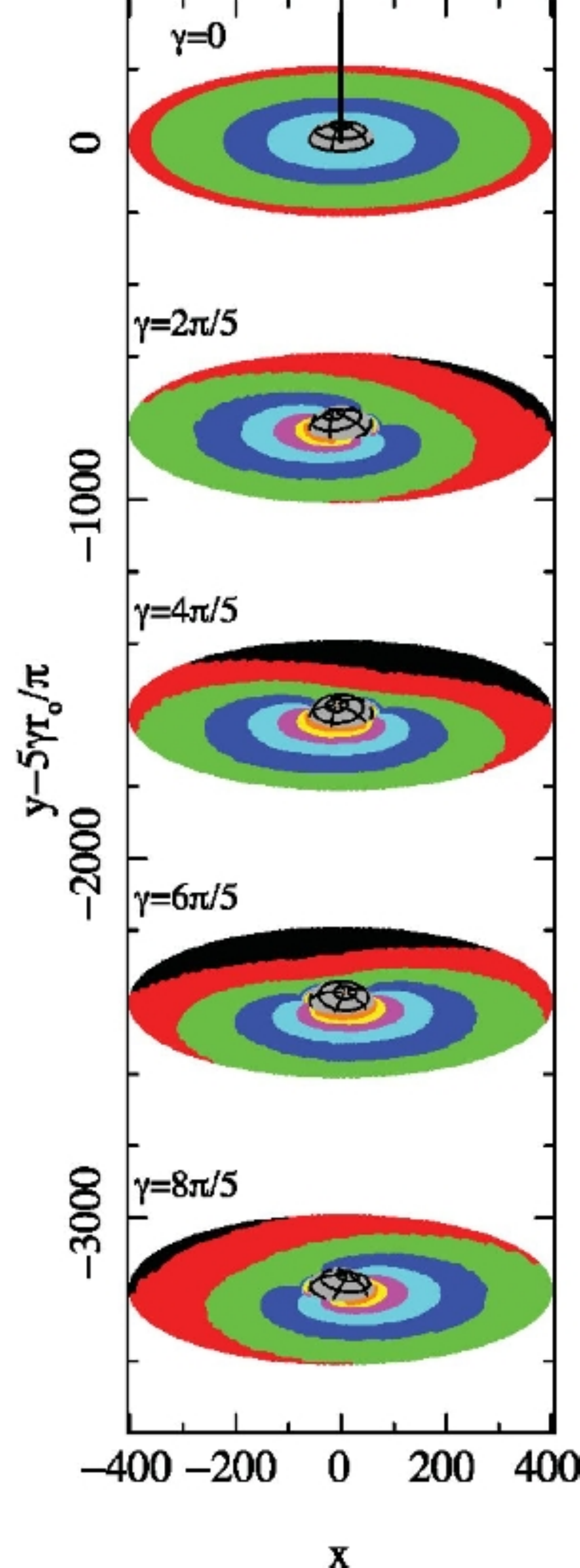}
\end{center}
\caption{Lense-Thirring precession of a hot inner flow is predicted
  illuminate different azimuths of the disk as a function of QPO
  phase\citep{IngramDone2011,IngramDone2012}.  This figure depicts a hot inner flow in gray with wire
  mesh, and the colors illustrate the intensity of illumination on the
  disk as the inner flow precesses.}
\end{figure}

Different models exist for the evolution of accretion flows across an
outburst, and numerous different models have been proposed to describe
the ``low/hard'' state in particular and/or $L << L_{Edd}$ regimes
generally (e.g. \citealt{Esin1997, beloborodov99, BlandfordBegelman1999, Markoff2001J1118, Taametal2008,MeyerHofmeisterLiu09}). 
 As the mass accretion rate
through the disk falls, cooling will eventually fail and the inner
disk will give rise to an inner hot accretion flow.  The Eddington
fraction at which this happens is uncertain, and it is possible that
it may vary even for a given source, depending on whether the
accretion rate is rising or falling.  Positive evidence of an absent
geometrical component is difficult to obtain, but some recent evidence
suggests that disks may truncate at or below $0.001~L_{Edd}$ (\citealt{tomsick09gx,reisj1118, reislhs})

The truncated disk model, in which the thin disk only extends down to
some radius $r_{\rm o}$, can potentially explain the evolution of the
SED \citep{Esin1997,donereview2007} and the
correlated evolution of the power spectra \citep{IngramDone2011,IngramDone2012}. 
 Interior to $r_{\rm o}$ is a large scale height,
optically thin accretion flow which acts as the
Comptonizing corona.  As the source flux increases, the truncation
radius moves inwards, thus increasing the flux of disk photons
incident on the flow and softening the power-law emission while
simultaneously decreasing all characteristic time scales associated
with $r_{\rm o}$.  The evolution of both the energy spectra and power
spectra imply that $r_{\rm o}$ moves from $\sim 60 -6$ (in units of
$R_{\rm g}=GM/c^2$) during the rise to outburst and back out again
during the fall back to quiescence.

This also gives a framework in which to incorporate the QPO and its
properties via Lense--Thirring precession.  This is a relativistic
effect that occurs because a spinning compact object drags spacetime
as it rotates.  The orbit of a test particle that is outside the plane of
black hole spin will therefore undergo precession because the starting
point of the orbit rotates around the compact object.  \citet{StellaVietri98} 
showed that the predicted frequency of a test mass at
the truncation radius is broadly consistent with the observed QPO
frequency and its evolution from $\sim 0.1-10$~Hz as $r_{\rm o}$ moves
inwards and the source spectrum softens. However, the energy spectrum
of the QPO is dominated by the Comptonized emission \citep{Sobolewska2006, Rodriguez04},
 requiring that the QPO mechanism
predominantly modulates the hot flow rather than the disk (although
the variability could be produced elsewhere before propagating into
the flow; \citealt{wilkinsonuttley09}). Instead, a global precession of the entire
hot flow modulates the Compton spectrum by the changing projected area
of the flow as it precesses.  This explains the QPO spectrum as well as
its frequency, giving a physical model for the origin of the
QPO which fits its known behaviour \citep{IngramDone09}.

One clear observational signature arising from this model, is that the
flow preferentially illuminates different azimuths of the disk, giving
rise to a periodically rocking of the reflected iron line between a
red and blue shift (see Figure 9).  There should be a
clear pattern to the QPO, where the rising QPO amplitude means that
the side of the disk coming towards us is illuminated giving a blue
line, while the falling QPO phase illuminates the disk rotating away
from us, giving a red line.

Indeed, earlier observational work was able to connect low--frequency
QPOs to variations in Fe K lines, and the results were found to be consistent
with Lense-Thirring precession.  \citet{MillerHoman05} found a link
between Fe K line properties and the ``phase'' of very strong 1--2~Hz
QPOs observed in GRS 1915$+$105 using {\it RXTE}.  Unfortunately, the
limited spectral resolution of {\it RXTE} prevented a detailed study of the
line properties.  However, based on that result, \citet{SchnittmanHomanMiller2006} 
explored a model consisting of a precessing inner disk
within a stable corona, and calculated the resulting line profile.
This model differs from the model discussed above in some details, but
the fact of competing theoretical frameworks and an encouraging
observational result highlight the need to explore such phenomena
using {\it ASTRO-H}.

This is not feasible to see with the SXS, as the upper limit of 35 c/s
is too low to build-up enough statistics in a reasonable exposure
time (see Figure 8).  This count rate is a factor 50 below the {\it XMM-Newton} count
rate seen in its high time resolution modes of a bright black hole
binary.  However, the HXI has no pileup issues.  It also has an
effective area (with 2 telescopes) that is comparable to the {\it RXTE}/PCA
and larger than that of {\it XMM-Newton}/EPIC-pn camera above 6~keV.
The energy resolution is comparable to the {\it RXTE}/PCA, especially when
taking into account the fact that all fast time modes of the PCA were
binned in energy resolution after the loss of the high gain antenna
early in the mission.  The HXI then has significant discovery space
(along with {\it NuSTAR}) in terms of the spectra of fast time variability
in black hole X-ray binaries.

\begin{figure}
\begin{center} 
\includegraphics[width=0.315\textwidth]{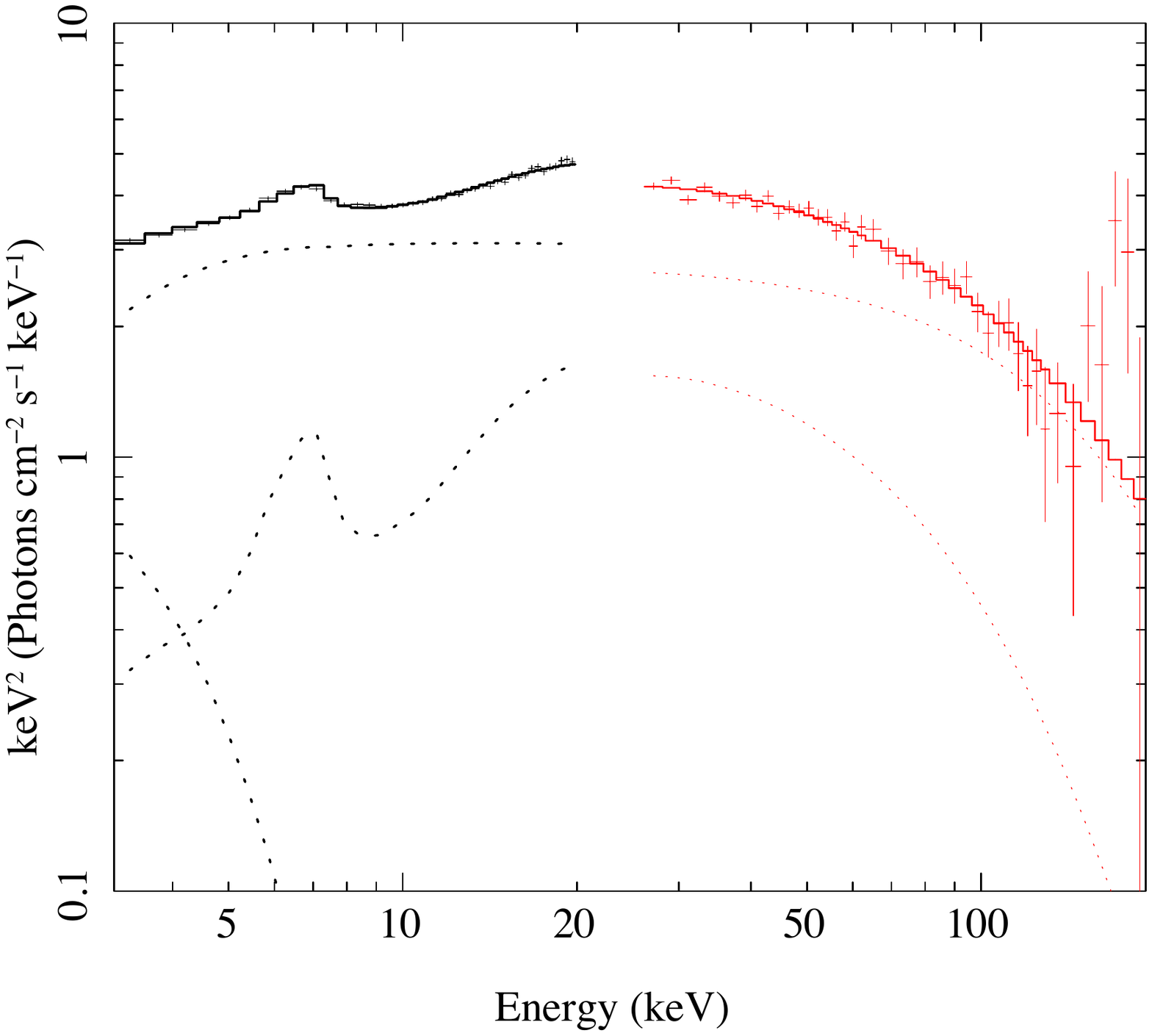}
\hspace{0.0cm}
\includegraphics[width=0.33\textwidth]{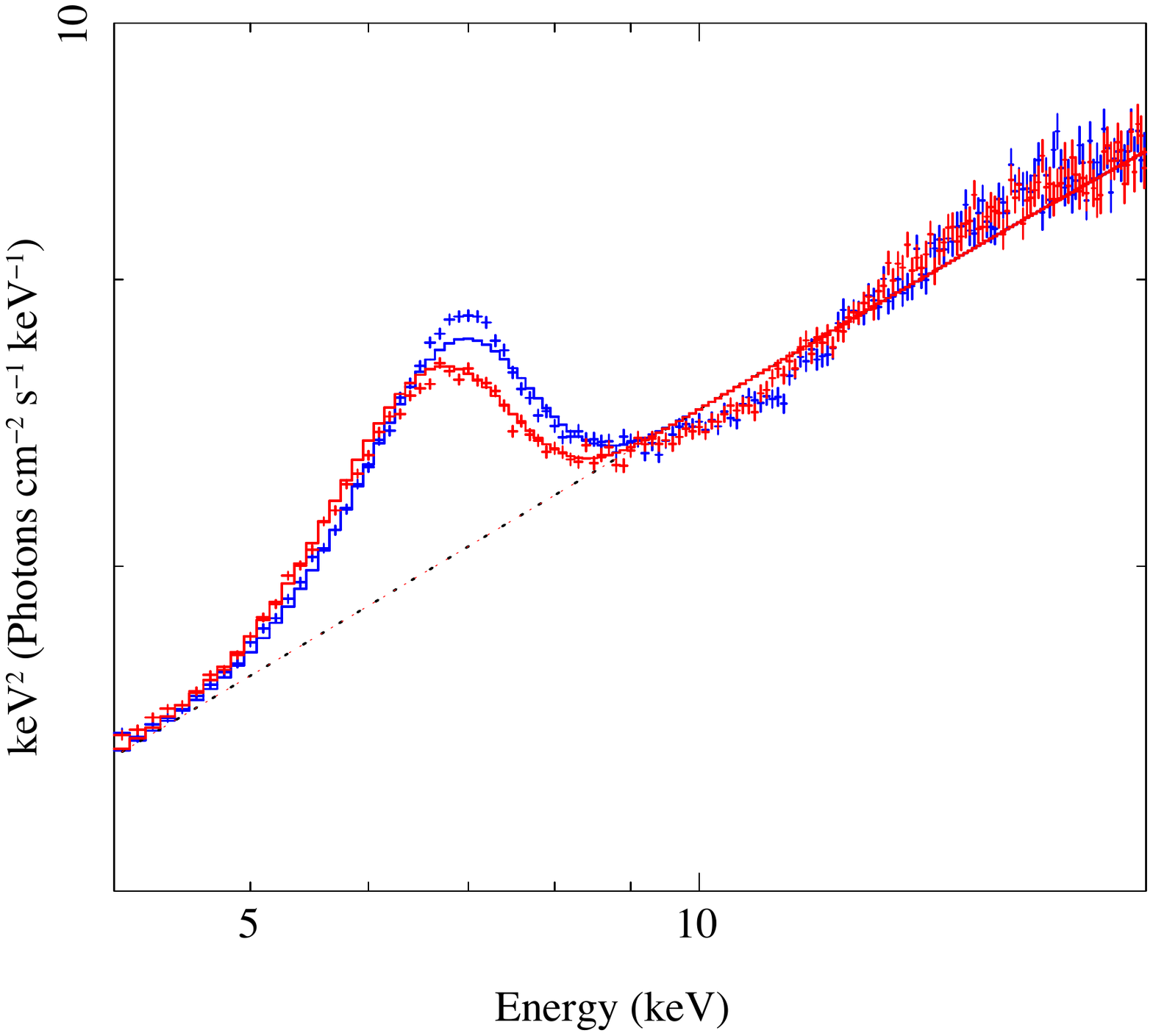}
\hspace{0.0cm}
\includegraphics[width=0.325\textwidth]{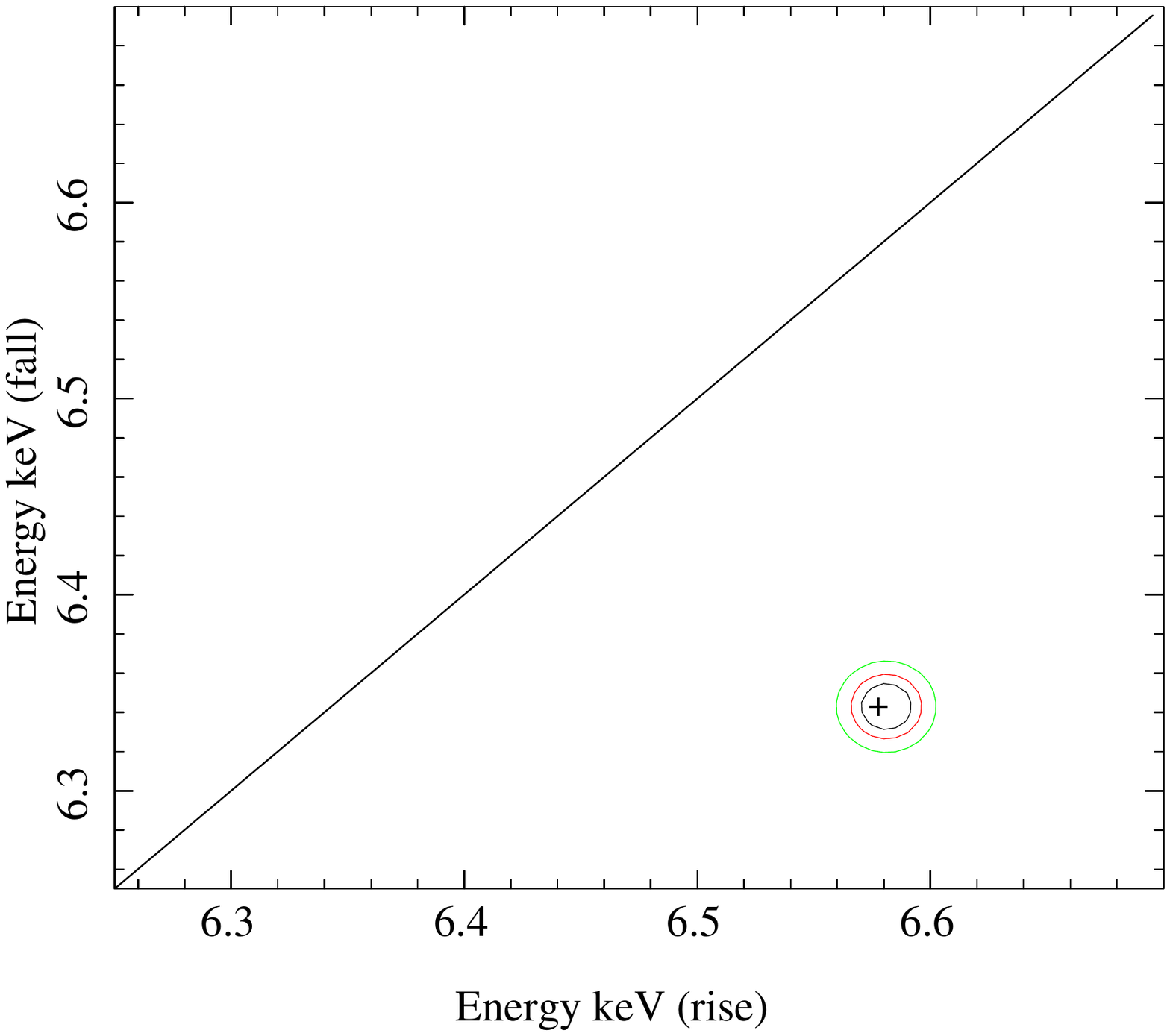}
\end{center}
\caption{a) The {\it RXTE}/PCA and HEXTE data for a bright hard intermediate
  state of GX339-4 with strong QPO are shown here.  b) A model
  simulated through the HXI and split into rising (blue) and falling
  (red) QPO phase is shown in this panel.  c) This panel depicts the statistical confidence at which changes in line profile due to the QPO
  precession illuminating different azimuths can be detected (A. Ingram, private communication).  The black, red, and green contours correspond to $\Delta \chi^{2} = 2.3, 4.61, {\rm and}~ 9.21$, respectively.}
\label{QPO_figure2} 
\end{figure}

Figure \ref{QPO_figure2}a shows a typical hard intermediate state
spectrum such as seen from GX339$-$4 with a strong QPO as seen by
{\it RXTE}.  Simulating this through the HXI gives a count rate of 300 c/s
for one telescope, so we double this to get the total count rate from
both detectors.  We construct a 100~ks HXI observation, as shown in
figure \ref{QPO_figure2}b.  Selecting data on rising phase and falling
phase of the QPO then gives significantly different spectra.  These
are shown as a ratio to a power-law in figure \ref{QPO_figure2}c.
This quasi-periodic shifting of the iron line peak energy with QPO
phase is a prediction of the Lense-Thirring precession model for the
low-frequency QPO that can be tested using the {\it ASTRO-H}/HXI.

\subsection{Shot noise studies}
Flux variability on $\sim$ms time scales (e.g., \citealt{Miyamoto1992}),
seen in spectrally hard states, has been studied in many ways (e.g., \citealt{Nowak1999gx, Poutanen2001, Pottschmidt2003cyg,Torii2011}).  However, the origin of such variability
is still uncertain, presumably owing to the difficulty of realizing
both high sensitivity and large effective area.  A distinctive
approach to understanding this variability is ``shot analysis''
\citep{Negoro94,Negoro95}.  The method is a time-domain
stacking analysis to determine the universal properties behind
non-periodic variability.  Shot analysis makes it possible to combine
timing analysis with spectral information in a straightforward manner.
This technique was adopted for observations of Cyg X-1 obtained with
{\it Ginga} \citep{ginga}, and was further explored with RXTE \citep{Focke05}.

\begin{figure*}[h]
\begin{center}
\vbox{}
\includegraphics[width=0.95\textwidth]{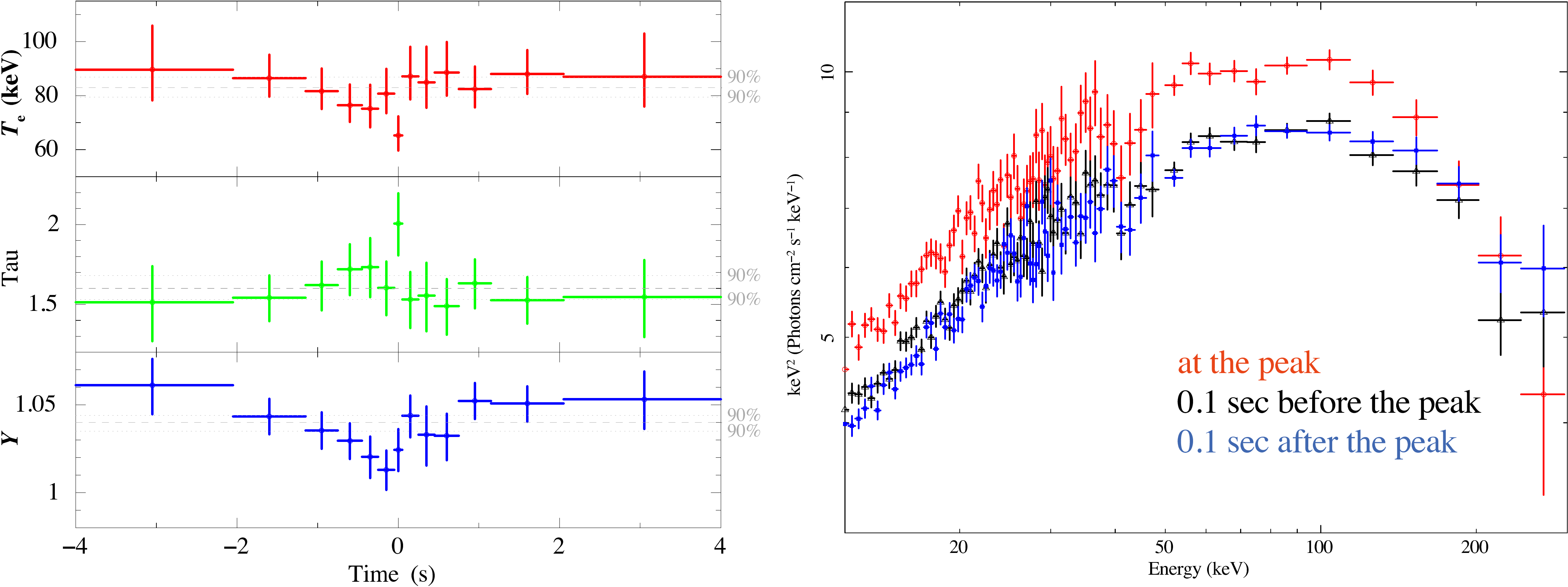}
\caption{Left: The time evolution of parameters obtained from
  Comptonization modeling of {\it Suzaku} spectra of Cygnus X-1 are
  shown here.  Fits were made to the 12--300 keV HXD spectra, in the
  13 shot phases.  From top to bottom, the electron temperature, the
  optical depth, and the Compton $y$ parameter are presented.  The
  90\% confidence range specified by the time-averaged spectrum is
  superposed by dotted and dashed lines.  Right: The
  background-subtracted $\nu F \nu$ HXD spectra, accumulated over
  different shot phases, are shown here\citep{2013ApJ...767L..34Y}.  The spectra were integrated
  from --0.25 to --0.05 seconds before the peak (black), from --0.05
  to 0.05 seconds around the peak (red), and from 0.05 to 0.25 seconds
  after the peak (blue).}
\label{yamada_figure1}
\end{center}
\end{figure*}

Shot analysis was recently applied to {\it Suzaku} data from Cygnus
X-1 in an effort to better understand stochastic phenomena in a
straightforward way \citep{2013ApJ...767L..34Y}.  In this work, the high energy
limit of the shot analysis was extended up to $\sim$ 200 keV, by
utilizing data from both the HXD and XIS cameras.  The interesting outcomes of this analysis include the following results:

\begin{itemize}
\item The shot feature is found at
least up to $\sim$ 200 keV, with high statistical significance. 
\item The shot profiles were found to be approximately symmetric,
  though the hardness changes progressively more asymmetrically toward
  higher energies of $E \gtrsim 100$ keV.
\item The 10--200 keV spectrum at the peak shows a lower
energy cut-off than the time-averaged spectrum.  
\item Within the framework of a single-zone Comptonization model for
  the cut-off, as a shot develops toward the peak, $y$ and
  $T_{\rm{e}}$ decrease, while $\tau$ and the flux increase.
  Immediately after the shot peak, $T_{\rm{e}}$ and $\tau$ (and hence
  $y$) suddenly return to their time-averaged values.  This fitting
  result and the spectrum at the peak, and that in 0.1 s before/after
  the peak are shown in figure \ref{yamada_figure1}.
\end{itemize}

The time constant of $\sim$1 s far exceeds the local (dynamical or
thermal) timescale of the innermost region, and should thus reflect
the motion of particular gas elements.  \citet{Manmoto96} proposed an
interesting explanation: that inward moving accreting blobs can cause
an increase in X-ray flux, and this is reflected as sonic wave when it
reaches the black hole \citep{Kato08book}.  Rapid heating, such as
might occur through magnetic reconnection, can explain the short
($\sim$ 0.1s) timescale of the shots, though the long ($\sim$ 1s)
timescale would be related to mass accretion time scale.  Further
observational studies are needed to completely understand the physics
causing the rapid variability.  For instance, shot profiles with
distinct features, such as polarization (Laurent et al.~2011),
$\gamma$-ray emission \citep{Ling1987}, or fine structures around Fe-K
lines, all of which will be precisely measured by {\it ASTRO-H}, could
provide a better understanding of the origin of the rapid flux
variability.

\section{Accretion flow evolution}

\subsection{Explorations of the low/hard state with {\it ASTRO-H}}
The canonical ``low/hard'' state of black hole X-ray binaries actually
encompass a large range in luminosity and accretion rate, with an
upper bound of a $few \times 10^{-2}~ L_{Edd}$.  Black holes
transiting through the low/hard state from X-ray quiescence, or back
into quiescence following an active period, can potentially be seen as
the evolutionary link between LLAGNs (often observed at $\sim10^{-6}
L_{Edd}$, similar to the super-massive black hole in M81) and Seyfert
AGN or even quasars accreting close to $L_{Edd}$.

It is generally agreed that the accretion flow properties and overall
geometry of the inner regions surrounding black holes differ markedly
between quiescent and quasar-like states.  Sources that are inferred
to accrete close to their Eddington limit reveal clear signatures of a
standard optically--thick but geometrically--thin accretion disk
extending close to the black hole.  In such phases, a compact, hard
X-ray emitting corona is also inferred through various means.  In
contrast, at very low fractions of the Eddington limit, a standard
thin disk is replaced with a radiatively--inefficient accretion flow.
So-called advection--dominated accretion flows (ADAFs, \citealt{NarayanYi1994,Esin1997}) are the best known flavor of
such models.  Some studies have found that ADAFs are convectively
unstable \citep{Quataert00CDAF}, and others have even found that the hot flow
may not be bound to the black hole (Blandford \& Begelman 1999).
Whatever the details, it is clear that at some point during the
low/hard state, the inner edge of the standard thin accretion disk
must truncate.  {\it ASTRO-H} will be able to incisively probe transient
black hole binaries during the decline or rise of the low/hard state,
down to luminosities $\sim10^{-3} L_{Edd}$.

{\it ASTRO-H} will provide continuous, sensitive coverage in the 5--80~keV
energy range via its HXI detector.  It will enable the precise
characterization of the disk reflection spectrum and a measure of the
overall strength of the Compton hump.  Combined with
simultaneous observations via the SGD -- spanning energies up to
600~keV -- {\it ASTRO-H} will be ideally suited to observe the expected
spectral rollover in the $100-200$~keV range due to thermal
Comptonization in addition to disk reflection, and it will be able to
do this fairly deep into the low/hard state.  Simulations indicate
that 100~ks observations of BHB at $\sim10^{-2} L_{Edd}$ will provide
reliable estimates of coronal temperatures, with errors no larger than
10\% for corona temperatures as large as 200~keV.

The unprecedented resolution and sensitivity of the SXS in energies up
to and beyond the Fe K region ($\sim$4--8~keV), especially when
combined with the HXI, will provide strong constraints on the
inclination and ionization state of the accretion disk.  These two
parameters are derived partly based on the overall strength of the
Compton hump at $\sim30$~keV and the depth of the Fe K absorption
edges in the 7.1--9.3~keV range.  Simulations show that {\it ASTRO-H} will
determine the ionization and inclination of systems like GX 339$-$4 at
$\sim10^{-2} L_{Edd}$ with 90\% statistical errors of $\leq$1\% and
$\leq$3\% respectively. For comparison, the equivalent precisions
obtained without the soft X-ray coverage, are 13\% and 22\%,
highlighting the immense importance of the broad-band coverage
provided by {\it ASTRO-H}.

The reprocessing of hard X-rays in the (relatively) cold accretion
disk results in the production of several fluorescence and
recombination lines; Fe K lines are merely the strongest in most
circumstances.  {\it ASTRO-H} will be sensitive to features from low-Z
elements (Figure 11), and their detection will further constrain the
ionization state of the accretion disk as the system evolves through
the low/hard state.  If the disk begins to recede at
$L\sim10^{-2}L_{Edd}$, {\it ASTRO-H} -- due to its broadband coverage and
highly sensitive soft X-ray detector -- will be in a position to map
this evolution with 90\% statistical errors of less than 10\% for an
inner accretion disk radius of 10 $R_{g}$ at
$L\sim1\times10^{-2}L_{Edd}$ and 35\% for a disk at 100 $R_{g}$ at
$1.4\times10^{-3}L_{Edd}$.

\begin{figure}
{\hspace*{1.5in}
 \rotatebox{0}{
{\includegraphics[width=8.5cm]{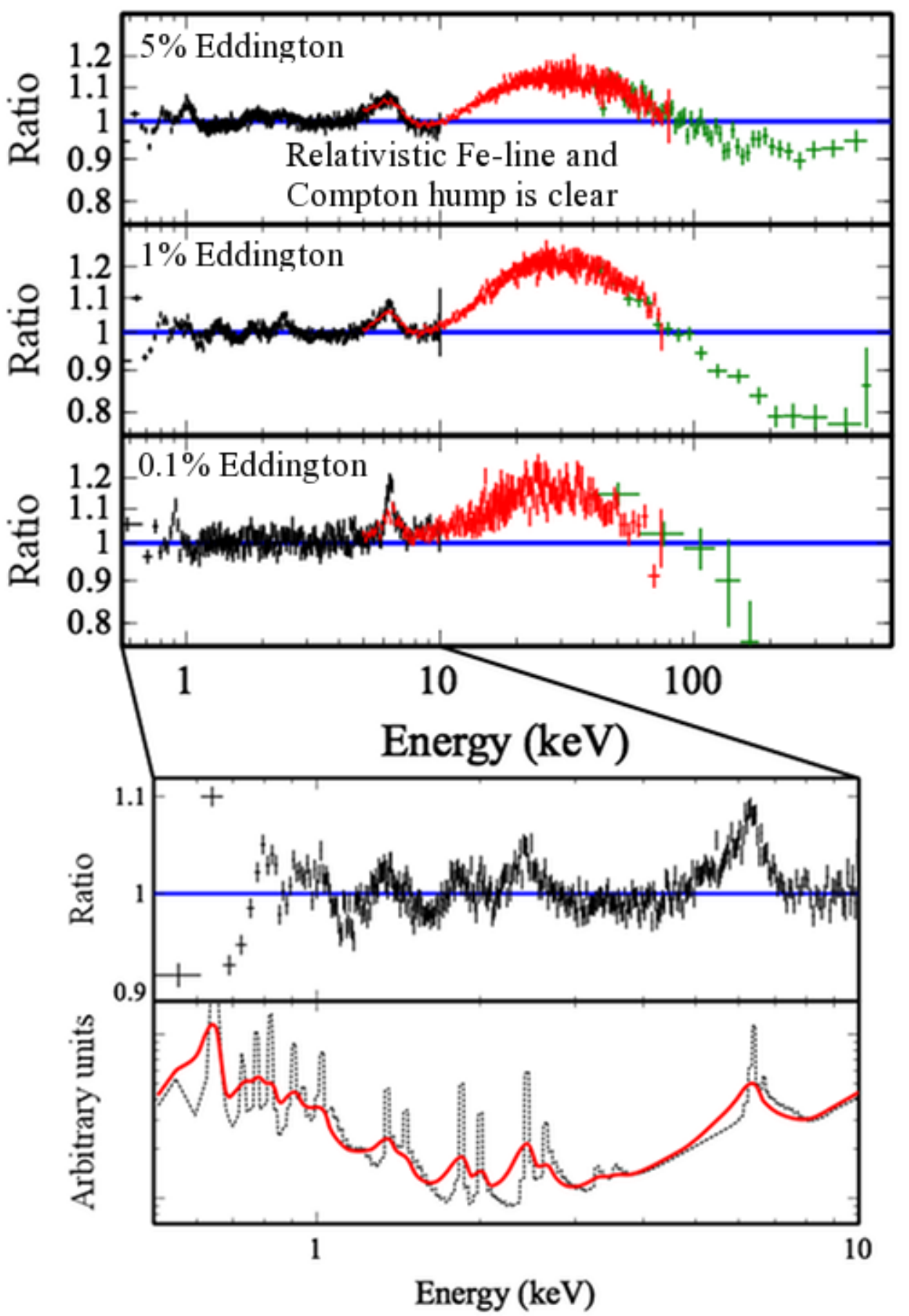}  
}}}\vspace*{-0.15cm}
\caption{Simulated data-to-model ratio of GX 339$-$4 at
  $L/L_{Edd} =5, 1$ and 0.1\%. The inner accretion disk is assumed to
  extend to the ISCO at 2.32 $R_{g}$ at $L/L_{Edd}=0.05$ and increase to
  10 $R_{g}$ and 100 $R_{g}$ as the luminosity decreases. In all cases, the
  simulated spectra were fit with an absorbed power-law together with
  a thermal disk component with the iron line and Compton hump
  energies excluded from the fit. These figures show that {\it ASTRO-H} will
  detect the clear signatures of reprocessed emission from which the
  radius of the accretion disk can be measured. The close up shows the
  data-to-model ratio for the system at 1\% Eddington (top) with the
  plot below showing the expected signatures from reflection before
  (black) and after (red) the effects of strong gravity.  }
\end{figure}

\begin{figure}
\begin{center}
\includegraphics[width=0.65\hsize]{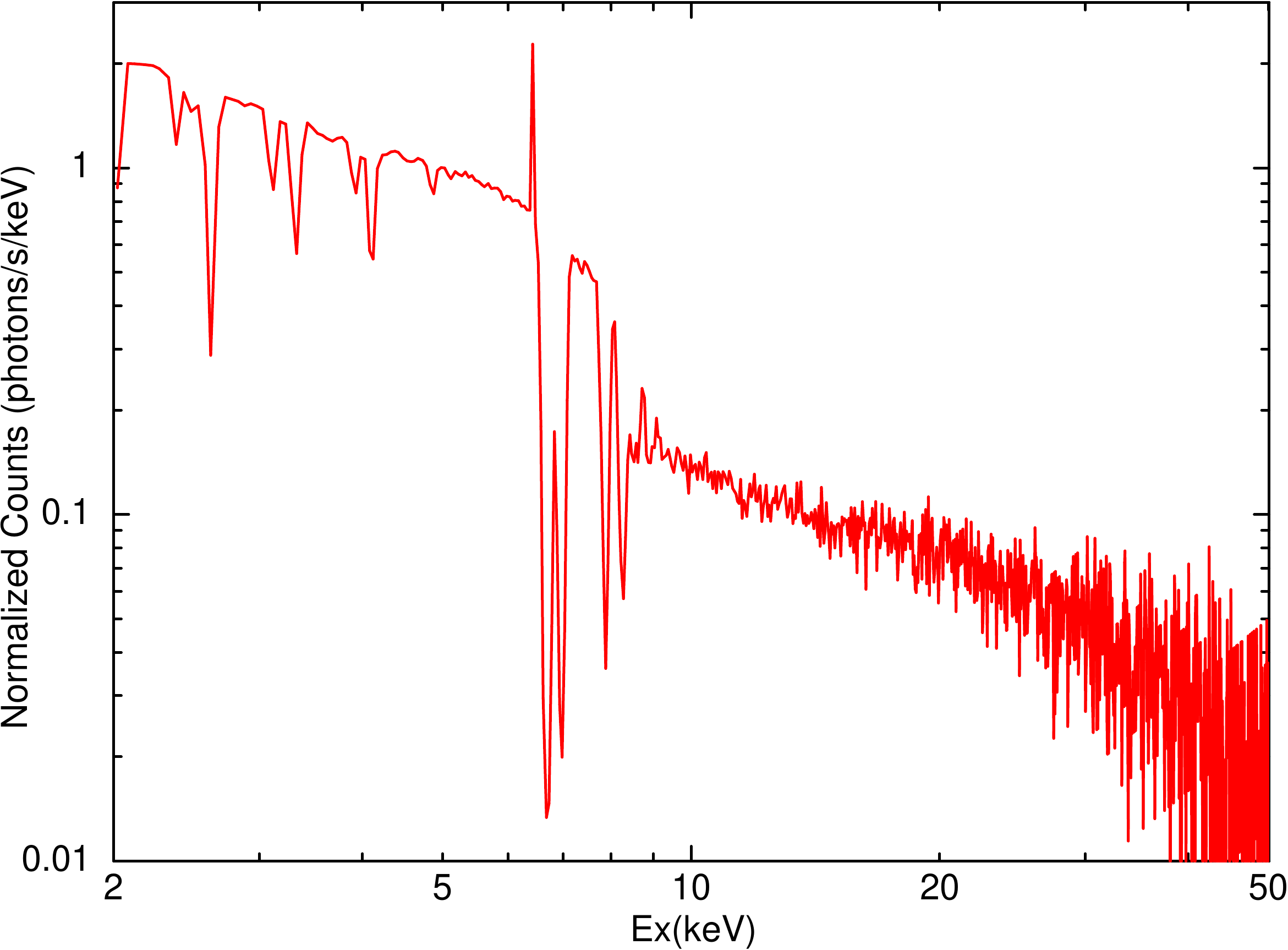}
\end{center}
\caption[corona] 
{ \label{fig:corona} 
Preliminary simulation of an emergent (model) spectrum from a
partially ionized corona. The corona is assumed to be a sphere of
Compton optical depth $\tau=5$ and $kT_{\rm e}=kT_{\rm i}$=10\,keV.
Soft photons are injected at its center. These parameters represent a
rather extreme case for demonstration, though a step like feature is
found at iron K band.}
\label{hayasida_figure1}
\end{figure}

\subsection{Tests of the hard X-ray corona with {\it ASTRO-H}}
A power-law spectral component is particularly prominent in the X-ray
spectra of black hole X-ray binaries in the ``low/hard'' and ``very high''
states.  Similar components are observed in the X-ray spectra of
Seyfert AGN.  Inverse Comptonization of soft photons from an
(optically thick) accretion disk by electrons in an (optically thin)
corona is one plausible means of generating power-law X-ray emission.
The corona is generally characterized with electron temperatures of
$10^{9}$\,K and the Compton $y$ parameter of 1, according to hard
X-ray observations of black hole X-ray binaries and a limited number
of Seyferts.

Very recently, a high energy cut-off was detected in the spectra of
several NLS1s (a class of radio quiet AGNs with low black hole masses
and high accretion rate), and some of the sources show electron
temperatures lower than 60~keV \citep[e.g.][]{Malizia2009}.  Changes in the electron temperature  were also recently
observed in the Seyfert 1.9 NGC 4151: $kT_e = 50$--70~keV was recorded
in a bright state, while $kT_e = 180$--230~keV was measured in a dim
state \citep{Lubiski10}.  These results are consistent with the
thermal Comptonization model, considering the Compton cooling is more
efficient for a larger accretion rate.
    
Extensive studies have been made to account for the X-ray power-law
component and its high energy cut-off in terms of a coupled
disk--plus--corona system (e.g.  \citealt{HaardtMaraschi91,Liu03corona, Kawabata10, MeyerHofmeister2012}).
A significant fraction of energy can be dissipated in the corona, even
when mass is accreted primarily through the disk.  Ions in corona are
typically assumed to be fully ionized in these studies.  According to
\citet{Liu2002corona}, the ion temperature is as high as $10^{11}$~K, two
orders of magnitude higher than the electron temperature at the
vicinity of a central black hole.  The ion temperature is $10^{12}$~K
in ADAF models (e.g. Narayan \& Yi 1994).  However, the low electron
temperatures derived from model--dependent fits to the spectra of some
sources are not in the range of these theoretical models.
Furthermore, considering the fact that evaporation and condensation of
materials between corona and disk must happen, there could be ions
with temperatures between the disk temperatures ($10^5-10^7$~K) and
the corona temperatures ($10^{9}-10^{11}$~K).  

The combined capabilities of the {\it ASTRO-H}/SXS and HXI may be able to
detect the presence of partially-ionized (or even neutral) ions within
the corona.  This may be of particular importance in low--temperature
coronae with high optical depth, like those noted in the above
discussion.  Rather than a smooth continuum, cooler components within
the corona may have the effect of causing atomic absorption features,
and a particularly pronounced step in the region of the Fe K edges
\citep{Hayashida2007PTh}.  Figure 12 shows a simulated {\it ASTRO-H}
spectrum based on the mixed corona model of Hayashida et al.\ (2007).

\subsection{X-ray Continuum Emission}
Establishing a complete picture of the geometry and emission
mechanisms in the central accretion flow onto black holes remains an
important observational goal.  The primary X-ray continuum in
Seyfert-I AGN, perhaps partially arising through Comptonization in
coronae near the black hole, has long been approximated by a single
power-law.  Novel techniques - such as the construction of difference
spectra - can reveal the nature of the primary continuum; deep
observations suggest that it obeys a power-law form to within 10\% in
the 3--10~keV band \citep{Vaug04}.  X-ray spectroscopy of
Seyferts has therefore focused on deviations from a power-law
continuum, including complex absorbers, reflection, and fluorescence
lines.  

However, broad--band {\it Suzaku} studies of several Seyfert-1s,
utilizing a novel variability-assisted spectroscopy method called the
C3PO (Count-Count Correlation with Positive Offset) method, has
suggested that the continuum may consist of multiple primary emission
components \citep{Noda2011a,Noda2011b}.  This means that the Compton
coronae constituting the central engines of AGN may consist of
multiple zones, as was suggested for some black hole X-ray binaries
(e.g.,\citealt{makishimacygx108}), in contrast to the conventional
belief of a single-zone corona.

Owing to the higher flux observed from Galactic stellar-mass black
holes, such techniques may be even more profitably employed to
understand continuum emission in these sources.  In this regard, the
broad-band spectral coverage and time resolution of the HXI and SGD
will be especially important (see Figure 13).  Even snapshot broadband spectra of
bright stellar-mass black holes with {\it ASTRO-H} can potentially
reveal any disagreement with simple, single-zone coronae.

\begin{figure}
 \begin{center}
 \vspace{-0.5cm}
  \includegraphics[width=80mm]{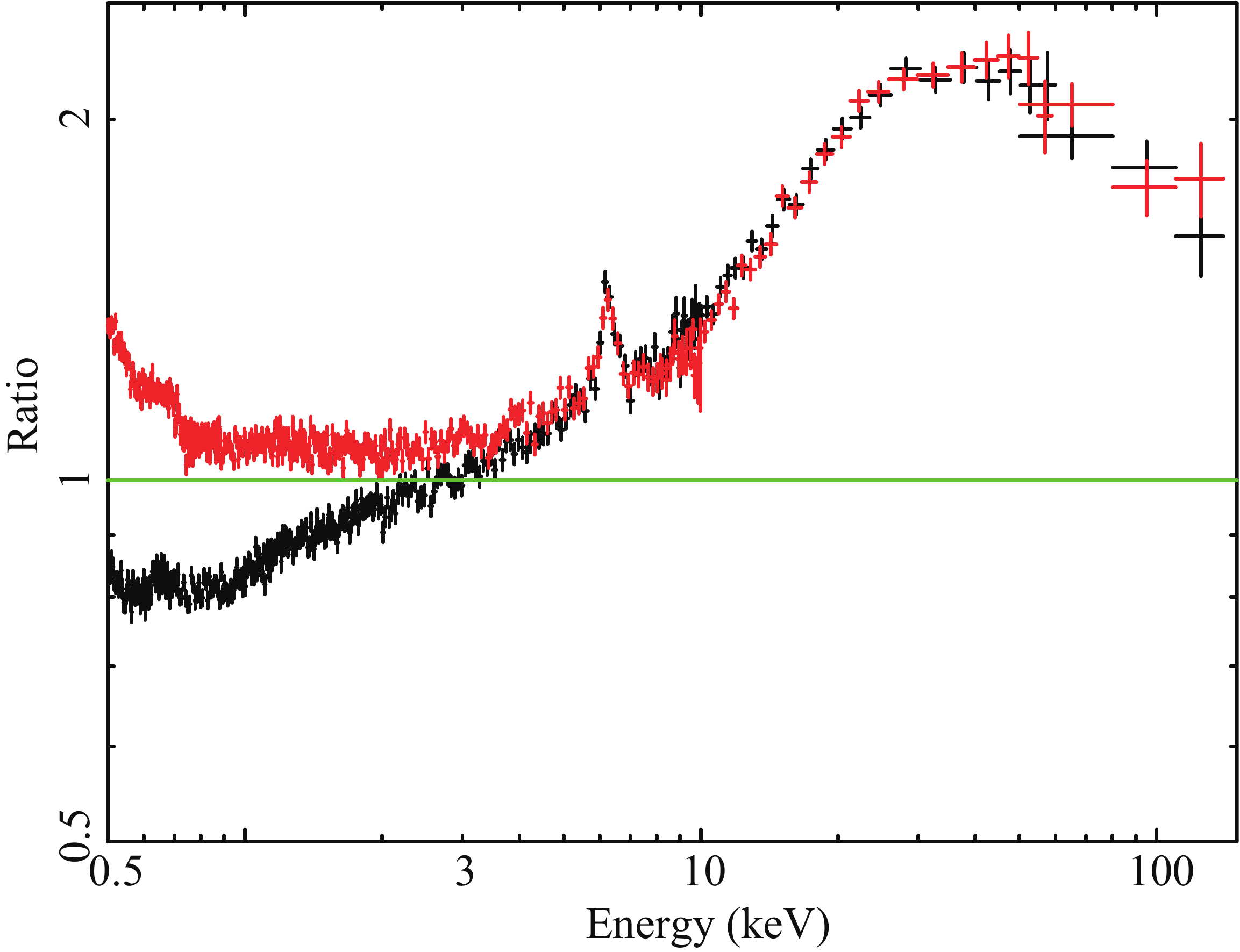}
 \end{center}
 \caption{Simulated {\it ASTRO-H} spectra of the type I Seyfert Mrk
   509 are shown here, in ratio to a $\Gamma=2$ power-law continuum.
   The spectra were generated assuming the two best-fit models for the
   {\it Suzaku} spectra of Mrk 509 obtained on 2006 April 25 and
   November 15 (Noda et al. 2011b).  An exposure of 100 ksec is
   assumed for each.  Spectra of this quality may help to better
   understand the nature of the X-ray continuum.}
 \label{noda_figure1}
\end{figure}

\section{Super-Eddington Accretion}
\subsection{Current Ideas}
Current observational knowledge of black hole accretion is largely
restricted to sub-Eddington regimes, in which
the (disk) luminosity is below the Eddington luminosity, 
$L_{\rm E} \sim 1.3 \times 10^{39}~ (M/10~M_\odot)$ erg~s$^{-1}$.
When the luminosity approaches the Eddington limit, the radiation
field exerts a strong pressure force against gas that is drawn inward
by gravitation, resulting in a powerful outflow.  The large mass
outflow rate affects the accretion flow structure, as well as the
local environment.  Thus, super-Eddington (or supercritical) accretion
is inhibited in a spherical geometry.

Supercritical accretion may be possible through an accretion disk,
however, since the radiation pressure force and gravitational force
are not direct opposition at all points.  Even in a supercritical disk
accretion geometry, radiation and matter are strongly coupled.  This
again causes a radiation-pressure driven outflow near the flow
surface, and photon trapping deep inside the accretion flow.  The
critical radius separating the inner, nearly isotropic outflow region
and the outer accretion region is called the spherization radius,
which increases as the accretion rate ($\dot M$) increases \citep{shakuraSunyaev73}.  
There is another critical radius, the photon trapping
radius.  Within this radius, photons are "trapped" within the flow,
and, hence, radiative efficiency is somewhat reduced (\citealt{Abramowicz88}, see also Chap. 10 of Kato et al. 2008).  Numerically, the
photon trapping radius is roughly equal to the spherization radius;
that is, these two effects work simultaneously.  The photon trapping
occurs also in neutron star accretion but only temporarily, since all
the photons produced within the accretion flow onto stars with solid a
surface should be eventually emitted.

To understand strong and complex matter-radiation interactions, global
radiation- hydrodynamic and/or radiation- MHD simulations have
recently been undertaken, and have predicted a number of features that
may be characteristic of a supercritical flow, such as mild beaming
(anisotropic radiation fields), the emergence of relativistic,
collimated jets, and loosely--collimated outflows with moderately high
velocity ($\sim 0.1 c$) and internal circulation \citep{Ohsuga2005,Ohsuga2009}. 
 As a result of the mild beaming, the maximum
apparent luminosities can exceed $10 L_{\rm E}$ for face-on observers.
The relativistic jets are accelerated by radiation pressure,and they
are collimated by magnetic force and/or pressure from a surrounding,
geometrically thick disk \citep{Takeuchi10}.  The outflow is of
particular importance, since it carries a large mass flux, ${\dot
  M}_{\rm outflow}\sim 10 L_{\rm E}/c^2$, momentum, $\sim L_{\rm
  E}/c$, and kinetic energy, $\sim 0.1 L_{\rm E}/c^2$.  The outflow
tends to have highly clumpy structure  \citep{Takeuchi13}, which
may cause variability.

These predicted features of super-Eddington flows are best studied in
X-rays.  The most important observational features that can be
pursued with {\it ASTRO-H} are (1) persistent relativistic jets, (2) line
absorption features by high-velocity outflow materials, and (3)
significant Comptonization by hot outflow gas in continuum spectra.

\subsection{SS 433 with {\it ASTRO-H}}
\begin{figure} 
\begin{center}
\includegraphics[width=\textwidth,bb=0 0 568 470,clip]{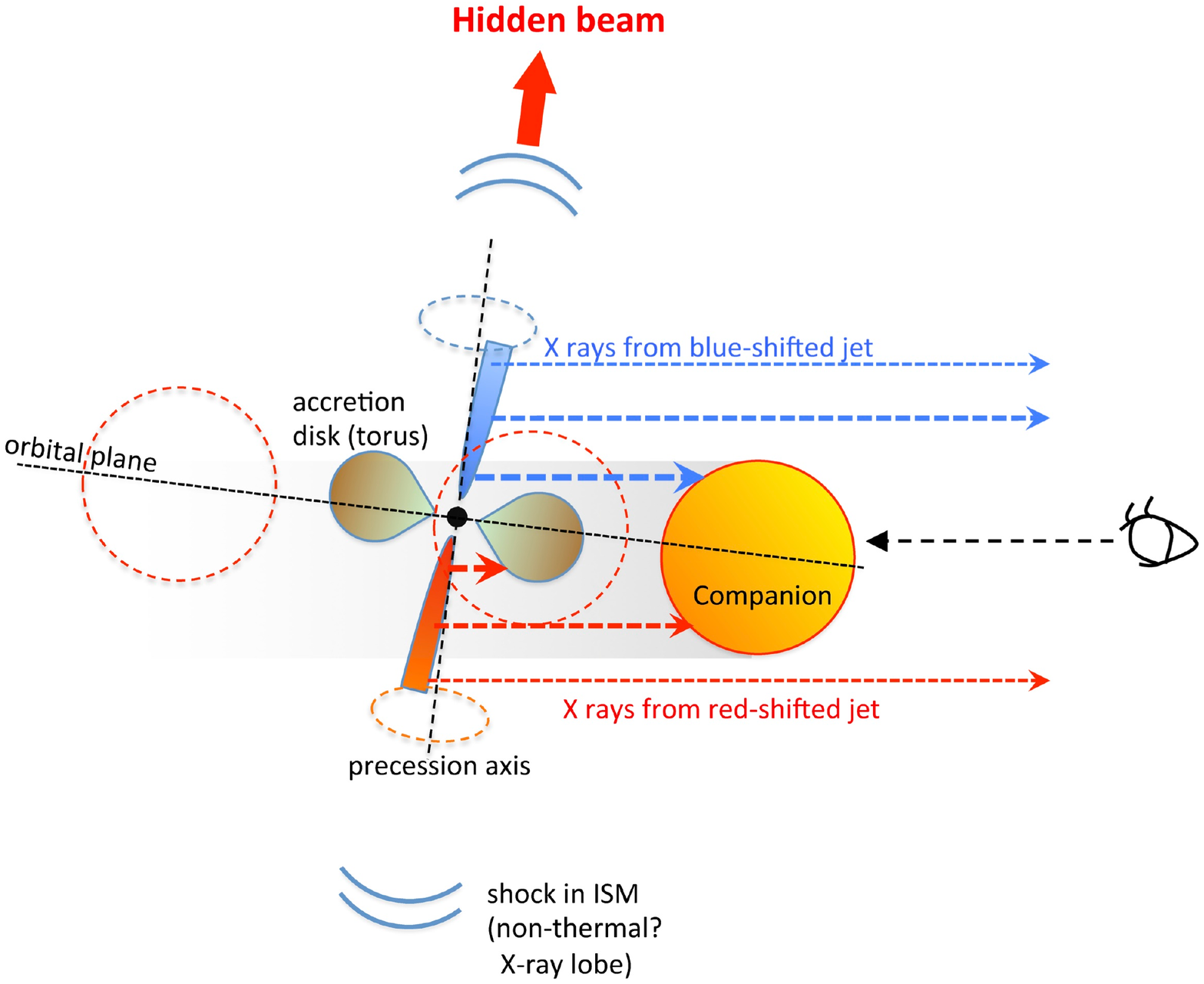}
\caption{The schematic view of the SS 433 system.}
\label{ss433_figure1}
\end{center}
\end{figure}

SS 433 is a 13.1-d binary with a pair of precessing jets with a
(nearly constant) speed of 0.26$c$.  It is the only source among all
the classes of astrophysical jets that shows unambiguous signature of
baryonic matter accelerated to a significant fraction of the speed of
light.  The jet is composed of normal hot plasma that gradually cools
as it travels, producing a multi-temperature thermal emission
spectrum.  The part of the jet near the base, with a temperature of
$kT>10$ keV, emits strong X-ray lines from heavy elements.  Further
out in the jet, at distances $> 10^{12}$ cm, the plasma emits
H$\alpha$ lines that vary in wavelength with the $\sim$ 163-d
precession period.  The geometry of the jet is well described by the
``kinetic model''; the inclination of the precession axis is
$i=79^{\circ}$ and the precession cone angle is $\theta=20^{\circ}$
(see Figure \ref{ss433_figure1}).

SS~433 is the only jet source that has yielded such detailed
information on the geometry and the thermal structure of the jet, and
therefore it represents a rare window on the physical processes
that drive jet production.  It may also hold clues to supercritical
accretion and ULXs: the size, density and thermal structure of the
X-ray emitting jet was spatially resolved by the eclipse mapping
technique using {\it ASCA} data \citep{Kotani1996}, and one of the most
important conclusions of the analysis is the kinetic power of the jet
is $L_{kin} \geq 10^{40}\; {\rm erg\; s}^{-1}$, far greater the
apparent X-ray luminosity ($L_{rad} \sim 10^{36} \;{\rm erg\;
    s}^{-1}$).

Observations of SS~433 with {\it ASTRO-H} will bear in the following problems:
\begin{itemize}
\item It is not yet clear how the jet of SS 433 is collimated into a narrow
cone angle ($<0.1$ rad), and then accelerated to a constant speed of
0.26$c$.  \citet{Namiki03} found the Fe K lines are broader than
lines from lighter elements in the {\it Chandra} HETG spectrum, hinting that
the jet is gradually collimated over a length of $> 10^{10}$ cm as it
cools.  Using the eclipse mapping on the {\it ASTRO-H} spectrum, it should
be possible to measure the radial velocity and width of emission lines as
a function of the position for each of the approaching and receding
jets.  The same data can greatly improve upon estimates of the
plasma condition, the jet power, and the geometry of the accretion
disk over the previous studies.

\item The striking disparity between the kinetic power of the jet
  $L_{kin} \geq 10^{40}\; {\rm erg\; s}^{-1}$ and the apparent X-ray
  luminosity ($L_{rad} \sim 10^{36} \;{\rm erg\; s}^{-1}$) may
  indicate that there are strong radiation beams co-aligned with the
  jets that never point to us.  The ionized absorption edges and
  fluorescence lines found in the {\it Suzaku} spectra of SS~433
  \citep{Kubota10b} suggest such beamed radiation may be present
  in the system.  If the putative beamed emission from this source is
  confirmed with {\it ASTRO-H}, it may have important implications for ULXs
  in nearby galaxies.

\item The nature of the compact object in SS~433 is not yet known.
  The mass of the compact object ($M_{\rm x}$) has been estimated
  based on the measured Keplerian velocities of the compact object and
  the donor star.  One current estimate, $1.9<M_{\rm x}/M_{\odot}<4.9$
  \citep{Kubota10a}, favors a small-mass black hole, but a heavy
  neutron star cannot be ruled out.  Measurements of the radial
  velocity amplitude of the stationary 6.4 keV Fe line, which may
  originate in the accretion disk, can improve the accuracy of mass
  constraints.
  
\item X-ray lobes \citep{Yamauchi94} are located at
  $\sim40^{\prime\prime}$ on the east and west sides of SS 433.  They
  are aligned with the axis of the precessing jet as determined in
  radio imaging.  It is natural to assume that they are powered by the
  jet, and represent the dissipation of the bulk kinetic energy of the
  jet in the shock, {\it i.e.} either the termination shock of the jet
  against the ISM or the internal shock of the colliding loops of the
  jet spiral.  The absence of emission lines in {\it ASCA} spectra of the
  lobes suggests that the emission is non-thermal, but this needs to
  be confirmed with the SXS and HXI.  The HXI image and spectra will
  clarify the emission process, and enable calorimetry of the jet.
  The detection of any thermal component in the shock would provide
  additional information on the dissipation of the jet.
\end{itemize}

\subsection{Models for ULXs}  
Ultraluminous X-ray sources (ULXs) are off-nuclear X-ray point-like
sources with large luminosities exceeding the Eddington luminosity of
stellar mass black holes (see, e.g., \citealt{MillerColbert04, Schartel2011, Walton2011ulx1}).  Despite extensive studies over three
decades, the nature of the central engine in ULXs remains an open
question.  However, it is likely that ULXs represent a new sort of
accretion flow: subcritical accretion onto intermediate-mass black
holes (IMBHs; $M \geq 100 M_\odot$; e.g. \citealt{Makishima2000, Godet09, Webb2010, Sutton2012ulx}), or supercritical accretion onto stellar-mass black holes
(e.g. \citealt{King01ULX, Watarai01}).  Of course, it is
possible that the class encompasses both source types.

Each hypothesis has pros and cons.  Some ULXs show blackbody-like
spectra whose temperatures are $kT \sim$0.2~keV, indicating large
innermost radii, and suggesting a black hole mass within the IMBH
range (e.g. Miller, Fabian, \& Miller 2004a), while
some others show rather high temperatures, $> 1$ keV, potentially
supporting the supercritical scenario (e.g. \citealt{Vierdayanti06}).
The discovery of very low frequency QPOs (Strohmayer \& Mushotzky
2009) also supports the IMBH hypothesis.  From the viewpoint of the
star formation theory, however, it is difficult to form an IMBH
(e.g. \citealt{Madhusudhan06ULX}).  Optical nebulae surround some ULXs
and have been used as bolometers; this technique suggests large
luminosities and also favors IMBHs (see, e.g., \citealt{Pakull2,Kaaret04ULX}).  Sustained fuelling of the accretion flow
is difficult in both scenarios.

In any case, these are new possibilities and their unique properties
must be explored both observationally and theoretically.  Studies with
{\it ASTRO-H} are expected to open new windows.  If the supercritical
accretion scenario is correct, we expect absorption features in the
SXS spectra produced by strong outflow associated with supercritical
flow.  Recent work has placed very strong limits on the equivalent
width of emission and absorption lines in Ho IX X-1 and NGC 1313 X-1
\citep{WaltonMillerReis2012ULX}.  The upper limit obtained for Ho IX X-1, just
30~eV, is within the range of absorption line strengths detected in
Galactic black hole X-ray binaries, and an order of magnitude below
the expected line strength of the mass outflow rate scales with the
inflow rate.  A deep, extremely sensitive SXS spectrum may finally be
able to detect absorption lines consistent with a super-Eddington
wind.  

\begin{figure}
 \begin{center}
 \vspace{-0.5cm}
  \includegraphics[width=70mm]{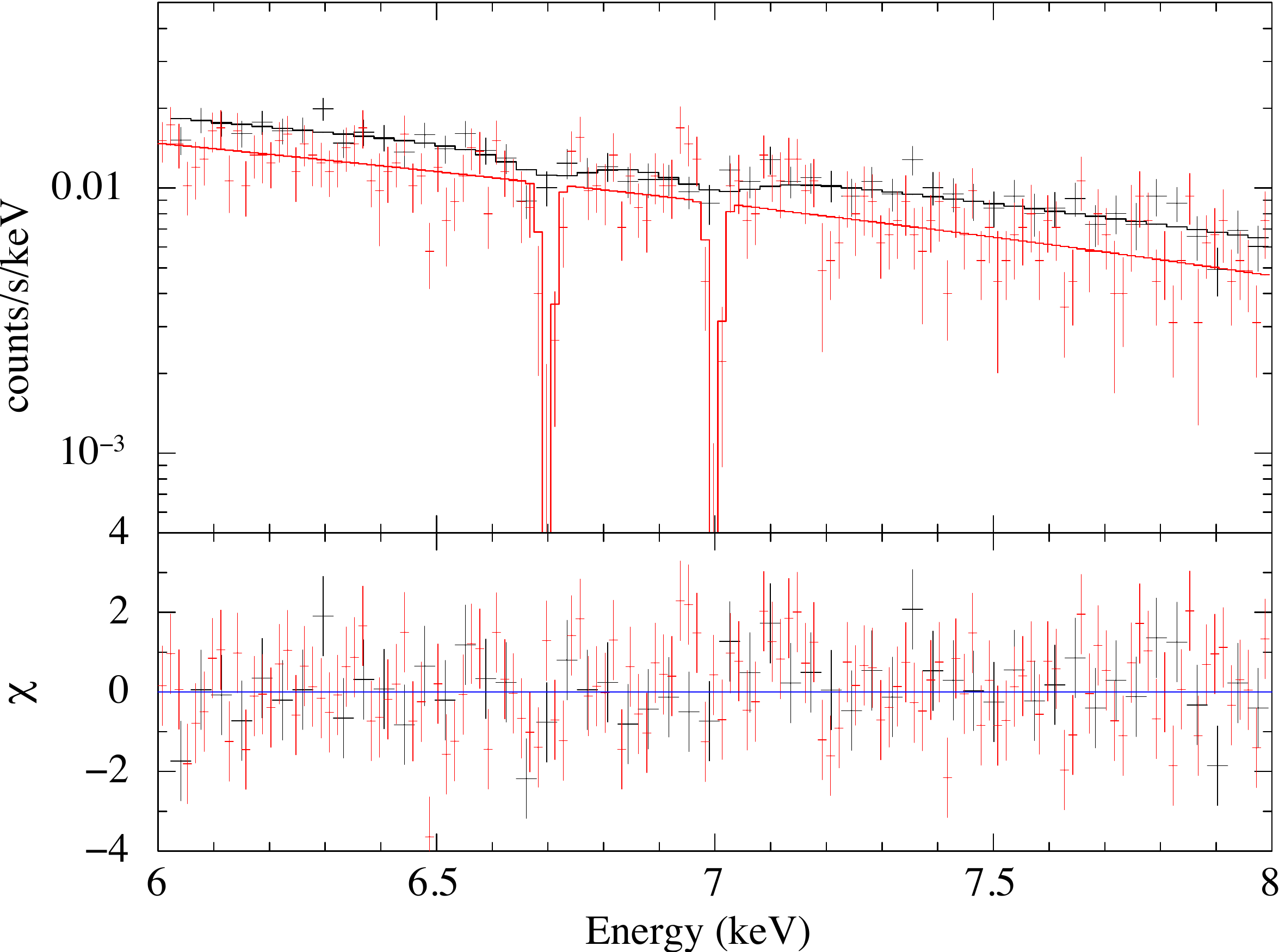}
  \includegraphics[width=70mm]{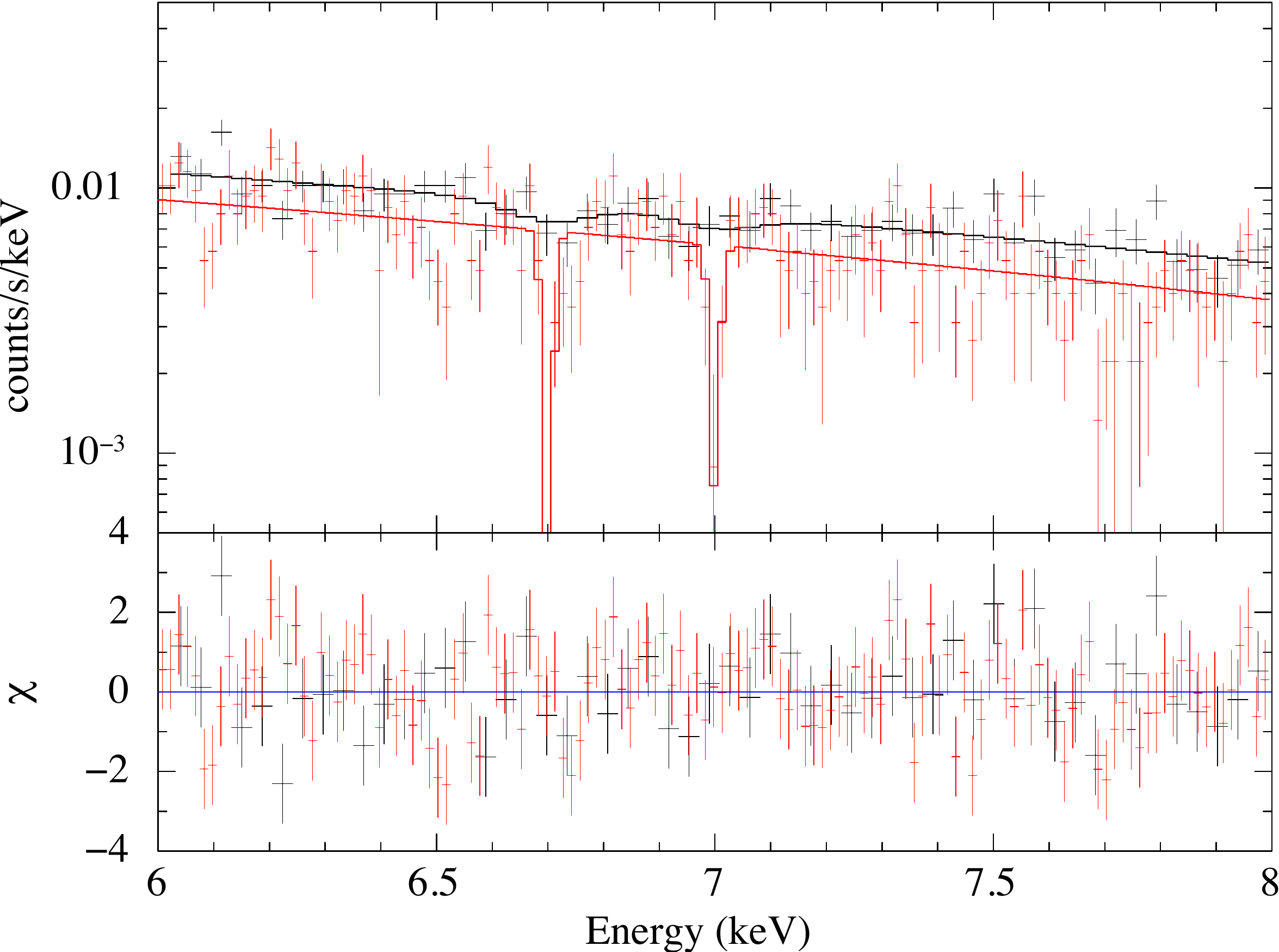}
 \end{center}
 \caption{
Simulated SXS (red) and HXI (black) spectra are shown above, based on
the high (left) and low (right) states observed in the ULX IC 342 X-1
by Yoshida et al.\ (2013).  The source luminosity was $1\times
10^{40}~ {\rm erg}~ {\rm s}^{-1}$, and $5\times 10^{39}~ {\rm erg}~
{\rm s}^{-1}$ in the high and low state, respectively.  In 150~ks of exposure, He-like and H-like Fe absorption lines with equivalent widths above 30~eV can be detected at more than 99.9\% confidence, and the velocity shifts of the lines can be well determined to within 250~km/s.  In the figures above, equivalent widths of 40~eV were assumed, and the spectra are binned for visual clarity.}
 \label{ULX_shidatsu}
\end{figure}

An example of the ability of {\it ASTRO-H} to decisively detect or reject
the outflows expected from super-Eddington accretion is shown in
Figure \ref{ULX_shidatsu}.  Two states of the well-known ULX IC 342
X-1 are simulated, based on the continua and flux levels reported by
Yoshida et al.\ (2013).  Narrow lines from He-like and H-like Fe can
be detected at greater than the 99.9\% confidence level in a 150~ks
exposure, assuming equivalent widths of 30~eV (comparable to current
limits).

It is also essential to obtain {\it broad} band spectra of ULXs (up to
a few hundreds of keV) with SXI and HXI for the first time to
establish similarities and differences between ULXs, Galactic black
hole binaries (BHBs), and AGNs.  Like Galactic BHBs, ULXs exhibit
spectral state transitions, however, the relationship between the
spectral states and luminosities, and with timing properties are
poorly investigated.

The spectra of some ULXs can be interpreted as the result of thermal Comptonization by
relatively low-temperature electrons ($kT_e \simeq 5$~keV), much lower than
those found in other black hole systems (with $\sim 100$ keV).
Gladstone et al. (2009) defined a state encompassing such spectra, the
``ultra-luminous'' state (see also \citealt{Vierdayanti2010}).  This spectral decomposition is in agreement with some simulations
\citep{Kawashima2012ULX}, and provides support for the
supercritical scenario.  If this scenario is
correct, we should see a spectral turnover in the hard X-ray range.

\begin{figure}
 \begin{center}
 \vspace{-0.5cm}
  \includegraphics[width=80mm]{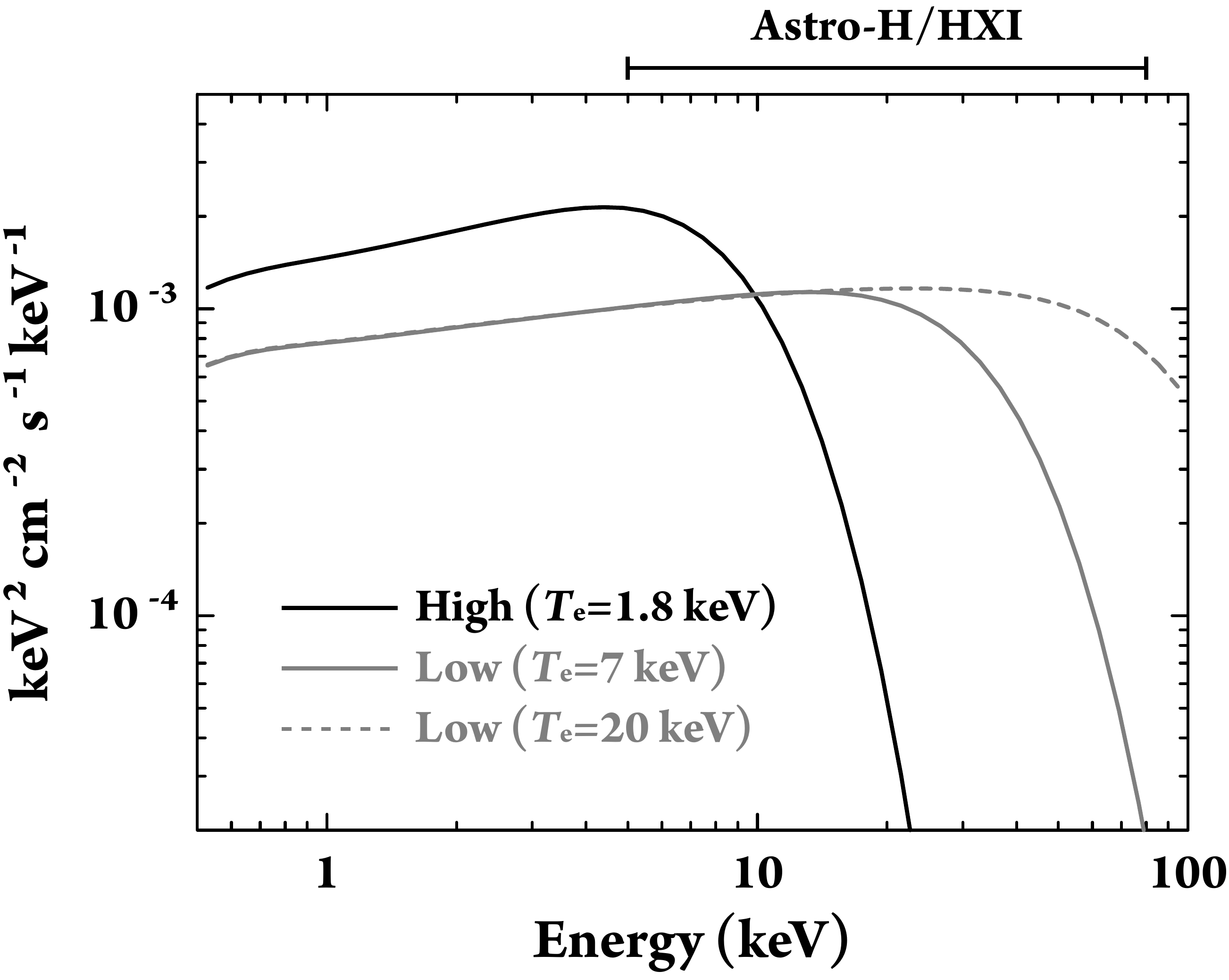}
 \end{center}
 \caption{
Expected spectral energy distributions (based on a Comptonization
model for ULXs; see \cite{2013PASJ...65...48Y}) with parameters of IC 342 X-1
during its power-law (PL) state: $(k T_{\rm e}, \tau) = (1.8~{\rm
  keV}, 8.5)$ by the black slid line, $(7~{\rm keV}, 3.6)$ by the gray
solid line, and $(20~{\rm keV}, 1.8)$ by the gray dashed line,
respectively.  The seed photon temperature of both states is assumed
to be 0.1 keV.  The energy range of the {\it ASTRO-H} is shown at the
top-right corner.  Adopted from Yoshida et al. (2013).}
 \label{ULX_figure1}
\end{figure}

The same is true for another state of ULXs, the power-law spectral
state.  Physical understanding of the power-law spectra of ULXs is a
potential key to a deeper understanding of the nature of ULXs, since
the supercritical scenario predicts a spectral turnover in the hard
X-ray band (Yoshida et al. 2013; see figure \ref{ULX_figure1}).  If
the power-law spectral component extends up to $\sim 100$ keV, ULXs
are more likely to be normal mode of accretion; that is, the IMBH
hypothesis is supported.  Clearly, the existence or absence of a break
in the power-law tail, will give us important information regarding
the physical situations of ULXs.


\begin{thebibliography}{}
\makeatletter
\relax
\def\mn@urlcharsother{\let\do\@makeother \do\$\do\&\do\#\do\^\do\_\do\%\do\~}
\def\mn@doi{\begingroup\mn@urlcharsother \@ifnextchar [ {\mn@doi@}
  {\mn@doi@[]}}
\def\mn@doi@[#1]#2{\def\@tempa{#1}\ifx\@tempa\@empty \href
  {http://dx.doi.org/#2} {doi:#2}\else \href {http://dx.doi.org/#2} {#1}\fi
  \endgroup}
\def\mn@eprint#1#2{\mn@eprint@#1:#2::\@nil}
\def\mn@eprint@arXiv#1{\href {http://arxiv.org/abs/#1} {{\tt arXiv:#1}}}
\def\mn@eprint@dblp#1{\href {http://dblp.uni-trier.de/rec/bibtex/#1.xml}
  {dblp:#1}}
\def\mn@eprint@#1:#2:#3:#4\@nil{\def\@tempa {#1}\def\@tempb {#2}\def\@tempc
  {#3}\ifx \@tempc \@empty \let \@tempc \@tempb \let \@tempb \@tempa \fi \ifx
  \@tempb \@empty \def\@tempb {arXiv}\fi \@ifundefined
  {mn@eprint@\@tempb}{\@tempb:\@tempc}{\expandafter \expandafter \csname
  mn@eprint@\@tempb\endcsname \expandafter{\@tempc}}}

\bibitem[\protect\citeauthoryear{{Abramowicz}, {Czerny}, {Lasota}  \&
  {Szuszkiewicz}}{{Abramowicz} et~al.}{1988}]{Abramowicz88}
{Abramowicz} M.~A.,  {Czerny} B.,  {Lasota} J.~P.,   {Szuszkiewicz} E.,  1988,
  \mn@doi [\apj] {10.1086/166683}, \href
  {http://adsabs.harvard.edu/abs/1988ApJ...332..646A} {332, 646}

\bibitem[\protect\citeauthoryear{{Allen}, {Dunn}, {Fabian}, {Taylor}  \&
  {Reynolds}}{{Allen} et~al.}{2006}]{Allen06}
{Allen} S.~W.,  {Dunn} R.~J.~H.,  {Fabian} A.~C.,  {Taylor} G.~B.,   {Reynolds}
  C.~S.,  2006, \mn@doi [\mnras] {10.1111/j.1365-2966.2006.10778.x}, \href
  {http://adsabs.harvard.edu/abs/2006MNRAS.372...21A} {372, 21}

\bibitem[\protect\citeauthoryear{{Atoyan} \& {Aharonian}}{{Atoyan} \&
  {Aharonian}}{1999}]{AtoyanAharonian99}
{Atoyan} A.~M.,  {Aharonian} F.~A.,  1999, \mn@doi [\mnras]
  {10.1046/j.1365-8711.1999.02172.x}, \href
  {http://adsabs.harvard.edu/abs/1999MNRAS.302..253A} {302, 253}

\bibitem[\protect\citeauthoryear{{Balbus} \& {Hawley}}{{Balbus} \&
  {Hawley}}{2002}]{balbus02}
{Balbus} S.~A.,  {Hawley} J.~F.,  2002, \mn@doi [\apj] {10.1086/340767}, \href
  {http://adsabs.harvard.edu/abs/2002ApJ...573..749B} {573, 749}

\bibitem[\protect\citeauthoryear{{Bardeen}, {Press}  \& {Teukolsky}}{{Bardeen}
  et~al.}{1972}]{Bardeenetal1972}
{Bardeen} J.~M.,  {Press} W.~H.,   {Teukolsky} S.~A.,  1972, \mn@doi [\apj]
  {10.1086/151796}, \href {http://adsabs.harvard.edu/abs/1972ApJ...178..347B}
  {178, 347}

\bibitem[\protect\citeauthoryear{{Barr}, {White}  \& {Page}}{{Barr}
  et~al.}{1985}]{Barrwhitepage85}
{Barr} P.,  {White} N.~E.,   {Page} C.~G.,  1985, \mnras, \href
  {http://adsabs.harvard.edu/abs/1985MNRAS.216P..65B} {216, 65P}

\bibitem[\protect\citeauthoryear{{Beckwith} \& {Done}}{{Beckwith} \&
  {Done}}{2004}]{BeckwithDone2004}
{Beckwith} K.,  {Done} C.,  2004, \mn@doi [\mnras]
  {10.1111/j.1365-2966.2004.07955.x}, \href
  {http://adsabs.harvard.edu/abs/2004MNRAS.352..353B} {352, 353}

\bibitem[\protect\citeauthoryear{{Begelman}, {McKee}  \& {Shields}}{{Begelman}
  et~al.}{1983}]{begelman83}
{Begelman} M.~C.,  {McKee} C.~F.,   {Shields} G.~A.,  1983, \mn@doi [\apj]
  {10.1086/161178}, \href {http://adsabs.harvard.edu/abs/1983ApJ...271...70B}
  {271, 70}

\bibitem[\protect\citeauthoryear{{Belloni}}{{Belloni}}{2010}]{Bellonibook2010}
{Belloni} T.~M.,  2010, in {T.~Belloni} ed.,  Lec. Notes in Physics, Berlin
  Springer Verlag Vol. 794, Lecture Notes in Physics, Berlin Springer Verlag.
  p.~53, \mn@eprint {arXiv} {0909.2474}, \mn@doi{10.1007/978-3-540-76937-8_3}

\bibitem[\protect\citeauthoryear{{Belloni} \& {Hasinger}}{{Belloni} \&
  {Hasinger}}{1990}]{BellonHasinger90i}
{Belloni} T.,  {Hasinger} G.,  1990, \aap, \href
  {http://adsabs.harvard.edu/abs/1990A%26A...227L..33B} {227, L33}

\bibitem[\protect\citeauthoryear{{Beloborodov}}{{Beloborodov}}{1999}]{beloborodov99}
{Beloborodov} A.~M.,  1999, \mn@doi [\apjl] {10.1086/311810}, \href
  {http://adsabs.harvard.edu/abs/1999ApJ...510L.123B} {510, L123}

\bibitem[\protect\citeauthoryear{{Berti} \& {Volonteri}}{{Berti} \&
  {Volonteri}}{2008}]{BertiVolonteri2008}
{Berti} E.,  {Volonteri} M.,  2008, \mn@doi [\apj] {10.1086/590379}, \href
  {http://adsabs.harvard.edu/abs/2008ApJ...684..822B} {684, 822}

\bibitem[\protect\citeauthoryear{{Bhattacharyya} \&
  {Strohmayer}}{{Bhattacharyya} \& {Strohmayer}}{2007}]{bhattacharyya07}
{Bhattacharyya} S.,  {Strohmayer} T.~E.,  2007, \mn@doi [\apjl]
  {10.1086/520844}, \href {http://adsabs.harvard.edu/abs/2007ApJ...664L.103B}
  {664, L103}

\bibitem[\protect\citeauthoryear{{Blaes}, {Davis}, {Hirose}, {Krolik}  \&
  {Stone}}{{Blaes} et~al.}{2006}]{magneticpressureblaes2006}
{Blaes} O.~M.,  {Davis} S.~W.,  {Hirose} S.,  {Krolik} J.~H.,   {Stone} J.~M.,
  2006, \mn@doi [\apj] {10.1086/503741}, \href
  {http://adsabs.harvard.edu/abs/2006ApJ...645.1402B} {645, 1402}

\bibitem[\protect\citeauthoryear{{Blandford} \& {Begelman}}{{Blandford} \&
  {Begelman}}{1999}]{BlandfordBegelman1999}
{Blandford} R.~D.,  {Begelman} M.~C.,  1999, \mn@doi [\mnras]
  {10.1046/j.1365-8711.1999.02358.x}, \href
  {http://adsabs.harvard.edu/abs/1999MNRAS.303L...1B} {303, L1}

\bibitem[\protect\citeauthoryear{{Blandford} \& {Payne}}{{Blandford} \&
  {Payne}}{1982}]{Blandfordpayne82}
{Blandford} R.~D.,  {Payne} D.~G.,  1982, \mnras, \href
  {http://adsabs.harvard.edu/abs/1982MNRAS.199..883B} {199, 883}

\bibitem[\protect\citeauthoryear{{Blandford} \& {Znajek}}{{Blandford} \&
  {Znajek}}{1977}]{BlandfordZnajek1977}
{Blandford} R.~D.,  {Znajek} R.~L.,  1977, \mnras, \href
  {http://adsabs.harvard.edu/abs/1977MNRAS.179..433B} {179, 433}

\bibitem[\protect\citeauthoryear{{Blum}, {Miller}, {Cackett}, {Yamaoka},
  {Takahashi}, {Raymond}, {Reynolds}  \& {Fabian}}{{Blum}
  et~al.}{2010}]{Blum2010}
{Blum} J.~L.,  {Miller} J.~M.,  {Cackett} E.,  {Yamaoka} K.,  {Takahashi} H.,
  {Raymond} J.,  {Reynolds} C.~S.,   {Fabian} A.~C.,  2010, \mn@doi [\apj]
  {10.1088/0004-637X/713/2/1244}, \href
  {http://adsabs.harvard.edu/abs/2010ApJ...713.1244B} {713, 1244}

\bibitem[\protect\citeauthoryear{{Boirin} \& {Parmar}}{{Boirin} \&
  {Parmar}}{2003}]{boirin03}
{Boirin} L.,  {Parmar} A.~N.,  2003, \mn@doi [\aap]
  {10.1051/0004-6361:20030975}, \href
  {http://adsabs.harvard.edu/abs/2003A%26A...407.1079B} {407, 1079}

\bibitem[\protect\citeauthoryear{{Boirin}, {Parmar}, {Barret}, {Paltani}  \&
  {Grindlay}}{{Boirin} et~al.}{2004}]{boirin04}
{Boirin} L.,  {Parmar} A.~N.,  {Barret} D.,  {Paltani} S.,   {Grindlay} J.~E.,
  2004, \mn@doi [\aap] {10.1051/0004-6361:20034550}, \href
  {http://adsabs.harvard.edu/abs/2004A%26A...418.1061B} {418, 1061}

\bibitem[\protect\citeauthoryear{{Cackett} et~al.,}{{Cackett}
  et~al.}{2008}]{cackett08}
{Cackett} E.~M.,  et~al., 2008, \mn@doi [\apj] {10.1086/524936}, \href
  {http://adsabs.harvard.edu/abs/2008ApJ...674..415C} {674, 415}

\bibitem[\protect\citeauthoryear{{Cackett} et~al.,}{{Cackett}
  et~al.}{2010}]{cackett10}
{Cackett} E.~M.,  et~al., 2010, \mn@doi [\apj] {10.1088/0004-637X/720/1/205},
  \href {http://adsabs.harvard.edu/abs/2010ApJ...720..205C} {720, 205}

\bibitem[\protect\citeauthoryear{{Cackett}, {Fabian}, {Zoghbi}, {Kara},
  {Reynolds}  \& {Uttley}}{{Cackett} et~al.}{2012}]{Cackett2012ESOlag}
{Cackett} E.~M.,  {Fabian} A.~C.,  {Zoghbi} A.,  {Kara} E.,  {Reynolds} C.,
  {Uttley} P.,  2012, ArXiv e-prints, \href
  {http://adsabs.harvard.edu/abs/2012arXiv1210.7874C} {}

\bibitem[\protect\citeauthoryear{{Calvet}, {Hartmann}  \& {Kenyon}}{{Calvet}
  et~al.}{1993}]{Calvet1993}
{Calvet} N.,  {Hartmann} L.,   {Kenyon} S.~J.,  1993, \mn@doi [\apj]
  {10.1086/172164}, \href {http://adsabs.harvard.edu/abs/1993ApJ...402..623C}
  {402, 623}

\bibitem[\protect\citeauthoryear{{Chartas}, {Kochanek}, {Dai}, {Poindexter}  \&
  {Garmire}}{{Chartas} et~al.}{2009}]{size1104}
{Chartas} G.,  {Kochanek} C.~S.,  {Dai} X.,  {Poindexter} S.,   {Garmire} G.,
  2009, \mn@doi [\apj] {10.1088/0004-637X/693/1/174}, \href
  {http://adsabs.harvard.edu/abs/2009ApJ...693..174C} {693, 174}

\bibitem[\protect\citeauthoryear{{Chartas}, {Kochanek}, {Dai}, {Moore},
  {Mosquera}  \& {Blackburne}}{{Chartas}
  et~al.}{2012}]{ChartasKochanek2012quasar}
{Chartas} G.,  {Kochanek} C.~S.,  {Dai} X.,  {Moore} D.,  {Mosquera} A.~M.,
  {Blackburne} J.~A.,  2012, ArXiv e-prints, \href
  {http://adsabs.harvard.edu/abs/2012arXiv1204.4480C} {}

\bibitem[\protect\citeauthoryear{{Church}, {Reed}, {Dotani},
  {Ba{\l}uci{\'n}ska-Church}  \& {Smale}}{{Church} et~al.}{2005}]{church05}
{Church} M.~J.,  {Reed} D.,  {Dotani} T.,  {Ba{\l}uci{\'n}ska-Church} M.,
  {Smale} A.~P.,  2005, \mn@doi [\mnras] {10.1111/j.1365-2966.2005.08728.x},
  \href {http://adsabs.harvard.edu/abs/2005MNRAS.359.1336C} {359, 1336}

\bibitem[\protect\citeauthoryear{{Coppi}}{{Coppi}}{1999}]{eqpair}
{Coppi} P.~S.,  1999, in {J.~Poutanen \& R.~Svensson} ed.,  Astronomical
  Society of the Pacific Conference Series Vol. 161, High Energy Processes in
  Accreting Black Holes. pp 375--+, \mn@eprint {} {arXiv:astro-ph/9903158}

\bibitem[\protect\citeauthoryear{{Crenshaw} \& {Kraemer}}{{Crenshaw} \&
  {Kraemer}}{2012}]{CrenshawKraemer2012}
{Crenshaw} D.~M.,  {Kraemer} S.~B.,  2012, \mn@doi [\apj]
  {10.1088/0004-637X/753/1/75}, \href
  {http://adsabs.harvard.edu/abs/2012ApJ...753...75C} {753, 75}

\bibitem[\protect\citeauthoryear{{Doeleman} et~al.,}{{Doeleman}
  et~al.}{2008}]{Doeleman2008Natur}
{Doeleman} S.~S.,  et~al., 2008, \mn@doi [\nat] {10.1038/nature07245}, \href
  {http://adsabs.harvard.edu/abs/2008Natur.455...78D} {455, 78}

\bibitem[\protect\citeauthoryear{{Doeleman} et~al.,}{{Doeleman}
  et~al.}{2012}]{Doeleman2012Sci}
{Doeleman} S.~S.,  et~al., 2012, \mn@doi [Science] {10.1126/science.1224768},
  \href {http://adsabs.harvard.edu/abs/2012Sci...338..355D} {338, 355}

\bibitem[\protect\citeauthoryear{{Done}, {Gierli{\'n}ski}  \& {Kubota}}{{Done}
  et~al.}{2007}]{donereview2007}
{Done} C.,  {Gierli{\'n}ski} M.,   {Kubota} A.,  2007, \mn@doi [\aapr]
  {10.1007/s00159-007-0006-1}, \href
  {http://adsabs.harvard.edu/abs/2007A%26ARv..15....1D} {15, 1}

\bibitem[\protect\citeauthoryear{{Dov{\v c}iak}, {Muleri}, {Goosmann}, {Karas}
  \& {Matt}}{{Dov{\v c}iak} et~al.}{2008}]{Dovciak08polorization}
{Dov{\v c}iak} M.,  {Muleri} F.,  {Goosmann} R.~W.,  {Karas} V.,   {Matt} G.,
  2008, \mn@doi [\mnras] {10.1111/j.1365-2966.2008.13872.x}, \href
  {http://adsabs.harvard.edu/abs/2008MNRAS.391...32D} {391, 32}

\bibitem[\protect\citeauthoryear{{Esin}, {McClintock}  \& {Narayan}}{{Esin}
  et~al.}{1997}]{Esin1997}
{Esin} A.~A.,  {McClintock} J.~E.,   {Narayan} R.,  1997, \mn@doi [\apj]
  {10.1086/304829}, \href {http://adsabs.harvard.edu/abs/1997ApJ...489..865E}
  {489, 865}

\bibitem[\protect\citeauthoryear{{Fabian}, {Sanders}, {Taylor}, {Allen},
  {Crawford}, {Johnstone}  \& {Iwasawa}}{{Fabian}
  et~al.}{2006}]{FabianSanders06cluster}
{Fabian} A.~C.,  {Sanders} J.~S.,  {Taylor} G.~B.,  {Allen} S.~W.,  {Crawford}
  C.~S.,  {Johnstone} R.~M.,   {Iwasawa} K.,  2006, \mn@doi [\mnras]
  {10.1111/j.1365-2966.2005.09896.x}, \href
  {http://adsabs.harvard.edu/abs/2006MNRAS.366..417F} {366, 417}

\bibitem[\protect\citeauthoryear{{Fabian} et~al.,}{{Fabian}
  et~al.}{2009}]{FabZog09}
{Fabian} A.~C.,  et~al., 2009, \mn@doi [\nat] {10.1038/nature08007}, \href
  {http://adsabs.harvard.edu/abs/2009Natur.459..540F} {459, 540}

\bibitem[\protect\citeauthoryear{{Fabian} et~al.,}{{Fabian}
  et~al.}{2012}]{Fabian2012iras13224}
{Fabian} A.~C.,  et~al., 2012, ArXiv e-prints, \href
  {http://adsabs.harvard.edu/abs/2012arXiv1208.5898F} {}

\bibitem[\protect\citeauthoryear{{Farrell}, {Webb}, {Barret}, {Godet}  \&
  {Rodrigues}}{{Farrell} et~al.}{2009}]{Farrell09Natur}
{Farrell} S.~A.,  {Webb} N.~A.,  {Barret} D.,  {Godet} O.,   {Rodrigues} J.~M.,
   2009, \mn@doi [\nat] {10.1038/nature08083}, \href
  {http://adsabs.harvard.edu/abs/2009Natur.460...73F} {460, 73}

\bibitem[\protect\citeauthoryear{{Fender}}{{Fender}}{2001}]{Fender2001jets}
{Fender} R.~P.,  2001, \mn@doi [\mnras] {10.1046/j.1365-8711.2001.04080.x},
  \href {http://adsabs.harvard.edu/abs/2001MNRAS.322...31F} {322, 31}

\bibitem[\protect\citeauthoryear{{Fender}, {Belloni}  \& {Gallo}}{{Fender}
  et~al.}{2004}]{fenderetal04}
{Fender} R.~P.,  {Belloni} T.~M.,   {Gallo} E.,  2004, \mn@doi [\mnras]
  {10.1111/j.1365-2966.2004.08384.x}, \href
  {http://adsabs.harvard.edu/abs/2004MNRAS.355.1105F} {355, 1105}

\bibitem[\protect\citeauthoryear{{Fender}, {Gallo}  \& {Russell}}{{Fender}
  et~al.}{2010}]{fender2010jets}
{Fender} R.~P.,  {Gallo} E.,   {Russell} D.,  2010, \mn@doi [\mnras]
  {10.1111/j.1365-2966.2010.16754.x}, \href
  {http://adsabs.harvard.edu/abs/2010MNRAS.406.1425F} {406, 1425}

\bibitem[\protect\citeauthoryear{{Ferland}, {Korista}, {Verner}, {Ferguson},
  {Kingdon}  \& {Verner}}{{Ferland} et~al.}{1998}]{1998PASP..110..761F}
{Ferland} G.~J.,  {Korista} K.~T.,  {Verner} D.~A.,  {Ferguson} J.~W.,
  {Kingdon} J.~B.,   {Verner} E.~M.,  1998, \mn@doi [\pasp] {10.1086/316190},
  \href {http://adsabs.harvard.edu/abs/1998PASP..110..761F} {110, 761}

\bibitem[\protect\citeauthoryear{{Focke}, {Wai}  \& {Swank}}{{Focke}
  et~al.}{2005}]{Focke05}
{Focke} W.~B.,  {Wai} L.~L.,   {Swank} J.~H.,  2005, \mn@doi [\apj]
  {10.1086/462407}, \href {http://adsabs.harvard.edu/abs/2005ApJ...633.1085F}
  {633, 1085}

\bibitem[\protect\citeauthoryear{{Gammie}, {Shapiro}  \& {McKinney}}{{Gammie}
  et~al.}{2004}]{gammie04}
{Gammie} C.~F.,  {Shapiro} S.~L.,   {McKinney} J.~C.,  2004, \mn@doi [\apj]
  {10.1086/380996}, \href {http://adsabs.harvard.edu/abs/2004ApJ...602..312G}
  {602, 312}

\bibitem[\protect\citeauthoryear{{Georganopoulos}, {Aharonian}  \&
  {Kirk}}{{Georganopoulos} et~al.}{2002}]{Georganopoulos02}
{Georganopoulos} M.,  {Aharonian} F.~A.,   {Kirk} J.~G.,  2002, \mn@doi [\aap]
  {10.1051/0004-6361:20020567}, \href
  {http://adsabs.harvard.edu/abs/2002A%26A...388L..25G} {388, L25}

\bibitem[\protect\citeauthoryear{{George} \& {Fabian}}{{George} \&
  {Fabian}}{1991}]{George91}
{George} I.~M.,  {Fabian} A.~C.,  1991, \mnras, \href
  {http://adsabs.harvard.edu/abs/1991MNRAS.249..352G} {249, 352}

\bibitem[\protect\citeauthoryear{{Gierli{\'n}ski}, {Middleton}, {Ward}  \&
  {Done}}{{Gierli{\'n}ski} et~al.}{2008}]{rej1034qpo}
{Gierli{\'n}ski} M.,  {Middleton} M.,  {Ward} M.,   {Done} C.,  2008, \mn@doi
  [\nat] {10.1038/nature07277}, \href
  {http://adsabs.harvard.edu/abs/2008Natur.455..369G} {455, 369}

\bibitem[\protect\citeauthoryear{{Gladstone}, {Roberts}  \& {Done}}{{Gladstone}
  et~al.}{2009}]{Gladstone09ULS}
{Gladstone} J.~C.,  {Roberts} T.~P.,   {Done} C.,  2009, \mn@doi [\mnras]
  {10.1111/j.1365-2966.2009.15123.x}, \href
  {http://adsabs.harvard.edu/abs/2009MNRAS.397.1836G} {397, 1836}

\bibitem[\protect\citeauthoryear{{Godet}, {Barret}, {Webb}, {Farrell}  \&
  {Gehrels}}{{Godet} et~al.}{2009}]{Godet09}
{Godet} O.,  {Barret} D.,  {Webb} N.~A.,  {Farrell} S.~A.,   {Gehrels} N.,
  2009, \mn@doi [\apjl] {10.1088/0004-637X/705/2/L109}, \href
  {http://adsabs.harvard.edu/abs/2009ApJ...705L.109G} {705, L109}

\bibitem[\protect\citeauthoryear{{Haardt} \& {Maraschi}}{{Haardt} \&
  {Maraschi}}{1991}]{HaardtMaraschi91}
{Haardt} F.,  {Maraschi} L.,  1991, \mn@doi [\apjl] {10.1086/186171}, \href
  {http://adsabs.harvard.edu/abs/1991ApJ...380L..51H} {380, L51}

\bibitem[\protect\citeauthoryear{{Hawley} \& {Krolik}}{{Hawley} \&
  {Krolik}}{2001}]{hawley01}
{Hawley} J.~F.,  {Krolik} J.~H.,  2001, \mn@doi [\apj] {10.1086/318678}, \href
  {http://adsabs.harvard.edu/abs/2001ApJ...548..348H} {548, 348}

\bibitem[\protect\citeauthoryear{{Hayashida} et~al.,}{{Hayashida}
  et~al.}{2007}]{Hayashida2007PTh}
{Hayashida} K.,  et~al., 2007, \mn@doi [Progress of Theoretical Physics
  Supplement] {10.1143/PTPS.169.269}, \href
  {http://adsabs.harvard.edu/abs/2007PThPS.169..269H} {169, 269}

\bibitem[\protect\citeauthoryear{{Hiemstra}, {M{\'e}ndez}, {Done}, {D{\'{\i}}az
  Trigo}, {Altamirano}  \& {Casella}}{{Hiemstra} et~al.}{2011}]{hiemstra1652}
{Hiemstra} B.,  {M{\'e}ndez} M.,  {Done} C.,  {D{\'{\i}}az Trigo} M.,
  {Altamirano} D.,   {Casella} P.,  2011, \mn@doi [\mnras]
  {10.1111/j.1365-2966.2010.17661.x}, \href
  {http://adsabs.harvard.edu/abs/2011MNRAS.411..137H} {411, 137}

\bibitem[\protect\citeauthoryear{{Hjellming} \& {Rupen}}{{Hjellming} \&
  {Rupen}}{1995}]{HjellmingRupen1995j1655}
{Hjellming} R.~M.,  {Rupen} M.~P.,  1995, \mn@doi [\nat] {10.1038/375464a0},
  \href {http://adsabs.harvard.edu/abs/1995Natur.375..464H} {375, 464}

\bibitem[\protect\citeauthoryear{{Ingram} \& {Done}}{{Ingram} \&
  {Done}}{2011}]{IngramDone2011}
{Ingram} A.,  {Done} C.,  2011, \mn@doi [\mnras]
  {10.1111/j.1365-2966.2011.18860.x}, \href
  {http://adsabs.harvard.edu/abs/2011MNRAS.415.2323I} {415, 2323}

\bibitem[\protect\citeauthoryear{{Ingram} \& {Done}}{{Ingram} \&
  {Done}}{2012}]{IngramDone2012}
{Ingram} A.,  {Done} C.,  2012, \mn@doi [\mnras]
  {10.1111/j.1365-2966.2011.19885.x}, \href
  {http://adsabs.harvard.edu/abs/2012MNRAS.419.2369I} {419, 2369}

\bibitem[\protect\citeauthoryear{{Ingram}, {Done}  \& {Fragile}}{{Ingram}
  et~al.}{2009}]{IngramDone09}
{Ingram} A.,  {Done} C.,   {Fragile} P.~C.,  2009, \mn@doi [\mnras]
  {10.1111/j.1745-3933.2009.00693.x}, \href
  {http://adsabs.harvard.edu/abs/2009MNRAS.397L.101I} {397, L101}

\bibitem[\protect\citeauthoryear{{Kaaret}, {Ward}  \& {Zezas}}{{Kaaret}
  et~al.}{2004}]{Kaaret04ULX}
{Kaaret} P.,  {Ward} M.~J.,   {Zezas} A.,  2004, \mn@doi [\mnras]
  {10.1111/j.1365-2966.2004.08020.x}, \href
  {http://adsabs.harvard.edu/abs/2004MNRAS.351L..83K} {351, L83}

\bibitem[\protect\citeauthoryear{{Kallman} \& {McCray}}{{Kallman} \&
  {McCray}}{1982}]{1982ApJS...50..263K}
{Kallman} T.~R.,  {McCray} R.,  1982, \mn@doi [\apjs] {10.1086/190828}, \href
  {http://adsabs.harvard.edu/abs/1982ApJS...50..263K} {50, 263}

\bibitem[\protect\citeauthoryear{{Kallman}, {Angelini}, {Boroson}  \&
  {Cottam}}{{Kallman} et~al.}{2003}]{kallman03}
{Kallman} T.~R.,  {Angelini} L.,  {Boroson} B.,   {Cottam} J.,  2003, \mn@doi
  [\apj] {10.1086/345475}, \href
  {http://adsabs.harvard.edu/abs/2003ApJ...583..861K} {583, 861}

\bibitem[\protect\citeauthoryear{{Kallman}, {Bautista}, {Goriely}, {Mendoza},
  {Miller}, {Palmeri}, {Quinet}  \& {Raymond}}{{Kallman}
  et~al.}{2009}]{kallman09}
{Kallman} T.~R.,  {Bautista} M.~A.,  {Goriely} S.,  {Mendoza} C.,  {Miller}
  J.~M.,  {Palmeri} P.,  {Quinet} P.,   {Raymond} J.,  2009, \mn@doi [\apj]
  {10.1088/0004-637X/701/2/865}, \href
  {http://adsabs.harvard.edu/abs/2009ApJ...701..865K} {701, 865}

\bibitem[\protect\citeauthoryear{{Kato}, {Fukue}  \& {Mineshige}}{{Kato}
  et~al.}{2008}]{Kato08book}
{Kato} S.,  {Fukue} J.,   {Mineshige} S.,  2008, {Black-Hole Accretion Disks
  --- Towards a New Paradigm ---}

\bibitem[\protect\citeauthoryear{{Kawabata} \& {Mineshige}}{{Kawabata} \&
  {Mineshige}}{2010}]{Kawabata10}
{Kawabata} R.,  {Mineshige} S.,  2010, \pasj, \href
  {http://adsabs.harvard.edu/abs/2010PASJ...62..621K} {62, 621}

\bibitem[\protect\citeauthoryear{{Kawashima}, {Ohsuga}, {Mineshige}, {Yoshida},
  {Heinzeller}  \& {Matsumoto}}{{Kawashima} et~al.}{2012}]{Kawashima2012ULX}
{Kawashima} T.,  {Ohsuga} K.,  {Mineshige} S.,  {Yoshida} T.,  {Heinzeller} D.,
    {Matsumoto} R.,  2012, \mn@doi [\apj] {10.1088/0004-637X/752/1/18}, \href
  {http://adsabs.harvard.edu/abs/2012ApJ...752...18K} {752, 18}

\bibitem[\protect\citeauthoryear{{King}, {Davies}, {Ward}, {Fabbiano}  \&
  {Elvis}}{{King} et~al.}{2001}]{King01ULX}
{King} A.~R.,  {Davies} M.~B.,  {Ward} M.~J.,  {Fabbiano} G.,   {Elvis} M.,
  2001, \mn@doi [\apjl] {10.1086/320343}, \href
  {http://adsabs.harvard.edu/abs/2001ApJ...552L.109K} {552, L109}

\bibitem[\protect\citeauthoryear{{King} et~al.,}{{King}
  et~al.}{2011}]{KingMiller2011}
{King} A.~L.,  et~al., 2011, \mn@doi [\apj] {10.1088/0004-637X/729/1/19}, \href
  {http://adsabs.harvard.edu/abs/2011ApJ...729...19K} {729, 19}

\bibitem[\protect\citeauthoryear{{King}, {Miller}  \& {Raymond}}{{King}
  et~al.}{2012a}]{KingMillerRaymond2012}
{King} A.~L.,  {Miller} J.~M.,   {Raymond} J.,  2012a, \mn@doi [\apj]
  {10.1088/0004-637X/746/1/2}, \href
  {http://adsabs.harvard.edu/abs/2012ApJ...746....2K} {746, 2}

\bibitem[\protect\citeauthoryear{{King} et~al.,}{{King}
  et~al.}{2012b}]{KingMiller2012}
{King} A.~L.,  et~al., 2012b, \mn@doi [\apjl] {10.1088/2041-8205/746/2/L20},
  \href {http://adsabs.harvard.edu/abs/2012ApJ...746L..20K} {746, L20}

\bibitem[\protect\citeauthoryear{{King} et~al.,}{{King}
  et~al.}{2013}]{KingMiller2013}
{King} A.~L.,  et~al., 2013, \mn@doi [\apj] {10.1088/0004-637X/762/2/103},
  \href {http://adsabs.harvard.edu/abs/2013ApJ...762..103K} {762, 103}

\bibitem[\protect\citeauthoryear{{Kotani}, {Kawai}, {Matsuoka}  \&
  {Brinkmann}}{{Kotani} et~al.}{1996}]{Kotani1996}
{Kotani} T.,  {Kawai} N.,  {Matsuoka} M.,   {Brinkmann} W.,  1996, \pasj, \href
  {http://adsabs.harvard.edu/abs/1996PASJ...48..619K} {48, 619}

\bibitem[\protect\citeauthoryear{{Kotani}, {Ebisawa}, {Dotani}, {Inoue},
  {Nagase}, {Tanaka}  \& {Ueda}}{{Kotani} et~al.}{2000}]{kotani00}
{Kotani} T.,  {Ebisawa} K.,  {Dotani} T.,  {Inoue} H.,  {Nagase} F.,  {Tanaka}
  Y.,   {Ueda} Y.,  2000, \mn@doi [\apj] {10.1086/309200}, \href
  {http://adsabs.harvard.edu/abs/2000ApJ...539..413K} {539, 413}

\bibitem[\protect\citeauthoryear{{Kraemer} et~al.,}{{Kraemer}
  et~al.}{2005}]{Kraemer2005}
{Kraemer} S.~B.,  et~al., 2005, \mn@doi [\apj] {10.1086/466522}, \href
  {http://adsabs.harvard.edu/abs/2005ApJ...633..693K} {633, 693}

\bibitem[\protect\citeauthoryear{{Kubota} et~al.,}{{Kubota}
  et~al.}{2007}]{kubota07}
{Kubota} A.,  et~al., 2007, \pasj, \href
  {http://adsabs.harvard.edu/abs/2007PASJ...59S.185K} {59, 185}

\bibitem[\protect\citeauthoryear{{Kubota} et~al.,}{{Kubota}
  et~al.}{2010a}]{Kubota10a}
{Kubota} K.,  et~al., 2010a, \pasj, \href
  {http://adsabs.harvard.edu/abs/2010PASJ...62..323K} {62, 323}

\bibitem[\protect\citeauthoryear{{Kubota}, {Ueda}, {Fabrika}, {Medvedev},
  {Barsukova}, {Sholukhova}  \& {Goranskij}}{{Kubota}
  et~al.}{2010b}]{Kubota10b}
{Kubota} K.,  {Ueda} Y.,  {Fabrika} S.,  {Medvedev} A.,  {Barsukova} E.~A.,
  {Sholukhova} O.,   {Goranskij} V.~P.,  2010b, \mn@doi [\apj]
  {10.1088/0004-637X/709/2/1374}, \href
  {http://adsabs.harvard.edu/abs/2010ApJ...709.1374K} {709, 1374}

\bibitem[\protect\citeauthoryear{{Laurent}, {Rodriguez}, {Wilms}, {Cadolle
  Bel}, {Pottschmidt}  \& {Grinberg}}{{Laurent} et~al.}{2011}]{Laurent2011Sci}
{Laurent} P.,  {Rodriguez} J.,  {Wilms} J.,  {Cadolle Bel} M.,  {Pottschmidt}
  K.,   {Grinberg} V.,  2011, \mn@doi [Science] {10.1126/science.1200848},
  \href {http://adsabs.harvard.edu/abs/2011Sci...332..438L} {332, 438}

\bibitem[\protect\citeauthoryear{{Lee}, {Reynolds}, {Remillard}, {Schulz},
  {Blackman}  \& {Fabian}}{{Lee} et~al.}{2002}]{lee02}
{Lee} J.~C.,  {Reynolds} C.~S.,  {Remillard} R.,  {Schulz} N.~S.,  {Blackman}
  E.~G.,   {Fabian} A.~C.,  2002, \mn@doi [\apj] {10.1086/338588}, \href
  {http://adsabs.harvard.edu/abs/2002ApJ...567.1102L} {567, 1102}

\bibitem[\protect\citeauthoryear{{Levinson} \& {Blandford}}{{Levinson} \&
  {Blandford}}{1996}]{LevinsonBlandford96}
{Levinson} A.,  {Blandford} R.,  1996, \mn@doi [\apjl] {10.1086/309851}, \href
  {http://adsabs.harvard.edu/abs/1996ApJ...456L..29L} {456, L29}

\bibitem[\protect\citeauthoryear{{Ling}, {Mahoney}, {Wheaton}  \&
  {Jacobson}}{{Ling} et~al.}{1987}]{Ling1987}
{Ling} J.~C.,  {Mahoney} W.~A.,  {Wheaton} W.~A.,   {Jacobson} A.~S.,  1987,
  \mn@doi [\apjl] {10.1086/185017}, \href
  {http://adsabs.harvard.edu/abs/1987ApJ...321L.117L} {321, L117}

\bibitem[\protect\citeauthoryear{{Liu}, {Mineshige}, {Meyer},
  {Meyer-Hofmeister}  \& {Kawaguchi}}{{Liu} et~al.}{2002}]{Liu2002corona}
{Liu} B.~F.,  {Mineshige} S.,  {Meyer} F.,  {Meyer-Hofmeister} E.,
  {Kawaguchi} T.,  2002, \mn@doi [\apj] {10.1086/341138}, \href
  {http://adsabs.harvard.edu/abs/2002ApJ...575..117L} {575, 117}

\bibitem[\protect\citeauthoryear{{Liu}, {Mineshige}  \& {Ohsuga}}{{Liu}
  et~al.}{2003}]{Liu03corona}
{Liu} B.~F.,  {Mineshige} S.,   {Ohsuga} K.,  2003, \mn@doi [\apj]
  {10.1086/368282}, \href {http://adsabs.harvard.edu/abs/2003ApJ...587..571L}
  {587, 571}

\bibitem[\protect\citeauthoryear{{Lubi{\'n}ski}, {Zdziarski}, {Walter},
  {Paltani}, {Beckmann}, {Soldi}, {Ferrigno}  \& {Courvoisier}}{{Lubi{\'n}ski}
  et~al.}{2010}]{Lubiski10}
{Lubi{\'n}ski} P.,  {Zdziarski} A.~A.,  {Walter} R.,  {Paltani} S.,  {Beckmann}
  V.,  {Soldi} S.,  {Ferrigno} C.,   {Courvoisier} T.~J.-L.,  2010, \mn@doi
  [\mnras] {10.1111/j.1365-2966.2010.17251.x}, \href
  {http://adsabs.harvard.edu/abs/2010MNRAS.408.1851L} {408, 1851}

\bibitem[\protect\citeauthoryear{{Luketic}, {Proga}, {Kallman}, {Raymond}  \&
  {Miller}}{{Luketic} et~al.}{2010}]{Luketic2010}
{Luketic} S.,  {Proga} D.,  {Kallman} T.~R.,  {Raymond} J.~C.,   {Miller}
  J.~M.,  2010, \mn@doi [\apj] {10.1088/0004-637X/719/1/515}, \href
  {http://adsabs.harvard.edu/abs/2010ApJ...719..515L} {719, 515}

\bibitem[\protect\citeauthoryear{{Lyubarskii}}{{Lyubarskii}}{1997}]{Lyubarskii97}
{Lyubarskii} Y.~E.,  1997, \mnras, \href
  {http://adsabs.harvard.edu/abs/1997MNRAS.292..679L} {292, 679}

\bibitem[\protect\citeauthoryear{{Maccarone}}{{Maccarone}}{2002}]{Maccarone2002misalignment}
{Maccarone} T.~J.,  2002, \mn@doi [\mnras] {10.1046/j.1365-8711.2002.05876.x},
  \href {http://adsabs.harvard.edu/abs/2002MNRAS.336.1371M} {336, 1371}

\bibitem[\protect\citeauthoryear{{Machida}, {Nakamura}  \&
  {Matsumoto}}{{Machida} et~al.}{2004}]{machida04}
{Machida} M.,  {Nakamura} K.,   {Matsumoto} R.,  2004, \pasj, \href
  {http://adsabs.harvard.edu/abs/2004PASJ...56..671M} {56, 671}

\bibitem[\protect\citeauthoryear{{Madhusudhan}, {Justham}, {Nelson}, {Paxton},
  {Pfahl}, {Podsiadlowski}  \& {Rappaport}}{{Madhusudhan}
  et~al.}{2006}]{Madhusudhan06ULX}
{Madhusudhan} N.,  {Justham} S.,  {Nelson} L.,  {Paxton} B.,  {Pfahl} E.,
  {Podsiadlowski} P.,   {Rappaport} S.,  2006, \mn@doi [\apj] {10.1086/500238},
  \href {http://adsabs.harvard.edu/abs/2006ApJ...640..918M} {640, 918}

\bibitem[\protect\citeauthoryear{{Magdziarz} \& {Zdziarski}}{{Magdziarz} \&
  {Zdziarski}}{1995}]{pexrav}
{Magdziarz} P.,  {Zdziarski} A.~A.,  1995, \mnras, \href
  {http://adsabs.harvard.edu/abs/1995MNRAS.273..837M} {273, 837}

\bibitem[\protect\citeauthoryear{{Mahadevan} \& {Quataert}}{{Mahadevan} \&
  {Quataert}}{1997}]{Mahadevan97}
{Mahadevan} R.,  {Quataert} E.,  1997, \mn@doi [\apj] {10.1086/304908}, \href
  {http://adsabs.harvard.edu/abs/1997ApJ...490..605M} {490, 605}

\bibitem[\protect\citeauthoryear{{Makishima} et~al.,}{{Makishima}
  et~al.}{2000}]{Makishima2000}
{Makishima} K.,  et~al., 2000, \mn@doi [\apj] {10.1086/308868}, \href
  {http://adsabs.harvard.edu/abs/2000ApJ...535..632M} {535, 632}

\bibitem[\protect\citeauthoryear{{Makishima} et~al.,}{{Makishima}
  et~al.}{2008}]{makishimacygx108}
{Makishima} K.,  et~al., 2008, \pasj, \href
  {http://adsabs.harvard.edu/abs/2008PASJ...60..585M} {60, 585}

\bibitem[\protect\citeauthoryear{{Malizia}, {Stephen}, {Bassani}, {Bird},
  {Panessa}  \& {Ubertini}}{{Malizia} et~al.}{2009}]{Malizia2009}
{Malizia} A.,  {Stephen} J.~B.,  {Bassani} L.,  {Bird} A.~J.,  {Panessa} F.,
  {Ubertini} P.,  2009, \mn@doi [\mnras] {10.1111/j.1365-2966.2009.15330.x},
  \href {http://adsabs.harvard.edu/abs/2009MNRAS.399..944M} {399, 944}

\bibitem[\protect\citeauthoryear{{Manmoto}, {Takeuchi}, {Mineshige},
  {Matsumoto}  \& {Negoro}}{{Manmoto} et~al.}{1996}]{Manmoto96}
{Manmoto} T.,  {Takeuchi} M.,  {Mineshige} S.,  {Matsumoto} R.,   {Negoro} H.,
  1996, \mn@doi [\apjl] {10.1086/310097}, \href
  {http://adsabs.harvard.edu/abs/1996ApJ...464L.135M} {464, L135}

\bibitem[\protect\citeauthoryear{{Markoff}, {Falcke}  \& {Fender}}{{Markoff}
  et~al.}{2001}]{Markoff2001J1118}
{Markoff} S.,  {Falcke} H.,   {Fender} R.,  2001, \mn@doi [\aap]
  {10.1051/0004-6361:20010420}, \href
  {http://adsabs.harvard.edu/abs/2001A%26A...372L..25M} {372, L25}

\bibitem[\protect\citeauthoryear{{Matsumoto}, {Tsuru}, {Koyama}, {Awaki},
  {Canizares}, {Kawai}, {Matsushita}  \& {Kawabe}}{{Matsumoto}
  et~al.}{2001}]{2001ApJ...547L..25M}
{Matsumoto} H.,  {Tsuru} T.~G.,  {Koyama} K.,  {Awaki} H.,  {Canizares} C.~R.,
  {Kawai} N.,  {Matsushita} S.,   {Kawabe} R.,  2001, \mn@doi [\apjl]
  {10.1086/318878}, \href {http://adsabs.harvard.edu/abs/2001ApJ...547L..25M}
  {547, L25}

\bibitem[\protect\citeauthoryear{{Mauche} \& {Raymond}}{{Mauche} \&
  {Raymond}}{2000}]{MaucheRaymond2000}
{Mauche} C.~W.,  {Raymond} J.~C.,  2000, \mn@doi [\apj] {10.1086/309489}, \href
  {http://adsabs.harvard.edu/abs/2000ApJ...541..924M} {541, 924}

\bibitem[\protect\citeauthoryear{{McClintock}, {Shafee}, {Narayan},
  {Remillard}, {Davis}  \& {Li}}{{McClintock} et~al.}{2006}]{mcclintock06}
{McClintock} J.~E.,  {Shafee} R.,  {Narayan} R.,  {Remillard} R.~A.,  {Davis}
  S.~W.,   {Li} L.,  2006, \mn@doi [\apj] {10.1086/508457}, \href
  {http://adsabs.harvard.edu/abs/2006ApJ...652..518M} {652, 518}

\bibitem[\protect\citeauthoryear{{McClintock} et~al.,}{{McClintock}
  et~al.}{2011}]{McClintock2011review}
{McClintock} J.~E.,  et~al., 2011, \mn@doi [Classical and Quantum Gravity]
  {10.1088/0264-9381/28/11/114009}, \href
  {http://adsabs.harvard.edu/abs/2011CQGra..28k4009M} {28, 114009}

\bibitem[\protect\citeauthoryear{{McKinney}, {Tchekhovskoy}  \&
  {Blandford}}{{McKinney} et~al.}{2012}]{McKinneyqpo2012}
{McKinney} J.~C.,  {Tchekhovskoy} A.,   {Blandford} R.~D.,  2012, ArXiv
  e-prints, \href {http://adsabs.harvard.edu/abs/2012arXiv1201.4163M} {}

\bibitem[\protect\citeauthoryear{{McNamara}, {Kazemzadeh}, {Rafferty},
  {B{\^i}rzan}, {Nulsen}, {Kirkpatrick}  \& {Wise}}{{McNamara}
  et~al.}{2009}]{McNamara2009}
{McNamara} B.~R.,  {Kazemzadeh} F.,  {Rafferty} D.~A.,  {B{\^i}rzan} L.,
  {Nulsen} P.~E.~J.,  {Kirkpatrick} C.~C.,   {Wise} M.~W.,  2009, \mn@doi
  [\apj] {10.1088/0004-637X/698/1/594}, \href
  {http://adsabs.harvard.edu/abs/2009ApJ...698..594M} {698, 594}

\bibitem[\protect\citeauthoryear{{Merloni} \& {Heinz}}{{Merloni} \&
  {Heinz}}{2007}]{2007MNRAS.381..589M}
{Merloni} A.,  {Heinz} S.,  2007, \mn@doi [\mnras]
  {10.1111/j.1365-2966.2007.12253.x}, \href
  {http://adsabs.harvard.edu/abs/2007MNRAS.381..589M} {381, 589}

\bibitem[\protect\citeauthoryear{{Merloni}, {Fabian}  \& {Ross}}{{Merloni}
  et~al.}{2000}]{merlonifabianross00}
{Merloni} A.,  {Fabian} A.~C.,   {Ross} R.~R.,  2000, \mn@doi [\mnras]
  {10.1046/j.1365-8711.2000.03226.x}, \href
  {http://adsabs.harvard.edu/abs/2000MNRAS.313..193M} {313, 193}

\bibitem[\protect\citeauthoryear{{Meyer-Hofmeister}, {Liu}  \&
  {Meyer}}{{Meyer-Hofmeister} et~al.}{2009}]{MeyerHofmeisterLiu09}
{Meyer-Hofmeister} E.,  {Liu} B.~F.,   {Meyer} F.,  2009, \mn@doi [\aap]
  {10.1051/0004-6361/200913044}, \href
  {http://adsabs.harvard.edu/abs/2009A%26A...508..329M} {508, 329}

\bibitem[\protect\citeauthoryear{{Meyer-Hofmeister}, {Liu}  \&
  {Meyer}}{{Meyer-Hofmeister} et~al.}{2012}]{MeyerHofmeister2012}
{Meyer-Hofmeister} E.,  {Liu} B.~F.,   {Meyer} F.,  2012, \mn@doi [\aap]
  {10.1051/0004-6361/201219245}, \href
  {http://adsabs.harvard.edu/abs/2012A%26A...544A..87M} {544, A87}

\bibitem[\protect\citeauthoryear{{Miller}}{{Miller}}{2007}]{miller07review}
{Miller} J.~M.,  2007, \mn@doi [\araa]
  {10.1146/annurev.astro.45.051806.110555}, \href
  {http://adsabs.harvard.edu/abs/2007ARA%26A..45..441M} {45, 441}

\bibitem[\protect\citeauthoryear{{Miller} \& {Colbert}}{{Miller} \&
  {Colbert}}{2004}]{MillerColbert04}
{Miller} M.~C.,  {Colbert} E.~J.~M.,  2004, \mn@doi [International Journal of
  Modern Physics D] {10.1142/S0218271804004426}, \href
  {http://adsabs.harvard.edu/abs/2004IJMPD..13....1M} {13, 1}

\bibitem[\protect\citeauthoryear{{Miller} \& {Homan}}{{Miller} \&
  {Homan}}{2005}]{MillerHoman05}
{Miller} J.~M.,  {Homan} J.,  2005, \mn@doi [\apjl] {10.1086/427775}, \href
  {http://adsabs.harvard.edu/abs/2005ApJ...618L.107M} {618, L107}

\bibitem[\protect\citeauthoryear{{Miller-Jones}, {Jonker}, {Dhawan}, {Brisken},
  {Rupen}, {Nelemans}  \& {Gallo}}{{Miller-Jones}
  et~al.}{2009}]{MillerJonesv404cyg2009}
{Miller-Jones} J.~C.~A.,  {Jonker} P.~G.,  {Dhawan} V.,  {Brisken} W.,  {Rupen}
  M.~P.,  {Nelemans} G.,   {Gallo} E.,  2009, \mn@doi [\apjl]
  {10.1088/0004-637X/706/2/L230}, \href
  {http://adsabs.harvard.edu/abs/2009ApJ...706L.230M} {706, L230}

\bibitem[\protect\citeauthoryear{{Miller} et~al.,}{{Miller}
  et~al.}{2002}]{miller02j1650}
{Miller} J.~M.,  et~al., 2002, \mn@doi [\apjl] {10.1086/341099}, \href
  {http://adsabs.harvard.edu/abs/2002ApJ...570L..69M} {570, L69}

\bibitem[\protect\citeauthoryear{{Miller}, {Raymond}, {Fabian}, {Steeghs},
  {Homan}, {Reynolds}, {van der Klis}  \& {Wijnands}}{{Miller}
  et~al.}{2006a}]{Miller06Natur}
{Miller} J.~M.,  {Raymond} J.,  {Fabian} A.,  {Steeghs} D.,  {Homan} J.,
  {Reynolds} C.,  {van der Klis} M.,   {Wijnands} R.,  2006a, \mn@doi [\nat]
  {10.1038/nature04912}, \href
  {http://adsabs.harvard.edu/abs/2006Natur.441..953M} {441, 953}

\bibitem[\protect\citeauthoryear{{Miller} et~al.,}{{Miller}
  et~al.}{2006b}]{miller06b}
{Miller} J.~M.,  et~al., 2006b, \mn@doi [\apj] {10.1086/504673}, \href
  {http://adsabs.harvard.edu/abs/2006ApJ...646..394M} {646, 394}

\bibitem[\protect\citeauthoryear{{Miller}, {Raymond}, {Reynolds}, {Fabian},
  {Kallman}  \& {Homan}}{{Miller} et~al.}{2008}]{Miller2008j1655wind}
{Miller} J.~M.,  {Raymond} J.,  {Reynolds} C.~S.,  {Fabian} A.~C.,  {Kallman}
  T.~R.,   {Homan} J.,  2008, \mn@doi [\apj] {10.1086/588521}, \href
  {http://adsabs.harvard.edu/abs/2008ApJ...680.1359M} {680, 1359}

\bibitem[\protect\citeauthoryear{{Miller}, {Reynolds}, {Fabian}, {Miniutti}  \&
  {Gallo}}{{Miller} et~al.}{2009}]{miller09spin}
{Miller} J.~M.,  {Reynolds} C.~S.,  {Fabian} A.~C.,  {Miniutti} G.,   {Gallo}
  L.~C.,  2009, \mn@doi [\apj] {10.1088/0004-637X/697/1/900}, \href
  {http://adsabs.harvard.edu/abs/2009ApJ...697..900M} {697, 900}

\bibitem[\protect\citeauthoryear{{Miller}, {Miller}  \& {Reynolds}}{{Miller}
  et~al.}{2011}]{millersn11}
{Miller} J.~M.,  {Miller} M.~C.,   {Reynolds} C.~S.,  2011, \mn@doi [\apjl]
  {10.1088/2041-8205/731/1/L5}, \href
  {http://adsabs.harvard.edu/abs/2011ApJ...731L...5M} {731, L5+}

\bibitem[\protect\citeauthoryear{{Mineshige}, {Takeuchi}  \&
  {Nishimori}}{{Mineshige} et~al.}{1994}]{Mineshige1994}
{Mineshige} S.,  {Takeuchi} M.,   {Nishimori} H.,  1994, \mn@doi [\apjl]
  {10.1086/187610}, \href {http://adsabs.harvard.edu/abs/1994ApJ...435L.125M}
  {435, L125}

\bibitem[\protect\citeauthoryear{{Miyamoto}, {Kitamoto}, {Iga}, {Negoro}  \&
  {Terada}}{{Miyamoto} et~al.}{1992}]{Miyamoto1992}
{Miyamoto} S.,  {Kitamoto} S.,  {Iga} S.,  {Negoro} H.,   {Terada} K.,  1992,
  \mn@doi [\apjl] {10.1086/186389}, \href
  {http://adsabs.harvard.edu/abs/1992ApJ...391L..21M} {391, L21}

\bibitem[\protect\citeauthoryear{{Morgan}, {Kochanek}, {Dai}, {Morgan}  \&
  {Falco}}{{Morgan} et~al.}{2008}]{sizepg1115}
{Morgan} C.~W.,  {Kochanek} C.~S.,  {Dai} X.,  {Morgan} N.~D.,   {Falco} E.~E.,
   2008, \mn@doi [\apj] {10.1086/592767}, \href
  {http://adsabs.harvard.edu/abs/2008ApJ...689..755M} {689, 755}

\bibitem[\protect\citeauthoryear{{Mosquera}, {Kochanek}, {Chen}, {Dai},
  {Blackburne}  \& {Chartas}}{{Mosquera} et~al.}{2013}]{size2237}
{Mosquera} A.~M.,  {Kochanek} C.~S.,  {Chen} B.,  {Dai} X.,  {Blackburne}
  J.~A.,   {Chartas} G.,  2013, ArXiv e-prints, \href
  {http://adsabs.harvard.edu/abs/2013arXiv1301.5009M} {}

\bibitem[\protect\citeauthoryear{{Namiki}, {Kawai}, {Kotani}  \&
  {Makishima}}{{Namiki} et~al.}{2003}]{Namiki03}
{Namiki} M.,  {Kawai} N.,  {Kotani} T.,   {Makishima} K.,  2003, \pasj, \href
  {http://adsabs.harvard.edu/abs/2003PASJ...55..281N} {55, 281}

\bibitem[\protect\citeauthoryear{{Narayan} \& {McClintock}}{{Narayan} \&
  {McClintock}}{2012}]{NarayanMcClintock2012}
{Narayan} R.,  {McClintock} J.~E.,  2012, \mn@doi [\mnras]
  {10.1111/j.1745-3933.2011.01181.x}, \href
  {http://adsabs.harvard.edu/abs/2012MNRAS.419L..69N} {419, L69}

\bibitem[\protect\citeauthoryear{{Narayan} \& {Yi}}{{Narayan} \&
  {Yi}}{1994}]{NarayanYi1994}
{Narayan} R.,  {Yi} I.,  1994, \mn@doi [\apjl] {10.1086/187381}, \href
  {http://adsabs.harvard.edu/abs/1994ApJ...428L..13N} {428, L13}

\bibitem[\protect\citeauthoryear{{Nayakshin} \& {Kallman}}{{Nayakshin} \&
  {Kallman}}{2001}]{NayakshinKallman2001}
{Nayakshin} S.,  {Kallman} T.~R.,  2001, \mn@doi [\apj] {10.1086/318250}, \href
  {http://adsabs.harvard.edu/abs/2001ApJ...546..406N} {546, 406}

\bibitem[\protect\citeauthoryear{{Negoro}, {Miyamoto}  \& {Kitamoto}}{{Negoro}
  et~al.}{1994}]{Negoro94}
{Negoro} H.,  {Miyamoto} S.,   {Kitamoto} S.,  1994, \mn@doi [\apjl]
  {10.1086/187253}, \href {http://adsabs.harvard.edu/abs/1994ApJ...423L.127N}
  {423, L127}

\bibitem[\protect\citeauthoryear{{Negoro}, {Kitamoto}, {Takeuchi}  \&
  {Mineshige}}{{Negoro} et~al.}{1995}]{Negoro95}
{Negoro} H.,  {Kitamoto} S.,  {Takeuchi} M.,   {Mineshige} S.,  1995, \mn@doi
  [\apjl] {10.1086/309704}, \href
  {http://adsabs.harvard.edu/abs/1995ApJ...452L..49N} {452, L49}

\bibitem[\protect\citeauthoryear{{Neilsen} \& {Lee}}{{Neilsen} \&
  {Lee}}{2009}]{NeilsenLee2009}
{Neilsen} J.,  {Lee} J.~C.,  2009, \mn@doi [\nat] {10.1038/nature07680}, \href
  {http://adsabs.harvard.edu/abs/2009Natur.458..481N} {458, 481}

\bibitem[\protect\citeauthoryear{{Neilsen}, {Remillard}  \& {Lee}}{{Neilsen}
  et~al.}{2011}]{NeilsenRemillardLee2011}
{Neilsen} J.,  {Remillard} R.~A.,   {Lee} J.~C.,  2011, \mn@doi [\apj]
  {10.1088/0004-637X/737/2/69}, \href
  {http://adsabs.harvard.edu/abs/2011ApJ...737...69N} {737, 69}

\bibitem[\protect\citeauthoryear{{Noble}, {Krolik}  \& {Hawley}}{{Noble}
  et~al.}{2009}]{NobleKrolik09}
{Noble} S.~C.,  {Krolik} J.~H.,   {Hawley} J.~F.,  2009, \mn@doi [\apj]
  {10.1088/0004-637X/692/1/411}, \href
  {http://adsabs.harvard.edu/abs/2009ApJ...692..411N} {692, 411}

\bibitem[\protect\citeauthoryear{{Noda}, {Makishima}, {Uehara}, {Yamada}  \&
  {Nakazawa}}{{Noda} et~al.}{2011a}]{Noda2011a}
{Noda} H.,  {Makishima} K.,  {Uehara} Y.,  {Yamada} S.,   {Nakazawa} K.,
  2011a, \pasj, \href {http://adsabs.harvard.edu/abs/2011PASJ...63..449N} {63,
  449}

\bibitem[\protect\citeauthoryear{{Noda}, {Makishima}, {Yamada}, {Torii},
  {Sakurai}  \& {Nakazawa}}{{Noda} et~al.}{2011b}]{Noda2011b}
{Noda} H.,  {Makishima} K.,  {Yamada} S.,  {Torii} S.,  {Sakurai} S.,
  {Nakazawa} K.,  2011b, \pasj, \href
  {http://adsabs.harvard.edu/abs/2011PASJ...63S.925N} {63, 925}

\bibitem[\protect\citeauthoryear{{Nomura}, {Ohsuga}, {Wada}, {Susa}  \&
  {Misawa}}{{Nomura} et~al.}{2013}]{2013PASJ...65...40N}
{Nomura} M.,  {Ohsuga} K.,  {Wada} K.,  {Susa} H.,   {Misawa} T.,  2013,
  \mn@doi [\pasj] {10.1093/pasj/65.2.40}, \href
  {http://adsabs.harvard.edu/abs/2013PASJ...65...40N} {65, 40}

\bibitem[\protect\citeauthoryear{{Nowak}, {Wilms}  \& {Dove}}{{Nowak}
  et~al.}{1999}]{Nowak1999gx}
{Nowak} M.~A.,  {Wilms} J.,   {Dove} J.~B.,  1999, \mn@doi [\apj]
  {10.1086/307189}, \href {http://adsabs.harvard.edu/abs/1999ApJ...517..355N}
  {517, 355}

\bibitem[\protect\citeauthoryear{{O'Neill}, {Reynolds}, {Miller}  \&
  {Sorathia}}{{O'Neill} et~al.}{2011}]{ONeillreynolds2011}
{O'Neill} S.~M.,  {Reynolds} C.~S.,  {Miller} M.~C.,   {Sorathia} K.~A.,  2011,
  \mn@doi [\apj] {10.1088/0004-637X/736/2/107}, \href
  {http://adsabs.harvard.edu/abs/2011ApJ...736..107O} {736, 107}

\bibitem[\protect\citeauthoryear{{Ohsuga} \& {Mineshige}}{{Ohsuga} \&
  {Mineshige}}{2011}]{OhsugaMineshige11}
{Ohsuga} K.,  {Mineshige} S.,  2011, \mn@doi [\apj]
  {10.1088/0004-637X/736/1/2}, \href
  {http://adsabs.harvard.edu/abs/2011ApJ...736....2O} {736, 2}

\bibitem[\protect\citeauthoryear{{Ohsuga}, {Mori}, {Nakamoto}  \&
  {Mineshige}}{{Ohsuga} et~al.}{2005}]{Ohsuga2005}
{Ohsuga} K.,  {Mori} M.,  {Nakamoto} T.,   {Mineshige} S.,  2005, \mn@doi
  [\apj] {10.1086/430728}, \href
  {http://adsabs.harvard.edu/abs/2005ApJ...628..368O} {628, 368}

\bibitem[\protect\citeauthoryear{{Ohsuga}, {Mineshige}, {Mori}  \&
  {Kato}}{{Ohsuga} et~al.}{2009}]{Ohsuga2009}
{Ohsuga} K.,  {Mineshige} S.,  {Mori} M.,   {Kato} Y.,  2009, \pasj, \href
  {http://adsabs.harvard.edu/abs/2009PASJ...61L...7O} {61, L7}

\bibitem[\protect\citeauthoryear{{Pakull} \& {Mirioni}}{{Pakull} \&
  {Mirioni}}{2002}]{Pakull2}
{Pakull} M.~W.,  {Mirioni} L.,  2002, ArXiv Astrophysics e-prints, \href
  {http://adsabs.harvard.edu/abs/2002astro.ph..2488P} {}

\bibitem[\protect\citeauthoryear{{Papitto}, {Di Salvo}, {D'A{\`i}}, {Iaria},
  {Burderi}, {Riggio}, {Menna}  \& {Robba}}{{Papitto} et~al.}{2009}]{Papitto09}
{Papitto} A.,  {Di Salvo} T.,  {D'A{\`i}} A.,  {Iaria} R.,  {Burderi} L.,
  {Riggio} A.,  {Menna} M.~T.,   {Robba} N.~R.,  2009, \mn@doi [\aap]
  {10.1051/0004-6361:200811401}, \href
  {http://adsabs.harvard.edu/abs/2009A%26A...493L..39P} {493, L39}

\bibitem[\protect\citeauthoryear{{Pereyra}, {Kallman}  \& {Blondin}}{{Pereyra}
  et~al.}{2000}]{Pereyra00}
{Pereyra} N.~A.,  {Kallman} T.~R.,   {Blondin} J.~M.,  2000, \mn@doi [\apj]
  {10.1086/308527}, \href {http://adsabs.harvard.edu/abs/2000ApJ...532..563P}
  {532, 563}

\bibitem[\protect\citeauthoryear{{Ponti}, {Fender}, {Begelman}, {Dunn},
  {Neilsen}  \& {Coriat}}{{Ponti} et~al.}{2012}]{Ponti2012diskjet}
{Ponti} G.,  {Fender} R.~P.,  {Begelman} M.~C.,  {Dunn} R.~J.~H.,  {Neilsen}
  J.,   {Coriat} M.,  2012, \mn@doi [\mnras]
  {10.1111/j.1745-3933.2012.01224.x}, \href
  {http://adsabs.harvard.edu/abs/2012MNRAS.422L..11P} {422, L11}

\bibitem[\protect\citeauthoryear{{Pottschmidt} et~al.,}{{Pottschmidt}
  et~al.}{2003}]{Pottschmidt2003cyg}
{Pottschmidt} K.,  et~al., 2003, \mn@doi [\aap] {10.1051/0004-6361:20030906},
  \href {http://adsabs.harvard.edu/abs/2003A%26A...407.1039P} {407, 1039}

\bibitem[\protect\citeauthoryear{{Poutanen}}{{Poutanen}}{2001}]{Poutanen2001}
{Poutanen} J.,  2001, \mn@doi [Advances in Space Research]
  {10.1016/S0273-1177(01)00406-9}, \href
  {http://adsabs.harvard.edu/abs/2001AdSpR..28..267P} {28, 267}

\bibitem[\protect\citeauthoryear{{Proga}}{{Proga}}{2003}]{proga03}
{Proga} D.,  2003, \mn@doi [\apj] {10.1086/345897}, \href
  {http://adsabs.harvard.edu/abs/2003ApJ...585..406P} {585, 406}

\bibitem[\protect\citeauthoryear{{Proga} \& {Kallman}}{{Proga} \&
  {Kallman}}{2002}]{proga02}
{Proga} D.,  {Kallman} T.~R.,  2002, \mn@doi [\apj] {10.1086/324534}, \href
  {http://adsabs.harvard.edu/abs/2002ApJ...565..455P} {565, 455}

\bibitem[\protect\citeauthoryear{{Proga}, {Stone}  \& {Kallman}}{{Proga}
  et~al.}{2000}]{proga00}
{Proga} D.,  {Stone} J.~M.,   {Kallman} T.~R.,  2000, \mn@doi [\apj]
  {10.1086/317154}, \href {http://adsabs.harvard.edu/abs/2000ApJ...543..686P}
  {543, 686}

\bibitem[\protect\citeauthoryear{{Quataert} \& {Gruzinov}}{{Quataert} \&
  {Gruzinov}}{2000}]{Quataert00CDAF}
{Quataert} E.,  {Gruzinov} A.,  2000, \mn@doi [\apj] {10.1086/309267}, \href
  {http://adsabs.harvard.edu/abs/2000ApJ...539..809Q} {539, 809}

\bibitem[\protect\citeauthoryear{{Reis}, {Fabian}, {Ross}, {Miniutti}, {Miller}
   \& {Reynolds}}{{Reis} et~al.}{2008}]{reisgx}
{Reis} R.~C.,  {Fabian} A.~C.,  {Ross} R.~R.,  {Miniutti} G.,  {Miller} J.~M.,
   {Reynolds} C.,  2008, \mn@doi [\mnras] {10.1111/j.1365-2966.2008.13358.x},
  \href {http://adsabs.harvard.edu/abs/2008MNRAS.387.1489R} {387, 1489}

\bibitem[\protect\citeauthoryear{{Reis}, {Miller}  \& {Fabian}}{{Reis}
  et~al.}{2009a}]{reisj1118}
{Reis} R.~C.,  {Miller} J.~M.,   {Fabian} A.~C.,  2009a, \mn@doi [\mnras]
  {10.1111/j.1745-3933.2009.00640.x}, \href
  {http://adsabs.harvard.edu/abs/2009MNRAS.395L..52R} {395, L52}

\bibitem[\protect\citeauthoryear{{Reis}, {Fabian}, {Ross}  \& {Miller}}{{Reis}
  et~al.}{2009b}]{reisspin}
{Reis} R.~C.,  {Fabian} A.~C.,  {Ross} R.~R.,   {Miller} J.~M.,  2009b, \mn@doi
  [\mnras] {10.1111/j.1365-2966.2009.14622.x}, \href
  {http://adsabs.harvard.edu/abs/2009MNRAS.395.1257R} {395, 1257}

\bibitem[\protect\citeauthoryear{{Reis}, {Fabian}  \& {Miller}}{{Reis}
  et~al.}{2010}]{reislhs}
{Reis} R.~C.,  {Fabian} A.~C.,   {Miller} J.~M.,  2010, \mn@doi [\mnras]
  {10.1111/j.1365-2966.2009.15976.x}, \href
  {http://adsabs.harvard.edu/abs/2010MNRAS.402..836R} {402, 836}

\bibitem[\protect\citeauthoryear{{Reis} et~al.,}{{Reis}
  et~al.}{2011}]{reis1752}
{Reis} R.~C.,  et~al., 2011, \mn@doi [\mnras]
  {10.1111/j.1365-2966.2010.17628.x}, \href
  {http://adsabs.harvard.edu/abs/2011MNRAS.410.2497R} {410, 2497}

\bibitem[\protect\citeauthoryear{{Reis}, {Miller}, {Reynolds}, {G{\"u}ltekin},
  {Maitra}, {King}  \& {Strohmayer}}{{Reis} et~al.}{2012a}]{Reis2012qpo}
{Reis} R.~C.,  {Miller} J.~M.,  {Reynolds} M.~T.,  {G{\"u}ltekin} K.,  {Maitra}
  D.,  {King} A.~L.,   {Strohmayer} T.~E.,  2012a, \mn@doi [Science]
  {10.1126/science.1223940}, \href
  {http://adsabs.harvard.edu/abs/2012Sci...337..949R} {337, 949}

\bibitem[\protect\citeauthoryear{{Reis}, {Miller}, {Reynolds}, {Fabian}  \&
  {Walton}}{{Reis} et~al.}{2012b}]{reismaxi}
{Reis} R.~C.,  {Miller} J.~M.,  {Reynolds} M.~T.,  {Fabian} A.~C.,   {Walton}
  D.~J.,  2012b, \mn@doi [\apj] {10.1088/0004-637X/751/1/34}, \href
  {http://adsabs.harvard.edu/abs/2012ApJ...751...34R} {751, 34}

\bibitem[\protect\citeauthoryear{{Reynolds} \& {Fabian}}{{Reynolds} \&
  {Fabian}}{2008}]{reynoldsfabian08}
{Reynolds} C.~S.,  {Fabian} A.~C.,  2008, \mn@doi [\apj] {10.1086/527344},
  \href {http://adsabs.harvard.edu/abs/2008ApJ...675.1048R} {675, 1048}

\bibitem[\protect\citeauthoryear{{Reynolds} \& {Nowak}}{{Reynolds} \&
  {Nowak}}{2003}]{reynoldsnowak03}
{Reynolds} C.~S.,  {Nowak} M.~A.,  2003, \mn@doi [\physrep]
  {10.1016/S0370-1573(02)00584-7}, \href
  {http://adsabs.harvard.edu/abs/2003PhR...377..389R} {377, 389}

\bibitem[\protect\citeauthoryear{{Rodriguez}, {Corbel}, {Hannikainen},
  {Belloni}, {Paizis}  \& {Vilhu}}{{Rodriguez} et~al.}{2004}]{Rodriguez04}
{Rodriguez} J.,  {Corbel} S.,  {Hannikainen} D.~C.,  {Belloni} T.,  {Paizis}
  A.,   {Vilhu} O.,  2004, \mn@doi [\apj] {10.1086/423978}, \href
  {http://adsabs.harvard.edu/abs/2004ApJ...615..416R} {615, 416}

\bibitem[\protect\citeauthoryear{{Ross} \& {Fabian}}{{Ross} \&
  {Fabian}}{2007}]{refbhb}
{Ross} R.~R.,  {Fabian} A.~C.,  2007, \mn@doi [\mnras]
  {10.1111/j.1365-2966.2007.12339.x}, \href
  {http://adsabs.harvard.edu/abs/2007MNRAS.381.1697R} {381, 1697}

\bibitem[\protect\citeauthoryear{{Russell}, {Miller-Jones}, {Maccarone},
  {Yang}, {Fender}  \& {Lewis}}{{Russell} et~al.}{2011}]{Russell2011softstate}
{Russell} D.~M.,  {Miller-Jones} J.~C.~A.,  {Maccarone} T.~J.,  {Yang} Y.~J.,
  {Fender} R.~P.,   {Lewis} F.,  2011, \mn@doi [\apjl]
  {10.1088/2041-8205/739/1/L19}, \href
  {http://adsabs.harvard.edu/abs/2011ApJ...739L..19R} {739, L19}

\bibitem[\protect\citeauthoryear{{Russell}, {Gallo}  \& {Fender}}{{Russell}
  et~al.}{2013}]{RussellGallofender2013jets}
{Russell} D.~M.,  {Gallo} E.,   {Fender} R.~P.,  2013, ArXiv e-prints, \href
  {http://adsabs.harvard.edu/abs/2013arXiv1301.6771R} {}

\bibitem[\protect\citeauthoryear{{Schartel}}{{Schartel}}{2011}]{Schartel2011}
{Schartel} N.,  2011, Astronomische Nachrichten, \href
  {http://adsabs.harvard.edu/abs/2011AN....332..323S} {332, 323}

\bibitem[\protect\citeauthoryear{{Schnittman} \& {Krolik}}{{Schnittman} \&
  {Krolik}}{2009}]{SchnittmanKrolik09polarization}
{Schnittman} J.~D.,  {Krolik} J.~H.,  2009, \mn@doi [\apj]
  {10.1088/0004-637X/701/2/1175}, \href
  {http://adsabs.harvard.edu/abs/2009ApJ...701.1175S} {701, 1175}

\bibitem[\protect\citeauthoryear{{Schnittman}, {Homan}  \&
  {Miller}}{{Schnittman} et~al.}{2006}]{SchnittmanHomanMiller2006}
{Schnittman} J.~D.,  {Homan} J.,   {Miller} J.~M.,  2006, \mn@doi [\apj]
  {10.1086/500923}, \href {http://adsabs.harvard.edu/abs/2006ApJ...642..420S}
  {642, 420}

\bibitem[\protect\citeauthoryear{{Schnittman}, {Krolik}  \&
  {Noble}}{{Schnittman} et~al.}{2012}]{Schnittman2012states}
{Schnittman} J.~D.,  {Krolik} J.~H.,   {Noble} S.~C.,  2012, ArXiv e-prints,
  \href {http://adsabs.harvard.edu/abs/2012arXiv1207.2693S} {}

\bibitem[\protect\citeauthoryear{{Shafee}, {McClintock}, {Narayan}, {Davis},
  {Li}  \& {Remillard}}{{Shafee} et~al.}{2006}]{shafee06}
{Shafee} R.,  {McClintock} J.~E.,  {Narayan} R.,  {Davis} S.~W.,  {Li} L.,
  {Remillard} R.~A.,  2006, \mn@doi [\apjl] {10.1086/498938}, \href
  {http://adsabs.harvard.edu/abs/2006ApJ...636L.113S} {636, L113}

\bibitem[\protect\citeauthoryear{{Shafee}, {McKinney}, {Narayan},
  {Tchekhovskoy}, {Gammie}  \& {McClintock}}{{Shafee}
  et~al.}{2008}]{Shafee2008}
{Shafee} R.,  {McKinney} J.~C.,  {Narayan} R.,  {Tchekhovskoy} A.,  {Gammie}
  C.~F.,   {McClintock} J.~E.,  2008, \mn@doi [\apjl] {10.1086/593148}, \href
  {http://adsabs.harvard.edu/abs/2008ApJ...687L..25S} {687, L25}

\bibitem[\protect\citeauthoryear{{Shakura} \& {Sunyaev}}{{Shakura} \&
  {Sunyaev}}{1973}]{shakuraSunyaev73}
{Shakura} N.~I.,  {Sunyaev} R.~A.,  1973, \aap, \href
  {http://adsabs.harvard.edu/abs/1973A%26A....24..337S} {24, 337}

\bibitem[\protect\citeauthoryear{{Shidatsu} et~al.,}{{Shidatsu}
  et~al.}{2011}]{Shidatsu2011lhs}
{Shidatsu} M.,  et~al., 2011, \pasj, \href
  {http://adsabs.harvard.edu/abs/2011PASJ...63S.785S} {63, 785}

\bibitem[\protect\citeauthoryear{{Shimura} \& {Takahara}}{{Shimura} \&
  {Takahara}}{1995}]{ShimuraTakahara1995}
{Shimura} T.,  {Takahara} F.,  1995, \mn@doi [\apj] {10.1086/175740}, \href
  {http://adsabs.harvard.edu/abs/1995ApJ...445..780S} {445, 780}

\bibitem[\protect\citeauthoryear{{Sidoli}, {Oosterbroek}, {Parmar}, {Lumb}  \&
  {Erd}}{{Sidoli} et~al.}{2001}]{sidoli01}
{Sidoli} L.,  {Oosterbroek} T.,  {Parmar} A.~N.,  {Lumb} D.,   {Erd} C.,  2001,
  \mn@doi [\aap] {10.1051/0004-6361:20011322}, \href
  {http://adsabs.harvard.edu/abs/2001A%26A...379..540S} {379, 540}

\bibitem[\protect\citeauthoryear{{Sidoli}, {Parmar}, {Oosterbroek}  \&
  {Lumb}}{{Sidoli} et~al.}{2002}]{sidoli02}
{Sidoli} L.,  {Parmar} A.~N.,  {Oosterbroek} T.,   {Lumb} D.,  2002, \mn@doi
  [\aap] {10.1051/0004-6361:20020192}, \href
  {http://adsabs.harvard.edu/abs/2002A%26A...385..940S} {385, 940}

\bibitem[\protect\citeauthoryear{{Sikora} \& {Begelman}}{{Sikora} \&
  {Begelman}}{2013}]{2013ApJ...764L..24S}
{Sikora} M.,  {Begelman} M.~C.,  2013, \mn@doi [\apjl]
  {10.1088/2041-8205/764/2/L24}, \href
  {http://adsabs.harvard.edu/abs/2013ApJ...764L..24S} {764, L24}

\bibitem[\protect\citeauthoryear{{Sobolewska} \& {{\.Z}ycki}}{{Sobolewska} \&
  {{\.Z}ycki}}{2006}]{Sobolewska2006}
{Sobolewska} M.~A.,  {{\.Z}ycki} P.~T.,  2006, \mn@doi [\mnras]
  {10.1111/j.1365-2966.2006.10489.x}, \href
  {http://adsabs.harvard.edu/abs/2006MNRAS.370..405S} {370, 405}

\bibitem[\protect\citeauthoryear{{Steiner} et~al.,}{{Steiner}
  et~al.}{2011}]{Steiner2011}
{Steiner} J.~F.,  et~al., 2011, \mn@doi [\mnras]
  {10.1111/j.1365-2966.2011.19089.x}, \href
  {http://adsabs.harvard.edu/abs/2011MNRAS.tmp.1036S} {pp 1036--+}

\bibitem[\protect\citeauthoryear{{Steiner}, {McClintock}  \&
  {Narayan}}{{Steiner} et~al.}{2013}]{Steiner2013spinjet}
{Steiner} J.~F.,  {McClintock} J.~E.,   {Narayan} R.,  2013, \mn@doi [\apj]
  {10.1088/0004-637X/762/2/104}, \href
  {http://adsabs.harvard.edu/abs/2013ApJ...762..104S} {762, 104}

\bibitem[\protect\citeauthoryear{{Stella} \& {Vietri}}{{Stella} \&
  {Vietri}}{1998}]{StellaVietri98}
{Stella} L.,  {Vietri} M.,  1998, \mn@doi [\apjl] {10.1086/311075}, \href
  {http://adsabs.harvard.edu/abs/1998ApJ...492L..59S} {492, L59}

\bibitem[\protect\citeauthoryear{{Strohmayer} \& {Mushotzky}}{{Strohmayer} \&
  {Mushotzky}}{2003}]{ulxqpo22003}
{Strohmayer} T.~E.,  {Mushotzky} R.~F.,  2003, \mn@doi [\apjl]
  {10.1086/374732}, \href {http://adsabs.harvard.edu/abs/2003ApJ...586L..61S}
  {586, L61}

\bibitem[\protect\citeauthoryear{{Strohmayer} \& {Mushotzky}}{{Strohmayer} \&
  {Mushotzky}}{2009}]{StrohmayerMushotzky2009}
{Strohmayer} T.~E.,  {Mushotzky} R.~F.,  2009, \mn@doi [\apj]
  {10.1088/0004-637X/703/2/1386}, \href
  {http://adsabs.harvard.edu/abs/2009ApJ...703.1386S} {703, 1386}

\bibitem[\protect\citeauthoryear{{Sutton}, {Roberts}, {Walton}, {Gladstone}  \&
  {Scott}}{{Sutton} et~al.}{2012}]{Sutton2012ulx}
{Sutton} A.~D.,  {Roberts} T.~P.,  {Walton} D.~J.,  {Gladstone} J.~C.,
  {Scott} A.~E.,  2012, \mn@doi [\mnras] {10.1111/j.1365-2966.2012.20944.x},
  \href {http://adsabs.harvard.edu/abs/2012MNRAS.423.1154S} {423, 1154}

\bibitem[\protect\citeauthoryear{{Taam}, {Liu}, {Meyer}  \&
  {Meyer-Hofmeister}}{{Taam} et~al.}{2008}]{Taametal2008}
{Taam} R.~E.,  {Liu} B.~F.,  {Meyer} F.,   {Meyer-Hofmeister} E.,  2008,
  \mn@doi [\apj] {10.1086/591901}, \href
  {http://adsabs.harvard.edu/abs/2008ApJ...688..527T} {688, 527}

\bibitem[\protect\citeauthoryear{{Takahashi} et~al.,}{{Takahashi}
  et~al.}{2008}]{takahashi165508}
{Takahashi} H.,  et~al., 2008, \pasj, \href
  {http://adsabs.harvard.edu/abs/2008PASJ...60S..69T} {60, 69}

\bibitem[\protect\citeauthoryear{{Takahashi} et~al.,}{{Takahashi}
  et~al.}{2012}]{astroh2012}
{Takahashi} T.,  et~al., 2012, in Society of Photo-Optical Instrumentation
  Engineers (SPIE) Conference Series. , \mn@eprint {arXiv} {1210.4378},
  \mn@doi{10.1117/12.926190}

\bibitem[\protect\citeauthoryear{{Takeuchi}, {Ohsuga}  \&
  {Mineshige}}{{Takeuchi} et~al.}{2010}]{Takeuchi10}
{Takeuchi} S.,  {Ohsuga} K.,   {Mineshige} S.,  2010, \pasj, \href
  {http://adsabs.harvard.edu/abs/2010PASJ...62L..43T} {62, L43}

\bibitem[\protect\citeauthoryear{{Takeuchi}, {Ohsuga}  \&
  {Mineshige}}{{Takeuchi} et~al.}{2013}]{Takeuchi13}
{Takeuchi} S.,  {Ohsuga} K.,   {Mineshige} S.,  2013, \pasj, \href
  {http://adsabs.harvard.edu/abs/2013PASJ...65...88T} {65, 88}

\bibitem[\protect\citeauthoryear{{Tanaka} et~al.,}{{Tanaka}
  et~al.}{1995}]{tanaka1995}
{Tanaka} Y.,  et~al., 1995, \mn@doi [\nat] {10.1038/375659a0}, \href
  {http://adsabs.harvard.edu/abs/1995Natur.375..659T} {375, 659}

\bibitem[\protect\citeauthoryear{{Tombesi}, {Cappi}, {Reeves}, {Palumbo},
  {Yaqoob}, {Braito}  \& {Dadina}}{{Tombesi} et~al.}{2010a}]{Tombesi2010}
{Tombesi} F.,  {Cappi} M.,  {Reeves} J.~N.,  {Palumbo} G.~G.~C.,  {Yaqoob} T.,
  {Braito} V.,   {Dadina} M.,  2010a, \mn@doi [\aap]
  {10.1051/0004-6361/200913440}, \href
  {http://adsabs.harvard.edu/abs/2010A%26A...521A..57T} {521, A57}

\bibitem[\protect\citeauthoryear{{Tombesi}, {Sambruna}, {Reeves}, {Braito},
  {Ballo}, {Gofford}, {Cappi}  \& {Mushotzky}}{{Tombesi}
  et~al.}{2010b}]{Tombesi2010b}
{Tombesi} F.,  {Sambruna} R.~M.,  {Reeves} J.~N.,  {Braito} V.,  {Ballo} L.,
  {Gofford} J.,  {Cappi} M.,   {Mushotzky} R.~F.,  2010b, \mn@doi [\apj]
  {10.1088/0004-637X/719/1/700}, \href
  {http://adsabs.harvard.edu/abs/2010ApJ...719..700T} {719, 700}

\bibitem[\protect\citeauthoryear{{Tomsick}, {Yamaoka}, {Corbel}, {Kaaret},
  {Kalemci}  \& {Migliari}}{{Tomsick} et~al.}{2009}]{tomsick09gx}
{Tomsick} J.~A.,  {Yamaoka} K.,  {Corbel} S.,  {Kaaret} P.,  {Kalemci} E.,
  {Migliari} S.,  2009, \mn@doi [\apjl] {10.1088/0004-637X/707/1/L87}, \href
  {http://adsabs.harvard.edu/abs/2009ApJ...707L..87T} {707, L87}

\bibitem[\protect\citeauthoryear{{Torii} et~al.,}{{Torii}
  et~al.}{2011}]{Torii2011}
{Torii} S.,  et~al., 2011, \pasj, \href
  {http://adsabs.harvard.edu/abs/2011PASJ...63S.771T} {63, 771}

\bibitem[\protect\citeauthoryear{{Turner} et~al.,}{{Turner}
  et~al.}{1989}]{ginga}
{Turner} M.~J.~L.,  et~al., 1989, \pasj, \href
  {http://adsabs.harvard.edu/abs/1989PASJ...41..345T} {41, 345}

\bibitem[\protect\citeauthoryear{{Ueda}, {Inoue}, {Tanaka}, {Ebisawa},
  {Nagase}, {Kotani}  \& {Gehrels}}{{Ueda} et~al.}{1998}]{Ueda1998j1655}
{Ueda} Y.,  {Inoue} H.,  {Tanaka} Y.,  {Ebisawa} K.,  {Nagase} F.,  {Kotani}
  T.,   {Gehrels} N.,  1998, \mn@doi [\apj] {10.1086/305063}, \href
  {http://adsabs.harvard.edu/abs/1998ApJ...492..782U} {492, 782}

\bibitem[\protect\citeauthoryear{{Ueda}, {Asai}, {Yamaoka}, {Dotani}  \&
  {Inoue}}{{Ueda} et~al.}{2001}]{ueda01}
{Ueda} Y.,  {Asai} K.,  {Yamaoka} K.,  {Dotani} T.,   {Inoue} H.,  2001,
  \mn@doi [\apjl] {10.1086/323007}, \href
  {http://adsabs.harvard.edu/abs/2001ApJ...556L..87U} {556, L87}

\bibitem[\protect\citeauthoryear{{Ueda}, {Murakami}, {Yamaoka}, {Dotani}  \&
  {Ebisawa}}{{Ueda} et~al.}{2004}]{ueda04}
{Ueda} Y.,  {Murakami} H.,  {Yamaoka} K.,  {Dotani} T.,   {Ebisawa} K.,  2004,
  \mn@doi [\apj] {10.1086/420973}, \href
  {http://adsabs.harvard.edu/abs/2004ApJ...609..325U} {609, 325}

\bibitem[\protect\citeauthoryear{{Ueda}, {Yamaoka}  \& {Remillard}}{{Ueda}
  et~al.}{2009}]{ueda09}
{Ueda} Y.,  {Yamaoka} K.,   {Remillard} R.,  2009, \mn@doi [\apj]
  {10.1088/0004-637X/695/2/888}, \href
  {http://adsabs.harvard.edu/abs/2009ApJ...695..888U} {695, 888}

\bibitem[\protect\citeauthoryear{{Vaughan} \& {Fabian}}{{Vaughan} \&
  {Fabian}}{2004}]{Vaug04}
{Vaughan} S.,  {Fabian} A.~C.,  2004, \mn@doi [\mnras]
  {10.1111/j.1365-2966.2004.07456.x}, \href
  {http://adsabs.harvard.edu/abs/2004MNRAS.348.1415V} {348, 1415}

\bibitem[\protect\citeauthoryear{{Vierdayanti}, {Mineshige}, {Ebisawa}  \&
  {Kawaguchi}}{{Vierdayanti} et~al.}{2006a}]{2006PASJ...58..915V}
{Vierdayanti} K.,  {Mineshige} S.,  {Ebisawa} K.,   {Kawaguchi} T.,  2006a,
  \mn@doi [\pasj] {10.1093/pasj/58.5.915}, \href
  {http://adsabs.harvard.edu/abs/2006PASJ...58..915V} {58, 915}

\bibitem[\protect\citeauthoryear{{Vierdayanti}, {Mineshige}, {Ebisawa}  \&
  {Kawaguchi}}{{Vierdayanti} et~al.}{2006b}]{Vierdayanti06}
{Vierdayanti} K.,  {Mineshige} S.,  {Ebisawa} K.,   {Kawaguchi} T.,  2006b,
  \pasj, \href {http://adsabs.harvard.edu/abs/2006PASJ...58..915V} {58, 915}

\bibitem[\protect\citeauthoryear{{Vierdayanti}, {Done}, {Roberts}  \&
  {Mineshige}}{{Vierdayanti} et~al.}{2010}]{Vierdayanti2010}
{Vierdayanti} K.,  {Done} C.,  {Roberts} T.~P.,   {Mineshige} S.,  2010,
  \mn@doi [\mnras] {10.1111/j.1365-2966.2009.16210.x}, \href
  {http://adsabs.harvard.edu/abs/2010MNRAS.403.1206V} {403, 1206}

\bibitem[\protect\citeauthoryear{{Volonteri}, {Sikora}, {Lasota}  \&
  {Merloni}}{{Volonteri} et~al.}{2013}]{2013ApJ...775...94V}
{Volonteri} M.,  {Sikora} M.,  {Lasota} J.-P.,   {Merloni} A.,  2013, \mn@doi
  [\apj] {10.1088/0004-637X/775/2/94}, \href
  {http://adsabs.harvard.edu/abs/2013ApJ...775...94V} {775, 94}

\bibitem[\protect\citeauthoryear{{Walton}, {Roberts}, {Mateos}  \&
  {Heard}}{{Walton} et~al.}{2011}]{Walton2011ulx1}
{Walton} D.~J.,  {Roberts} T.~P.,  {Mateos} S.,   {Heard} V.,  2011, \mn@doi
  [\mnras] {10.1111/j.1365-2966.2011.19154.x}, \href
  {http://adsabs.harvard.edu/abs/2011MNRAS.416.1844W} {416, 1844}

\bibitem[\protect\citeauthoryear{{Walton}, {Miller}, {Reis}  \&
  {Fabian}}{{Walton} et~al.}{2012}]{WaltonMillerReis2012ULX}
{Walton} D.~J.,  {Miller} J.~M.,  {Reis} R.~C.,   {Fabian} A.~C.,  2012,
  \mn@doi [\mnras] {10.1111/j.1365-2966.2012.21727.x}, \href
  {http://adsabs.harvard.edu/abs/2012MNRAS.426..473W} {426, 473}

\bibitem[\protect\citeauthoryear{{Watarai}, {Mizuno}  \& {Mineshige}}{{Watarai}
  et~al.}{2001}]{Watarai01}
{Watarai} K.-y.,  {Mizuno} T.,   {Mineshige} S.,  2001, \mn@doi [\apjl]
  {10.1086/319125}, \href {http://adsabs.harvard.edu/abs/2001ApJ...549L..77W}
  {549, L77}

\bibitem[\protect\citeauthoryear{{Webb}, {Barret}, {Godet}, {Servillat},
  {Farrell}  \& {Oates}}{{Webb} et~al.}{2010}]{Webb2010}
{Webb} N.~A.,  {Barret} D.,  {Godet} O.,  {Servillat} M.,  {Farrell} S.~A.,
  {Oates} S.~R.,  2010, \mn@doi [\apjl] {10.1088/2041-8205/712/1/L107}, \href
  {http://adsabs.harvard.edu/abs/2010ApJ...712L.107W} {712, L107}

\bibitem[\protect\citeauthoryear{{Wilkins} \& {Fabian}}{{Wilkins} \&
  {Fabian}}{2012}]{wilkins2012}
{Wilkins} D.~R.,  {Fabian} A.~C.,  2012, \mn@doi [\mnras]
  {10.1111/j.1365-2966.2012.21308.x}, \href
  {http://adsabs.harvard.edu/abs/2012MNRAS.424.1284W} {424, 1284}

\bibitem[\protect\citeauthoryear{{Wilkinson} \& {Uttley}}{{Wilkinson} \&
  {Uttley}}{2009}]{wilkinsonuttley09}
{Wilkinson} T.,  {Uttley} P.,  2009, \mn@doi [\mnras]
  {10.1111/j.1365-2966.2009.15008.x}, \href
  {http://adsabs.harvard.edu/abs/2009MNRAS.397..666W} {397, 666}

\bibitem[\protect\citeauthoryear{{Woods}, {Klein}, {Castor}, {McKee}  \&
  {Bell}}{{Woods} et~al.}{1996}]{woods96}
{Woods} D.~T.,  {Klein} R.~I.,  {Castor} J.~I.,  {McKee} C.~F.,   {Bell} J.~B.,
   1996, \mn@doi [\apj] {10.1086/177101}, \href
  {http://adsabs.harvard.edu/abs/1996ApJ...461..767W} {461, 767}

\bibitem[\protect\citeauthoryear{{Yamada}, {Negoro}, {Torii}, {Noda},
  {Mineshige}  \& {Makishima}}{{Yamada} et~al.}{2013}]{2013ApJ...767L..34Y}
{Yamada} S.,  {Negoro} H.,  {Torii} S.,  {Noda} H.,  {Mineshige} S.,
  {Makishima} K.,  2013, \mn@doi [\apjl] {10.1088/2041-8205/767/2/L34}, \href
  {http://adsabs.harvard.edu/abs/2013ApJ...767L..34Y} {767, L34}

\bibitem[\protect\citeauthoryear{{Yamaoka}, {Ueda}, {Inoue}, {Nagase},
  {Ebisawa}, {Kotani}, {Tanaka}  \& {Zhang}}{{Yamaoka}
  et~al.}{2001}]{yamaoka01}
{Yamaoka} K.,  {Ueda} Y.,  {Inoue} H.,  {Nagase} F.,  {Ebisawa} K.,  {Kotani}
  T.,  {Tanaka} Y.,   {Zhang} S.~N.,  2001, \pasj, \href
  {http://adsabs.harvard.edu/abs/2001PASJ...53..179Y} {53, 179}

\bibitem[\protect\citeauthoryear{{Yamauchi}, {Kawai}  \& {Aoki}}{{Yamauchi}
  et~al.}{1994}]{Yamauchi94}
{Yamauchi} S.,  {Kawai} N.,   {Aoki} T.,  1994, \pasj, \href
  {http://adsabs.harvard.edu/abs/1994PASJ...46L.109Y} {46, L109}

\bibitem[\protect\citeauthoryear{{Yoshida}, {Isobe}, {Mineshige}, {Kubota},
  {Mizuno}  \& {Saitou}}{{Yoshida} et~al.}{2013}]{2013PASJ...65...48Y}
{Yoshida} T.,  {Isobe} N.,  {Mineshige} S.,  {Kubota} A.,  {Mizuno} T.,
  {Saitou} K.,  2013, \mn@doi [\pasj] {10.1093/pasj/65.2.48}, \href
  {http://adsabs.harvard.edu/abs/2013PASJ...65...48Y} {65, 48}

\bibitem[\protect\citeauthoryear{{Zdziarski}, {Sikora}, {Dubus}, {Yuan},
  {Cerutti}  \& {Ogorza{\l}ek}}{{Zdziarski} et~al.}{2012}]{Zdziarski2011}
{Zdziarski} A.~A.,  {Sikora} M.,  {Dubus} G.,  {Yuan} F.,  {Cerutti} B.,
  {Ogorza{\l}ek} A.,  2012, \mn@doi [\mnras]
  {10.1111/j.1365-2966.2012.20519.x}, \href
  {http://adsabs.harvard.edu/abs/2012MNRAS.421.2956Z} {421, 2956}

\bibitem[\protect\citeauthoryear{{Zoghbi}, {Fabian}, {Reynolds}  \&
  {Cackett}}{{Zoghbi} et~al.}{2012}]{Zoghbi2012lag}
{Zoghbi} A.,  {Fabian} A.~C.,  {Reynolds} C.~S.,   {Cackett} E.~M.,  2012,
  \mn@doi [\mnras] {10.1111/j.1365-2966.2012.20587.x}, \href
  {http://adsabs.harvard.edu/abs/2012MNRAS.422..129Z} {422, 129}

\bibitem[\protect\citeauthoryear{{van der Klis}}{{van der
  Klis}}{2006}]{vanderKlisbook}
{van der Klis} M.,  2006, {Rapid X-ray Variability}.
pp 39--112

\makeatother
\end{thebibliography}
\begin{multicols}{2}
{\footnotesize

\input{WP_05-06_BlackHoleSpinAccretion.bbl}

}
\end{multicols}

\end{document}